\documentclass{elsart}
\usepackage{ragged2e}
\usepackage{graphics}
\usepackage{graphicx}
\usepackage{epsfig}
\usepackage{amssymb}
\usepackage{lineno}
\usepackage{blindtext}
\usepackage{lineno}
\usepackage{lineno, blindtext}
\usepackage{mathtools}
\usepackage{xcolor}
\usepackage{hyperref}
\usepackage{microtype}
\usepackage{breakcites}
\usepackage{footnote}
\begin{document}
\begin{frontmatter}
\title{Classification of the mechanisms of wave energy dissipation in the nonlinear oscillations of coated and uncoated bubbles}
\author{A.J. Sojahrood \thanksref{a},\thanksref{b}} \footnote{Email: amin.jafarisojahrood@ryerson.ca},
\author{H. Haghi \thanksref{a},\thanksref{b}}
\author{R. Karshfian \thanksref{a},\thanksref{b}}
\author{and  M. C. Kolios \thanksref{a},\thanksref{b}}
\address[a]{Department of Physics, Ryerson University, Toronto, Canada}
\address[b]{Institute for Biomedical Engineering, Science and Technology (IBEST) a partnership between Ryerson University and St. Mike's Hospital, Toronto, Ontario, Canada}
\begin{abstract}
Acoustic waves are dissipated when they pass through bubbly media. Dissipation by bubbles takes place through thermal damping (Td), radiation damping (Rd) and damping due to the friction of the liquid (Ld) and friction of the coating (Cd). Knowledge of the contributions of the Td, Rd, Ld and Cd during nonlinear bubble oscillations will help in optimizing bubble and ultrasound exposure parameters for the relevant applications by maximizing a desirable parameter. In this work we investigate the mechanisms of dissipation in bubble oscillations and their contribution to the total damping ($W_{total}$) in various nonlinear regimes. By using bifurcation analysis, we have classified nonlinear dynamics of bubbles that are sonicated with their 3rd superharmonic (SuH) and 2nd SuH resonance frequency ($f_r$), pressure dependent resonance frequency ($PDf_r$), $f_r$, subharmonic (SH) resonance ($f_{sh}=2f_r$), pressure dependent SH resonance ($PDf_{sh}$) and $1/3$ order SH resonance. The corresponding Td, Rd, Ld, Cd, $W_{total}$, scattering to dissipation ratio (STDR), maximum wall velocity and maximum back-scattered pressure from non-destructive oscillations of bubbles were calculated and analyzed using the bifurcation diagrams. We classified different regimes of dissipation and provided parameter regions in which a particular parameter of interest (e.g. Rd) can be enhanced. Afterwards enhanced bubble activity is linked to some relevant applications in ultrasound. This paper represents the first comprehensive analysis of the nonlinear oscillations regimes and the corresponding damping mechanisms.       
\end{abstract}
\end{frontmatter}

\section{Introduction}
An ultrasonically excited bubble is a highly nonlinear oscillator in which  deterministic chaos manifests itself \cite{1,2,3}. When a high pressure acoustic field is generated in an aqueous medium, the rare faction cycle may exceed the attractive forces among liquid molecules generating cavitation bubbles. 
Bubbles begin oscillating and emit sound \cite{4,5,6}. The spectral components of the emitted sound consist of harmonics and subharmonics of the incident sound wave center frequency and broadband noise (Lauterborn $\&$ Holzfuss 1991 \cite{3}). The nonlinear frequency content of the emitted sound by bubbles has found its applications in contrast enhanced diagnostic ultrasound to visualize the vascular structure \cite{7,8,9} with superior contrast. Bubbles signatures are also used for monitoring treatments in therapeutic ultrasound \cite{10,11,12}.\\ The pressure emitted by collapsing bubbles may form a shock wave (Radek 1972; Vogel et al. 1986) \cite{5,13}, that can mechanically damage nearby nearby structures. Bubble oscillations generate micro-streaming in the liquid which results in shear stresses 
on the objects in its vicinity and micro-mixing in the liquid \cite{14,15}. The induced shear stresses and the emitted shock-waves has found their applications in industry (cleaning the micro-structures \cite{14,15,16}) and medicine (e.g. enhanced drug and gene delivery \cite{17,18,19}, blood brain barrier opening \cite{20,21} and shock wave lithotripsy and histotripsy \cite{22,23}).\\
Ultrasonically excited bubbles can focus and concentrate the acoustic energy from the macro-scale (acoustic wave) to the micro-scale and nano-scale \cite{19,24} generating extremely high
temperatures and pressures as the bubbles collapse.  This leads to molecular disassociation which triggers the production of highly reactive free radicals \cite{24,25,26} which then interact with other substances in the solution. 
This phenomenon has been shown useful in numerous industrial processing applications ranging from sonochemistry \cite{24,25,26} (chemical reaction rate enhancement and treatment of organic
compounds) to the food industry \cite{27} and medicine (sonodynamic therapy \cite{28}). Bubbles can focus and amplify the energy of the sound field by more than 11 orders of magnitude, which is sufficient not only to break chemical bonds but also to induce luminescence \cite{29}. Local sound amplification and enhanced dissipation of acoustic energy by bubbles been used to enhance the heating generated by ultrasound during  ultrasound thermal therapies and high intensity focused ultrasound (HIFU) tumor ablation \cite{30}.\\ 
Understanding and enhancing a specific type of bubble oscillatory pattern can help in enhancing the outcome of the relevant application. For example, in contrast enhanced ultrasound the goal is maximizing the radiated pressure by the bubbles while keeping the dissipation of energy due to bubble attenuation minimum \cite{31,32,33}. This will lead to enhanced contrast and better visualization of the target and eliminating the shadowing in ultrasonic images \cite{31,32,33}. Shadowing \cite{34,35} is caused by the dissipation of the ultrasonic energy by bubbles which leads to a weaker signal intensity from underlying tissue. In HIFU the goal is to increase the dissipation at the focus while reducing pre-focal shielding and energy dissipation by bubble oscillations. Here, knowledge of the pressure dependent dissipation effects and the advantage of the sharp pressure gradients of HIFU transducers facilitate the desired effect \cite{32,33,36}.\\ Bubbles dissipate the acoustic energy through radiation damping (Rd), thermal damping (Td), damping due to the viscosity of the liquid (Ld)  and damping due to the friction of the coating (Cd) \cite{37,38,39,40,41}. Despite the importance of detailed knowledge of the energy dissipation mechanisms in bubble oscillations, the majority of previous studies have been limited by linear approximations \cite{37,42,43,44,45}. Linear studies simplify the bubble oscillations to very small amplitudes at low excitation pressures (e.g. 1 kPa) \cite{37,42}. However, bubble oscillations are nonlinear and energy dissipation depends highly on the excitation pressure \cite{39,40,41,46}.
Moreover, the majority of the applications are based on sending ultrasound pulses of high pressure amplitude; thus, linear approximations are inappropriate to model bubble oscillations.\\Despite the importance of the knowledge on nonlinear energy dissipation by bubbles; however, there are only few recent studies that explored the pressure dependent effects on energy dissipation \cite{38,39,40,41,46,47}. Louisnard \cite{38} derived the pressure dependent energy equations by considering the conservation of mass and momentum in a bubbly media and used the Rayleigh-Plesset equation for bubble oscillations \cite{48}. He derived the dissipation equations for Ld and Td. His analysis showed that energy dissipation is pressure dependent and predictions of the linear model can be orders of magnitude smaller than the pressure dependent model.
Jamshidi $\&$ Brenner used Louisnard's approach and Keller-Miksis equation \cite{49} to incorporate the compressibility effects up to the first order of Mach number. They were able to derive Ld, Td and Rd. Their analysis showed that Rd has an important role in energy dissipation and as is typically done cannot be neglected \cite{39}. In our recent work, we showed that equations derived by Jamshidi $\&$ Brenner need to be corrected as their model predicts non-physical values for Rd near resonance and predictions of Rd are not consistent with the predictions of the scattered pressured energy (Sd)  by bubbles \cite{40}. We presented the corrected forms of Ld, Rd and Td. We showed that dissipation terms are highly pressure dependent and as pressure increases Rd may grow faster than Td and Ld; thus, there exist optimum pressure and frequency ranges where the scattering to dissipation ratio (STDR) can be maximized \cite{40,41}. Moreover, we showed that the STDR which can be used as standardization parameter to assess the efficacy of bubble oscillations \cite{40} in applications is pressure dependent. STDR should be used in conjunction with Rd and the maximum scattered (re-radiated) pressure by bubbles for a more complete assessment of a given control parameter for bubble oscillation optimization \cite{40}.\\ Using the same approach as in \cite{40}, we derived the nonlinear energy dissipation equations for a coated bubble \cite{41}. We analyzed the resonance power curves for free and encapsulated bubbles and showed that Td can be neglected for coated bubbles that have C3F8-like gas cores. We also showed that although Td is the dominant dissipation mechanism for large uncoated bubbles; at higher pressures Rd can supersede Td. Moreover, Cd is the strongest dissipation mechanism in the oscillations of the coated bubbles; pressure increase however, there are instances in which Rd is stronger than Ld and Td.\\ 
In this paper we provide a detailed analysis of the pressure dependent dissipation mechanisms by bubble oscillations and role of each of the dissipation components (Td, Ld,Rd and Cd) at various nonlinear regimes. Knowledge of the pressure dependent dissipation effects and the examination of each contributing component will help us better understand bubble related phenomena and enhance a desirable effect in bubble oscillations.\\ In this paper we have classified major nonlinear regimes of the oscillations for free and coated bubbles. In this regard, our recent comprehensive approach is used to analyze the bubble oscillations \cite{50} as a function of pressure. The major nonlinear regimes that are considered here are 2nd and 3rd SuH resonant oscillations, $3/2$, $5/2$ and $7/2$ UH regimes, pressure dependent resonance effects, excitation with linear resonance ($f_r$), pressure dependent resonance($PDf_r$), $1/2$, $1/3$ and higher order SH resonance. Afterwards, the pressure dependent dissipation mechanism and the role of each contributing factor to the total dissipation is analyzed in detail for each category of oscillations. STDR, maximum bubble wall velocity and maximum re-radiated pressure amplitude are analyzed for each regime. We show that depending on the specific oscillation regime, there is a exposure condition in which a particular parameter (e.g. maximum wall velocity, the maximum re-radiated pressure amplitude, Rd, $W_{total}$) can be maximized or minimized. These findings are then related to some of the current applications of bubbles.   
\section{Methods}
\subsection{Coated bubble model}
The dynamics of a coated bubble oscillator including compressibility effects to the first order of Mach-number can be modeled using the Keller-Miksis-Church-Hoff (KMCH) model \cite{41,42,49}:
\begin{equation}
\begin{gathered}
\rho \left[\left(1-\frac{\dot{R}}{c}\right)R\ddot{R}+3/2\dot{R}^2\left(1-\frac{\dot{R}}{3c}\right)\right]=\\
\left(1+\frac{\dot{R}}{c}+\frac{R}{c}\frac{d}{dt}\right)\left(P_g-\frac{4\mu_L\dot{R}}{R}-\frac{12\mu_{sh}\epsilon R_0^2\dot{R}}{R^4}-12G_s\epsilon R_0^2 \left(\frac{1}{R^3}-\frac{R_0}{R^4}\right)-P_0-P\right)
\end{gathered}
\end{equation}
Where $\rho$ and c are respectively  the density and sound speed of the medium, R is the radius at time t, $\dot{R}$ is the bubble wall velocity, $\ddot{R}$ is the bubble wall acceleration, $R_0$ is the initial radius of the bubble, $\mu$ and $\mu_{sh}$ are the viscosity of the liquid and shell (coating) respectively, $\epsilon$ is the thickness of the coating, $G_s$ is the shell shear modulus, $P_g$ is the gas pressure inside the bubble, $P_0$ is the atmospheric pressure (101.325 kPa) and P is the acoustic pressure given by $P=P_asin(2\pi ft)$ with $P_a$ and $f$ are respectively the excitation pressure and frequency.  In this paper for all of the simulations of the coated bubbles $G_s$=50 MPa and $\mu_{sh}=\frac{1.49(R_0(\mu m)-0.86)}{\theta (nm)}$ \cite{51} with $\theta=4 nm$. The gas inside the bubble was chosen to be C3F8 and the surrounding medium water.
\subsection{Uncoated Bubble model}
The dynamics of the uncoated bubble including the compressibility effects to the first order of Mach number can be modeled using Keller-Miksis (KM) equation\cite{49}:
\justifying
\begin{equation}
\rho[(1-\frac{\dot{R}}{c})R\ddot{R}+3/2\dot{R}^2(1-\frac{\dot{R}}{3c})]=(1+\frac{\dot{R}}{c})(G)+\frac{R}{c}\frac{d}{dt}(G)
\end{equation}
where $G=P_g-\frac{4\mu_L\dot{R}}{R}-\frac{2\sigma}{R}-P_0-P_A sin(2 \pi f t)$.\\
In this equation, R is radius at time t, $R_0$ is the initial bubble radius, $\dot{R}$ is the wall velocity of the bubble, $\ddot{R}$ is the wall acceleration,	$\rho{}$ is the liquid density (998 $\frac{kg}{m^3}$), c is the sound speed (1481 m/s), $P_g$ is the gas pressure, $\sigma{}$ is the surface tension (0.0725 $\frac{N}{m}$), $\mu{}$ is the liquid viscosity (0.001 Pa.s), and $P_A$ and \textit{f} are the amplitude and frequency of the applied acoustic pressure. The values in the parentheses are for pure water at 293 K. In this paper the gas inside the uncoated bubble is air and water is the host media.\\
\begin{table}
	\begin{tabular}{ |p{2cm}||p{3.5cm}|p{2cm}|p{2cm}|p{2cm}|  }
		\hline
		\multicolumn{5}{|c|}{Thermal parameters of the gases at 1 atm} \\
		\hline
		Gas type  & L($\frac{W}{mK}$) &$c_p$$(\frac{kJ}{kg K})$ &$c_v$ $(\frac{kJ}{kg K})$&$\rho_g$ $(\frac{kg}{m^3})$\\
		\hline
		Air \cite{53}   & $0.01165+C\times T^2$ &1.0049&   0.7187&1.025\\
		C3F8 \cite{54} &   0.012728  & 0.79   &0.7407&8.17\\
		\hline
	\end{tabular}
	\caption{Thermal properties of the gases used in simulations.{$^2$ C=$5.528\times10^{25}$ $\frac{W}{m K^2}$}.}
	\label{table:1}
\end{table}
\subsection{Thermal effects}
If thermal effects are considered, $P_g$ is given by Eq. 5 \cite{49,50,51,52,53}:
\begin{equation}
P_g=\frac{N_gKT}{\frac{4}{3}\pi R(t)^3-N_g B}
\end{equation}
Where $N_g$ is the total number of the gas molecules, $K$ is the Boltzman constant and B is the molecular co-volume of the gas inside the bubble. The average temperature inside the gas can be calculated using Eq. 6 \cite{49}:
\vspace{-2mm}
\begin{equation}
\dot{T}=\frac{4\pi R(t)^2}{C_v} \left(\frac{L\left(T_0-T\right)}{L_{th}}-\dot{R}P_g\right)
\end{equation}
where $C_v$ is the heat capacity at constant volume, $T_0$=$293 $K is the initial gas temperature, $L_{th}$ is the thickness of the thermal boundary layer. $L_{th}$ is given by $L_{th}=min(\sqrt{\frac{aR(t)}{|\dot{R(t)}|}},\frac{R(t)}{\pi})$ where $a$ is the thermal diffusivity of the gas which can be calculated using $a=\frac{L}{c_p \rho_g}$ where L is the gas thermal conductivity and $c_p$ is specific heat capacity at constant pressure and $\rho_g$ is the gas density.\\
Predictions of the full thermal model have been shown to be in good agreement with predictions of the models that incorporate the thermal effects using the PDEs \cite{55} that incorporate the temperature gradients within the bubble.
To calculate the radial oscillations of the coated bubble and uncoated bubble while including the thermal effects Eqs. 1 and Eq. 2 are respectively coupled with Eq. 3 and 4 and then solved using the ode45 solver of Matlab.
\subsection{Nonlinear terms of dissipation for the KMCH model}
We have derived the equations for the average power loss in the oscillations of the KMCH model \cite{41}:
\begin{equation}
\begin{dcases}
Td=\frac{-4\pi}{T}\int_{0}^{T}R^2\dot{R}P_g dt\\ \\
Ld=\frac{16\pi\mu_L}{T}\int_{0}^{T}R\dot{R}^2dt\\ \\
Cd=\frac{48\pi\mu_{sh}\varepsilon R_0^2}{T}\int_{0}^{T}\frac{\dot{R}^2}{R^2}dt\\ \\
Gd=\frac{48\pi G_s\varepsilon R_0^2}{T}\int_{0}^{T}\left(\frac{\dot{R}}{R}-\frac{R_0\dot{R}}{R^2}\right)dt\\ \\
\begin{gathered}
Rd=\frac{1}{T}\int_{0}^{T}\left(4\pi\left[\frac{R^2\dot{R}^2}{c}\left(P-P_g\right)+\frac{R^3\dot{R}}{c}\left(\dot{P}-\dot{P}_g\right)+\frac{4\mu_LR^2\dot{R}\ddot{R}}{c}\right.\right.\\
\left.+12\mu_{sh}\varepsilon R0^2 \left(\frac{\dot{R}\ddot{R}}{cR}-\frac{3\dot{R}^3}{c R^2}\right)+12G_s\varepsilon R0^2\left(\frac{-2\dot{R}^2}{cR}+\frac{3R_0\dot{R}^2}{cR^2}\right) \right]\\
\left. -\frac{\rho R^2\dot{R}^4}{2c}-\frac{\rho R^3 \dot{R}^2\ddot{R}}{c}\right)dt
\end{gathered}
\end{dcases}
\end{equation}
Where Td, Ld, Cd, Rd and Gd are the average dissipated powers due to thermal, Liquid viscosity , coating viscosity, re-radiation  and stiffness of the coating. In simulations we did not present the values for Gd since it is always zero for a full cycle. $T$ is the integration time and can be given as n/f where n=1,2...... . In this paper the integrals are performed over the last 20 cycles of a 500 cycles pulses to avoid the transient bubble behavior.
\subsection{Nonlinear terms of dissipation for the KM model}
We have derived the dissipation power terms of the KM model as follows \cite{40}:
\begin{equation}
\begin{dcases}
Td=\frac{-1}{T}\int_{0}^{T}\left(P_g\right)\frac{\partial V}{\partial t}dt\\ \\
Ld=\frac{16\pi\mu_L}{T}\int_{0}^{T}\left(R\dot{R}^2\right)dt\\ \\
\begin{gathered}
Rd=\frac{1}{T}\int_{0}^{T} \left[\frac{4\pi}{c}\left(R^2\dot{R}\left(\dot{R}P+R\dot{P}-1/2\rho \dot{R}^3-\rho R\dot{R}\ddot{R}\right)\right)\right.\\
\left.-\left(\frac{\dot{R}}{c}P_g+\frac{R}{c}\dot{P}_g\right)\frac{\partial V}{\partial t}+\frac{16\pi\mu_LR^2\dot{R}\ddot{R}}{c}\right]dt
\end{gathered}
\end{dcases}
\end{equation}
All the dissipated powers were calculated for the last 20 cycles of pulses with 500 cycles length. Simulations were carried out in Matlab using ODE45 with the highest possible relative and absolute tolerance. The  time steps for integration in each simulation were $\leq$ $\frac{10^{-4}}{f}$.  
\subsection{Bifurcation diagrams}
Bifurcation diagrams are valuable tools to analyze the dynamics of nonlinear systems where the qualitative and quantitative changes of the dynamics of the system can be investigated effectively over a wide range of the control parameters. In this paper, we employ a more comprehensive bifurcation analysis method introduced in \cite{50,56}.\\
\subsubsection{Poincaré section}
When dealing with systems responding to a driving force, to generate the points in the bifurcation diagrams vs. the control parameter, one option is to sample the R(t) curves using a specific point in each driving period. The approach can be summarized by:
\begin{equation}
P \equiv (R(\Theta))\{(R(t),  \dot{R}(t) ):\Theta= \frac{n}{f} \} \hspace{0.5cm} where \hspace{0.5cm} n=480,481...500
\end{equation}
Where $P$ denotes the points in the bifurcation diagram, $R$ and $\dot{R}$
are the time dependent radius and wall velocity of the bubble at a given
set of control parameters of ($R_{0}$, $P_{0}$, $P_{A}$, $c$, $k$, $\mu$, $G_s$, $\mu_{sh}$, $\theta$,
$\sigma$, $f$) and $\Theta$ is given by $\frac{n}{f}$.  Points on the bifurcation diagram are constructed by plotting the solution of $R(t)$ at time points that are multiples of the driving acoustic period. The results are plotted for $n=480-500$ to ensure a steady state solution has been reached.
\subsubsection{Method of peaks}
As a more general method, bifurcation points can be constructed by setting one of the phase space coordinates to zero:      
\begin{equation}
Q \equiv max(R)\{(R, \dot{R} ):\dot{R}= 0\}
\end{equation}
In this method, the steady state solution of the radial oscillations for each control parameter is considered. The maxima of the radial peaks ($\dot{R}=0$) are identified (determined within $n=480-500$ cycles of the stable oscillations) and are plotted versus the given control parameter in the bifurcation diagrams. 
The bifurcation diagrams of the normalized bubble oscillations ($R/R_0$) are calculated using both methods a) and b). When the two results are plotted alongside each other, it is easier to uncover more important details about the SuH and UH oscillations, as well as the SH and chaotic oscillations.\\
\section{Results and Discussion}
In this section various nonlinear oscillation regimes of coated and uncoated bubbles are introduced by visualizing the radial oscillations of the bubble as a function of pressure at various frequencies. Then we build a link between different nonlinear oscillation regimes and the dissipated powers.\\ In the simulations, the uncoated bubbles that enclose air have initial radii of 10 $\mu m$ and 2 $\mu m$. The bubble with $R_0=10 \mu m$ is chosen as it will have strong thermal damping due to its bigger size. The bubble with $R_0=2\mu m$ is chosen as viscous effects are strong due to its size (Results related to this case are presented in Appendix).\\ For the coated bubbles we investigated the bubbles with initial radii of 1 and $4 \mu m$. The bubble with $R_0=4 \mu m$ is probably the largest bubble that can be used in medical applications (as the capillaries have diameters around $8 \mu m$ \cite{8}). Therefore, this bubble has the highest possible size dependent Td (Results of this case are presented in Appendix). The bubble with $R_0=1 \mu m$ is also chosen as it is in the typical range of the contrast agents that are used in medical applications and viscous effects strongly influence its dynamics.
\subsection{Bifurcation structure and dissipation mechanisms of uncoated bubbles}
\subsubsection{The case of an uncoated air bubble with $R_0=10 \mu m$}
\begin{figure*}
	\begin{center}
		\scalebox{0.43}{\includegraphics{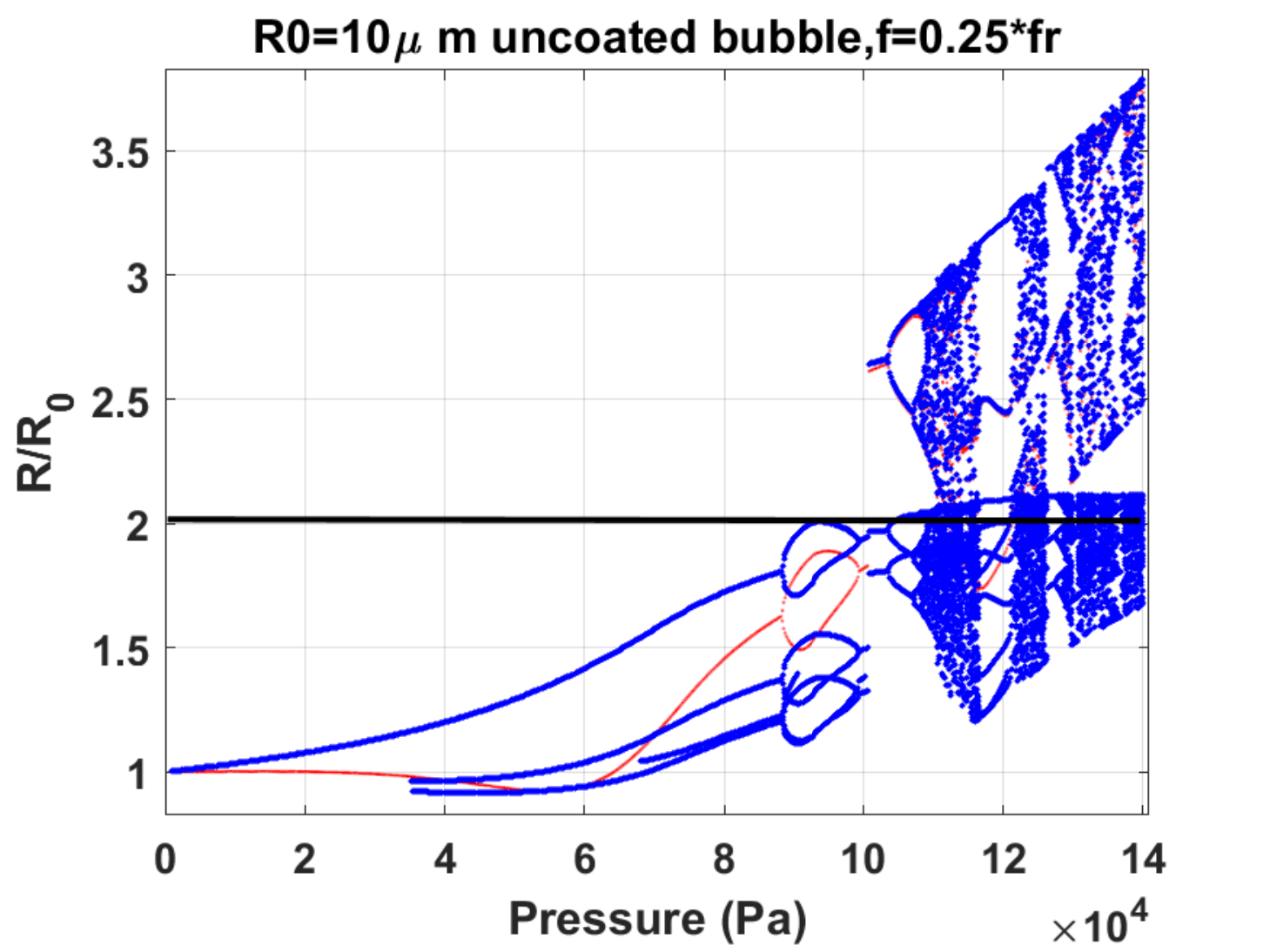}}  \scalebox{0.43}{\includegraphics{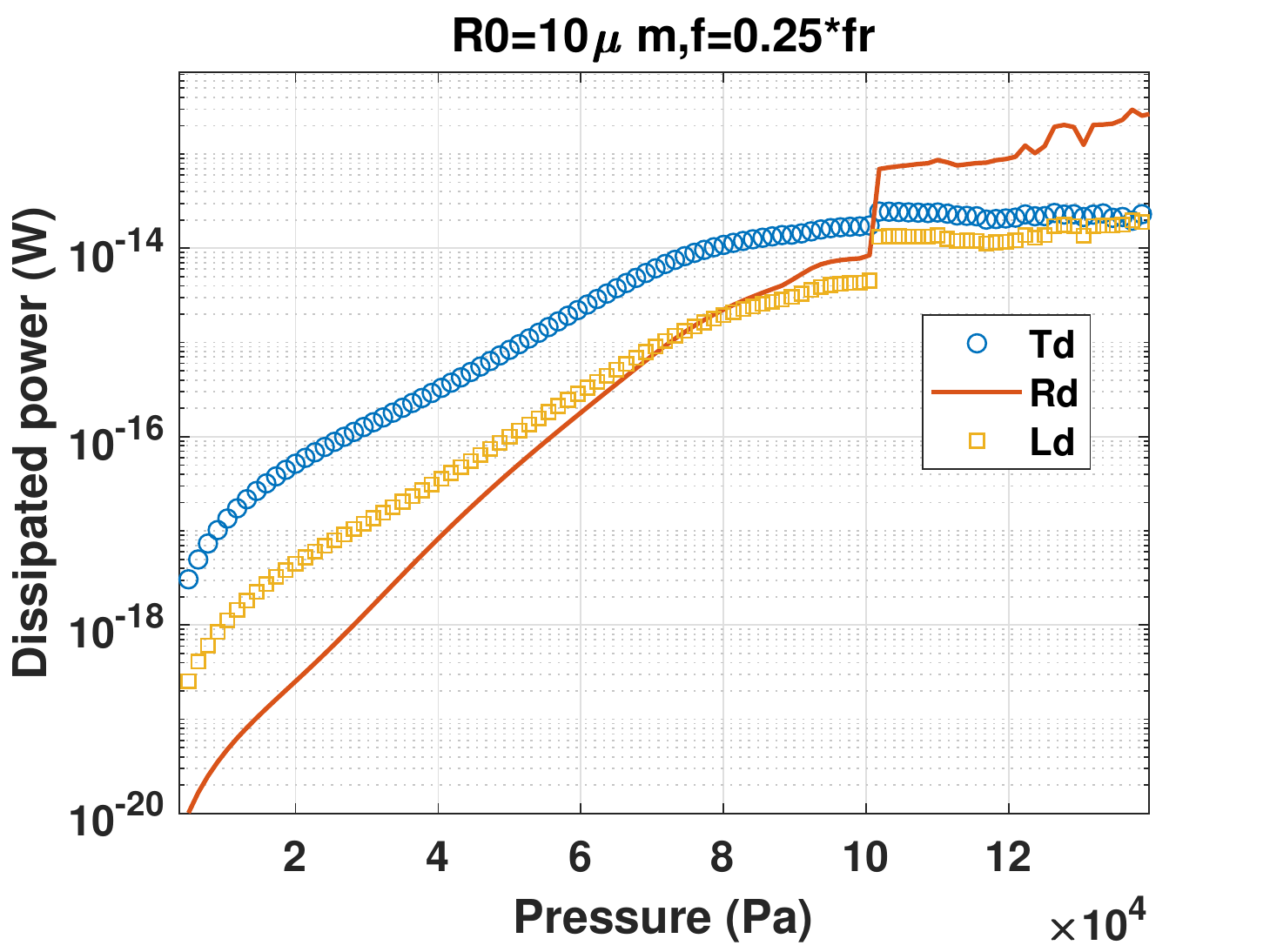}}\\
		\hspace{0.5cm} (a) \hspace{6cm} (b)\\
		\scalebox{0.43}{\includegraphics{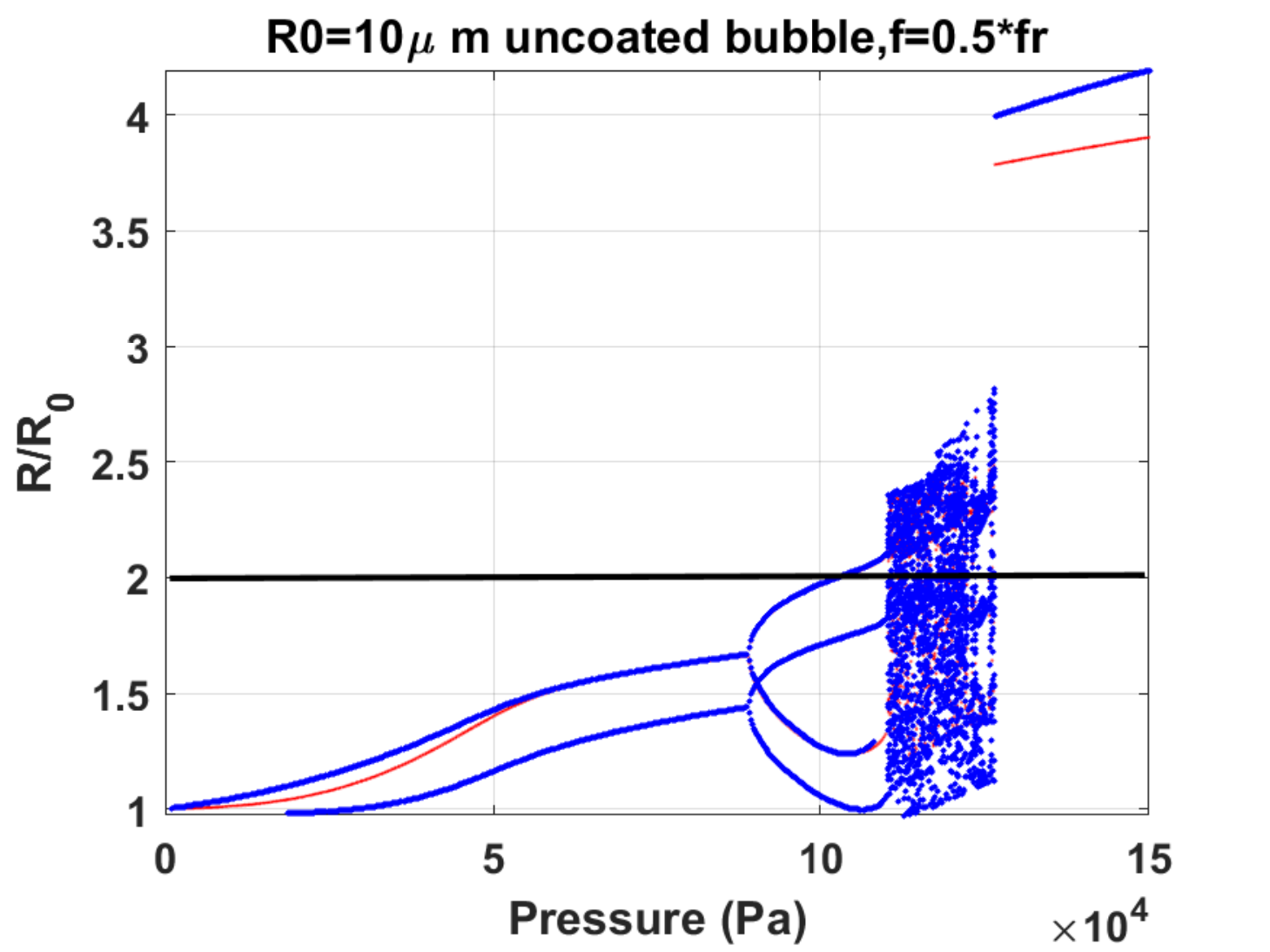}} \scalebox{0.43}{\includegraphics{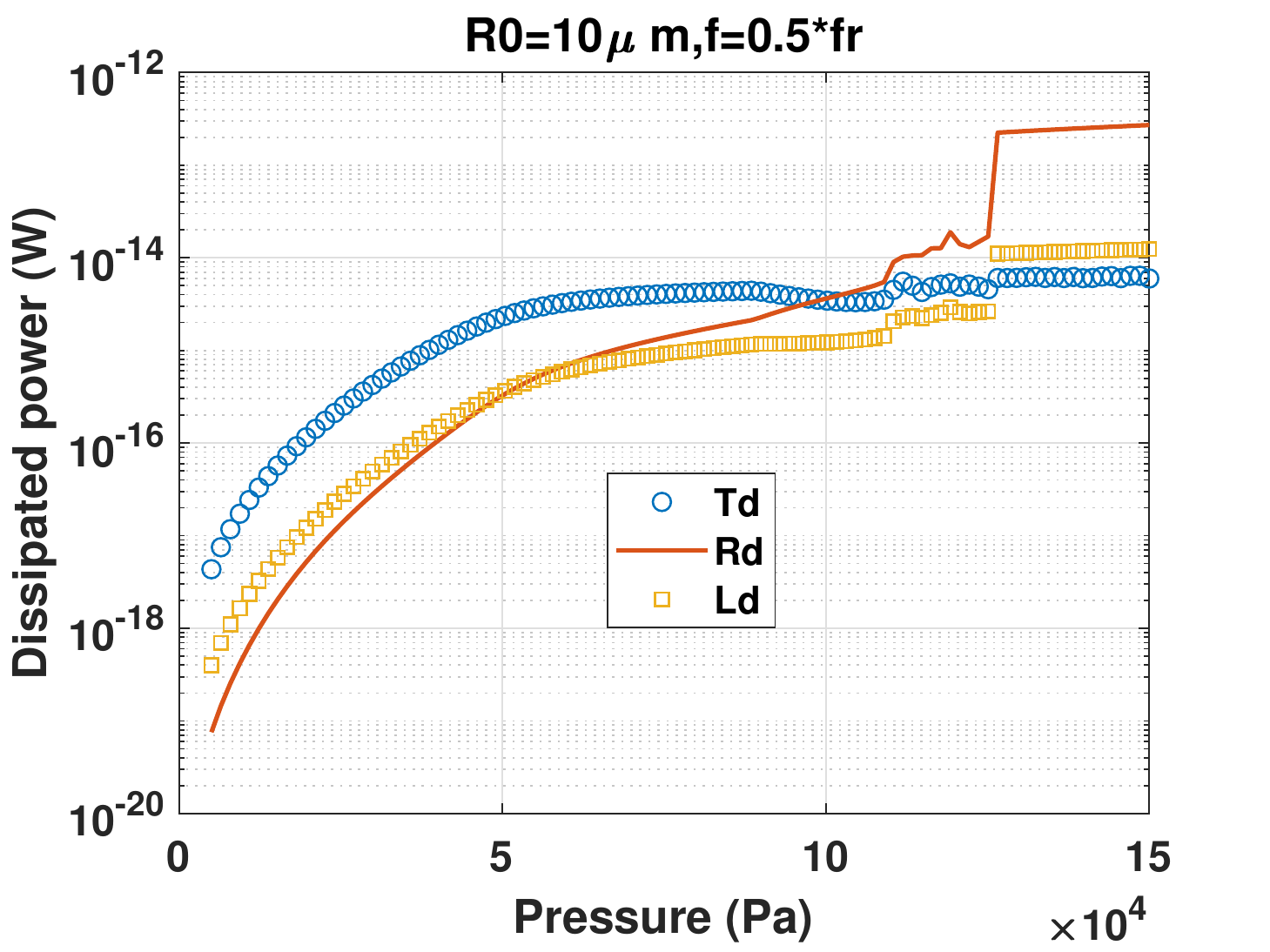}}\\
		\hspace{0.5cm} (c) \hspace{6cm} (d)\\
		\scalebox{0.43}{\includegraphics{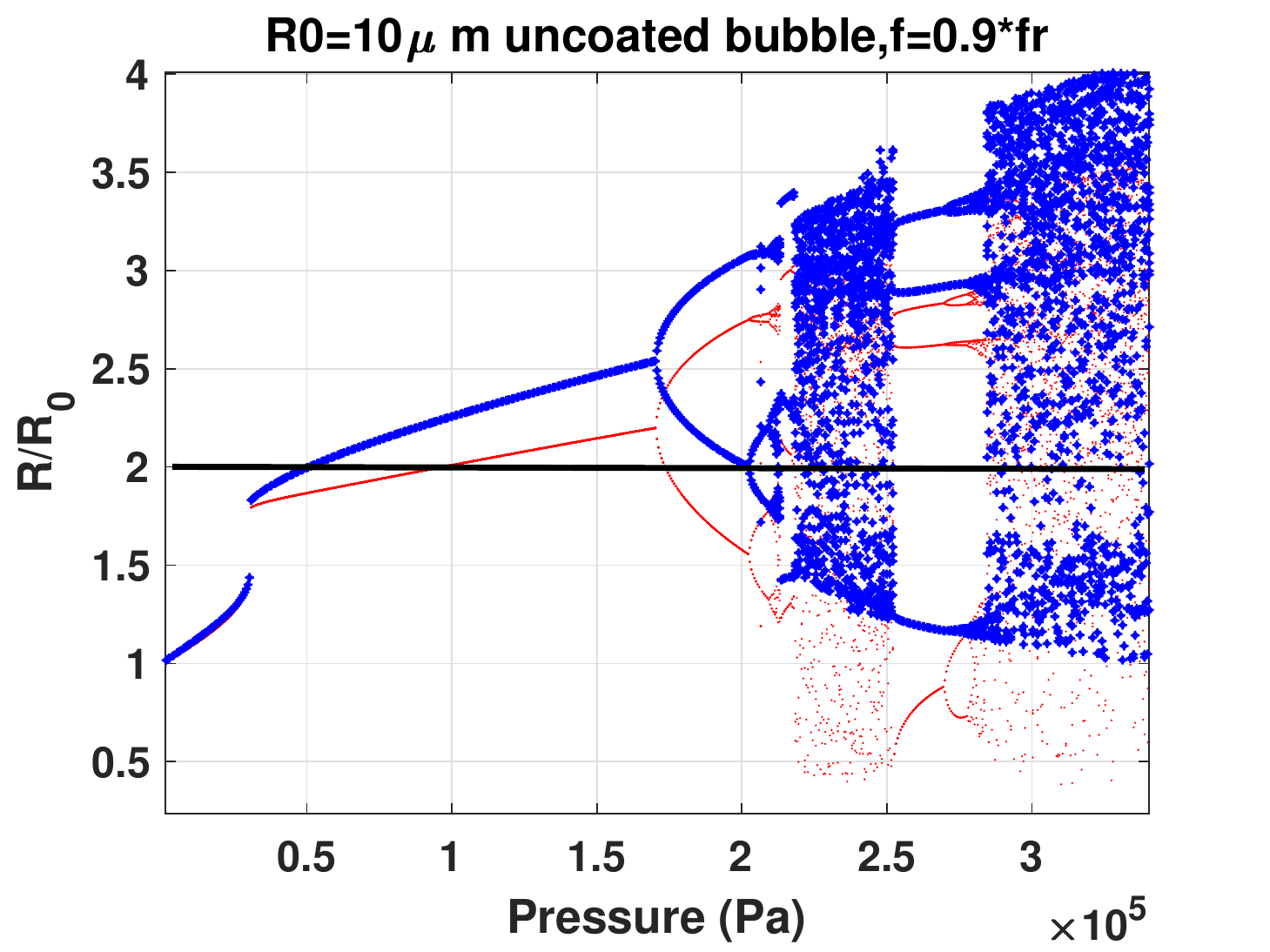}} \scalebox{0.43}{\includegraphics{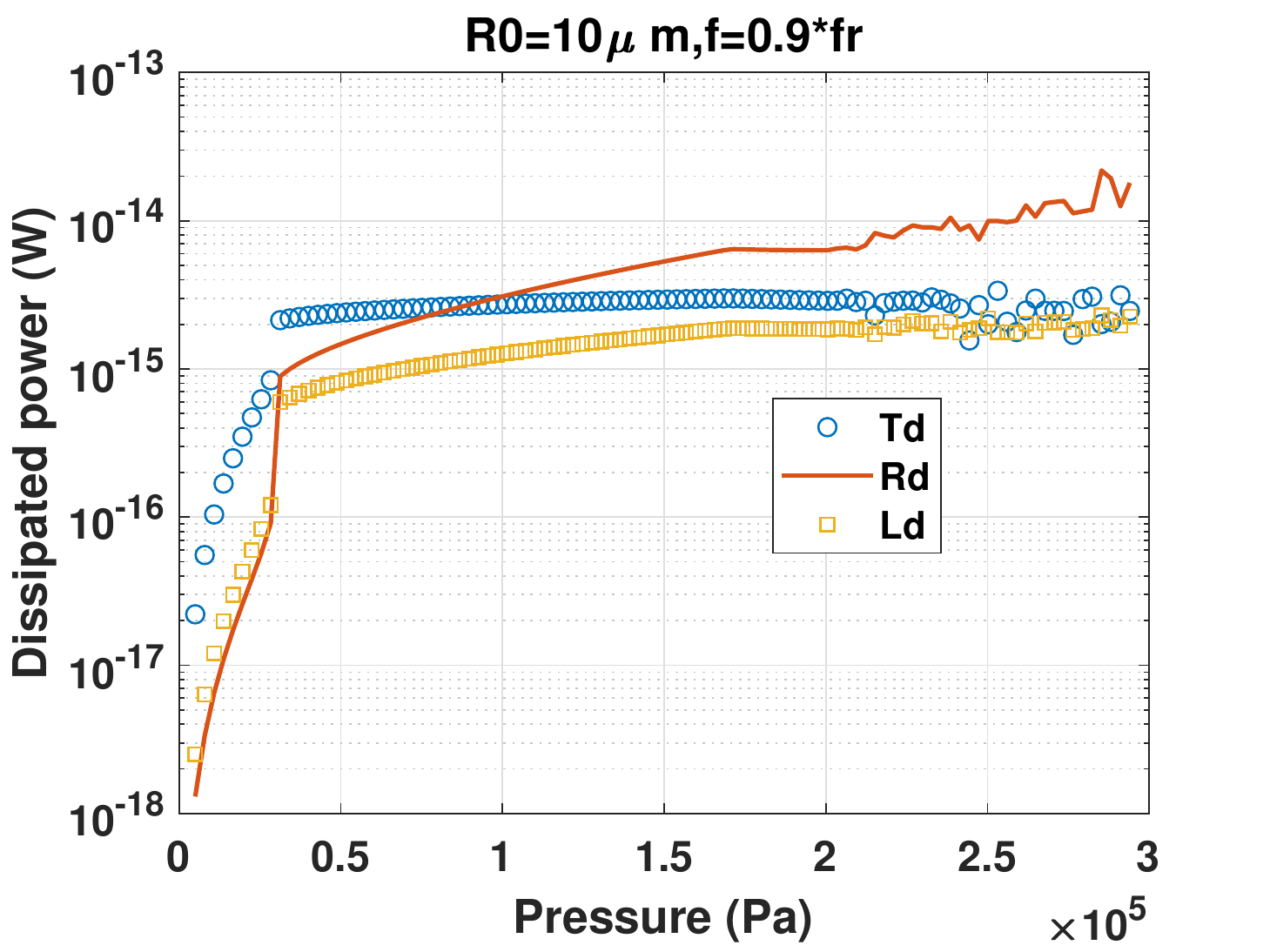}}\\
		\hspace{0.5cm} (e) \hspace{6cm} (f)\\
		\scalebox{0.43}{\includegraphics{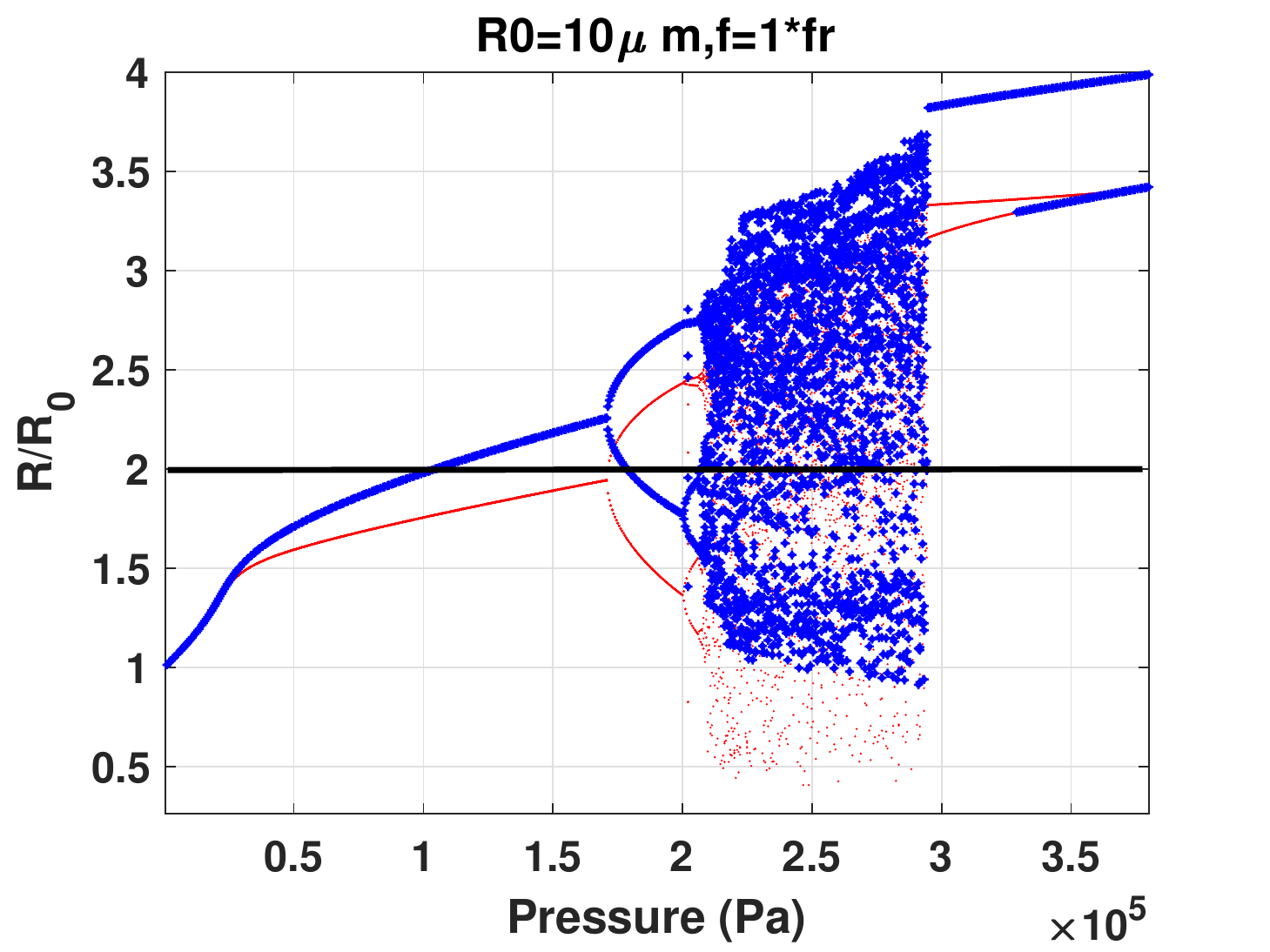}} \scalebox{0.43}{\includegraphics{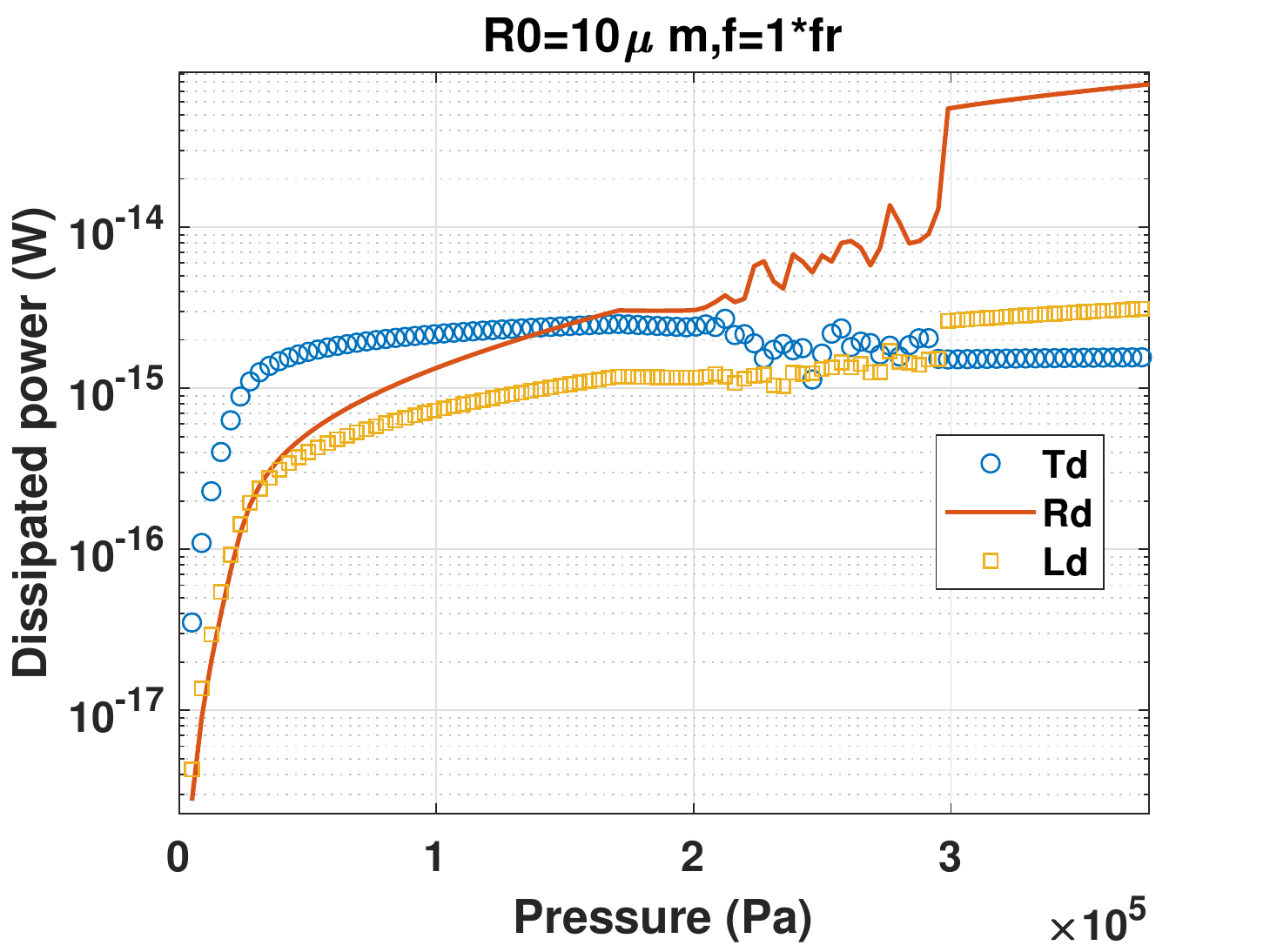}}\\
		\hspace{0.5cm} (g) \hspace{6cm} (h)\\
		\caption{Bifurcation structure (left column) and the dissipated power as a function of pressure (right column) of the oscillations of an uncoated air bubble with $R_0=10\mu m$ for $f= 0.25f_r$ (a-b)-$f=0.5f_r$ (c-d)-$f=0.9f_r$ (e-f) \& $f=f_r$ (g-h).}
	\end{center}
\end{figure*}
	Figure 1 shows the bifurcation structure of the normalized oscillations ($R/R_0$) as a function of acoustic pressure of an uncoated air bubble with $R_0=10 \mu m$ and the corresponding dissipated powers due to Ld, Td and Rd for ($0.25 f_r \leq f \leq 1$). Throughout the manuscript and in this Fig. the blue diagram is constructed using the method of maxima (2.6.2) and the red diagram is constructed through Poincaré analysis (2.6.1). When f=$0.25f_r$ (Fig. 1a) an increase in pressure results in the generation of 3rd order SuH oscillations at $P_a \approxeq 63 kPa$ (the blue curve shows three maxima for a period one oscillation (1 solution in the red graph)). The red curve undergoes a period doubling (Pd) bifurcation concomitant with a Pd in the blue graph at $P_a \approxeq 90 kPa$. This results in $7/2$ UH oscillations (P2 with 6 maxima). The period 2 (P2) oscillations undergo reverse Pd to period one (P1) oscillations with thee maxima at $P_a \approxeq 100 kPa$. With a slight pressure increase a saddle node bifurcation takes place to P1 oscillations with 3 maxima of higher amplitude. At this point the bubble may not sustain stable oscillations as $R/R_0>2$ (black horizontal line) \cite{57} (for further discussion on the minimum threshold for bubble destruction please refer to \cite{32}). Further pressure increase results in period doubling cascades to chaos. The corresponding power losses are presented in Fig. 1b. For $P_a \leq 63 kPa$, Rd is the weakest damping mechanism with Td the strongest mechanism (approximately 2 orders of magnitude larger). Rd grows faster than other damping mechanisms with increasing pressure and at $P_a \approxeq 90 kPa$ concomitant with the appearance of 3rd SuH oscillations, Rd becomes equal to Ld. Rd becomes stronger than Ld when UH oscillations occur; later, simultaneous with the saddle node bifurcation Rd undergoes a large increase and becomes the strongest damping mechanism. Td is the dominant mechanism for pressures below 100 kPa (the saddle node bifurcation) and at $\approx$ 130 kPa $Rd>Td=Ld$.\\
	When $f=0.5 f_r$ (Fig 1c); 2nd order SuH occurs in the oscillations of the bubble at $P_a \approxeq 20 kPa$; this manifests itself as a P1 oscillation (1 red line) with 2 maxima (two solutions for the blue curve). Radial oscillations grow with increasing pressure and at $P_a \approxeq 90 kPa$ the red curve undergoes a Pd which is coincident with a Pd for the blue curve; this results in $5/2$ order UH oscillations (P2 with 4 maxima). Oscillations become chaotic (sudden unset of chaos at $P_a \approxeq 110 kPa$ ); further at $P_a \approxeq 125 kPa$ a giant P1 resonance emerges out of chaos. Possible bubble destruction occurs at $\approx 110 kPa$ (black horizontal line ($R/R_0>2$)). For pressures below 50 kPa $Td>Ld>Rd$. Later, concomitant with saturation of 2nd order SuH oscillations at $\approx 50 kPa$ (red line becomes equal to one of the maxima indicating the wall velocity becomes in phase with the driving acoustic pressure). Rd becomes equal to Td and gets stronger than Ld during UH oscillations. Td is the dominant mechanism at $P_a < 90 kPa$; however, when UH oscillations are saturated, Rd supersedes Td and stays higher during the chaotic oscillation regime. Occurrence of the giant resonance is concomitant with a large increase in Rd as it becomes approximately two orders of magnitude higher than Td.\\
	When $f=0.9f_r$ (Fig. 1e), P1 oscillations (with 1 maxima) undergo a saddle node bifurcation to P1 oscillations of higher amplitude at $P_a \approxeq 40 kPa$. The bubble possibly is destroyed at  $Pa\approxeq 50 kPa$ (black horizontal line). Further increase in pressure results in Pd at 175 kPa; P2 oscillations undergo a cascade of Pds to chaos at 210 kPa. The corresponding dissipated power is presented in Fig. 1f.  For pressures below the saddle node (SN) bifurcation Td is the strongest damping mechanism (an order of magnitude larger) with $Td>Ld\approxeq Rd$. Concomitant with the SN, (note that at this pressure the wall velocity becomes in phase with the driving pressure) Rd becomes stronger than Ld and at 100 kPa it surpasses the initially larger Td. Further increase in pressure results in the fastest growth rate in Rd and the slowest growth rate in Td. Simultaneous with Pd and during majority of the P2 oscillation regime, Rd, Ld and Td stay approximately constant (this can be due the decrease in wall velocity concomitant with Pd when bubble is sonicated with a frequency near its resonance frequency \cite{32}). During chaotic oscillations $Rd>Td>Ld$ with fluctuations due to sporadic oscillations.\\
	For $f=f_r$ (Fig. 1g), at lower pressures ($P_a<25 kPa$) oscillations are P1 and the wall velocity is in phase with the driving acoustic force (blue and red curve are on top of each other). Further pressure increases result in possible bubble destruction at $P_a=100 kPa$ (black horizontal line meets the blue line $R/R_0>2$). At $P_a \approxeq 175 kPa$, Pd occurs and choas appears for  $205<P_a<295$ followed by the emergence of a giant P2 resonance. The corresponding power graph is presented in Fig. 1h. For $P_a<25 kPa$ where wall velocity is in phase with the driving pressure $Td>Rd=Ld$ and there is a very sharp growth for all the damping factors (possibly due to the resonant nature of oscillations). Rd becomes bigger than Ld above 25 kPa and grows faster than both Ld and Td until it becomes equal to Td at at $P_a \approxeq 150 kPa$. Rd becomes sightly higher than Td when Pd occurs; however, the occurrence of Pd decreases the rate of growth of the damping powers and they which stays relatively constant during P2 oscillations (due to possibly the decrease of the wall velocity during P2 oscillations when $f=f_r$ \cite{32,58}). Chaotic oscillations result in a slight decrease in Td but Rd keeps growing and at the giant resonance Rd undergoes a large increase and becomes approximately two orders of magnitude larger than the other damping factors. Occurrence of the P2 giant resonance is concomitant with a decrease in Td. The reduction in Td is concomitant with the occurrence of the giant resonance may lead to better sonochemical efficacy as higher temperatures are created while at the same time thermal conduction becomes more limited.   
	\begin{figure*}
		\begin{center}
			\scalebox{0.43}{\includegraphics{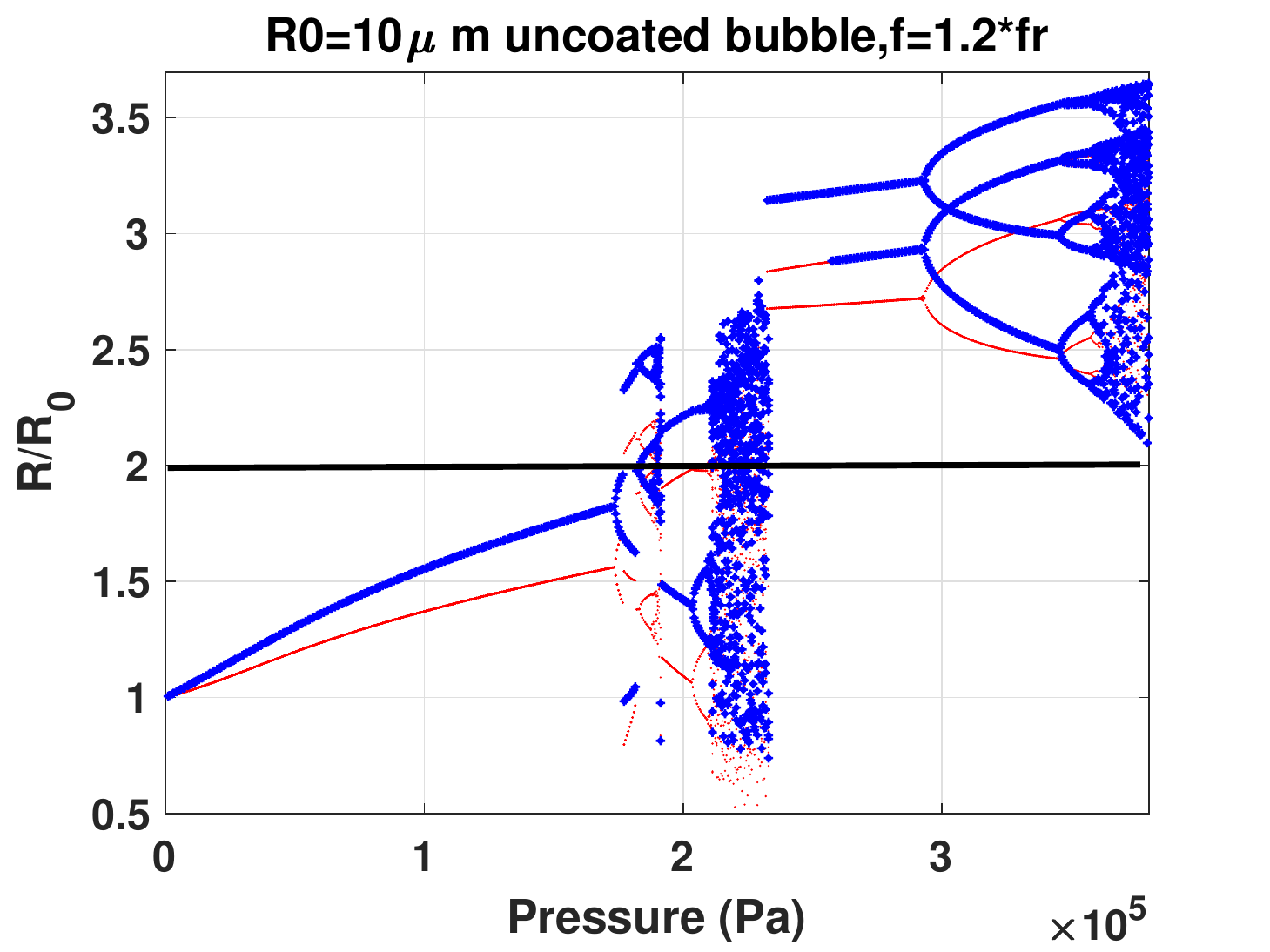}}  \scalebox{0.43}{\includegraphics{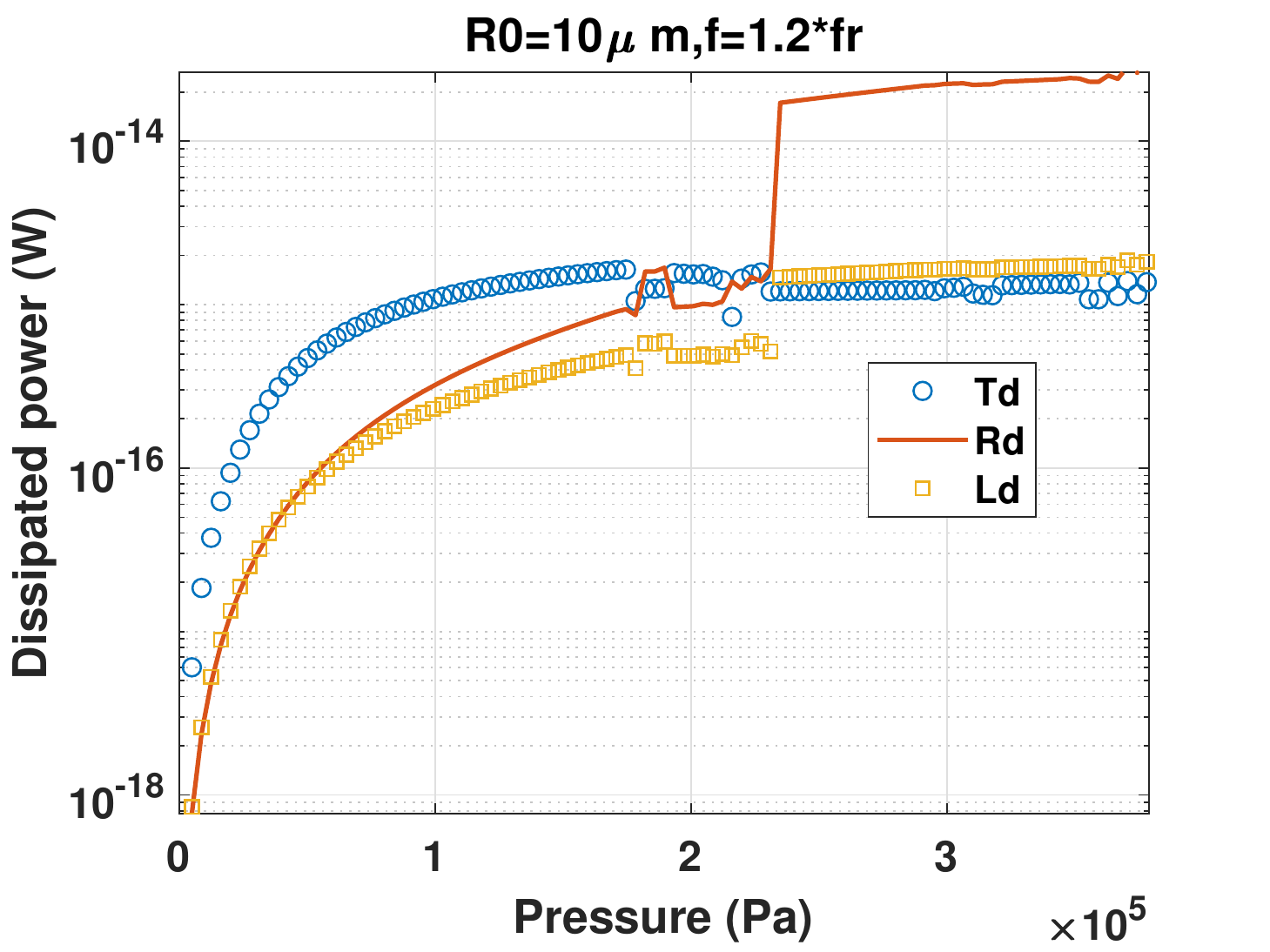}}\\
			\hspace{0.5cm} (a) \hspace{6cm} (b)\\
			\scalebox{0.43}{\includegraphics{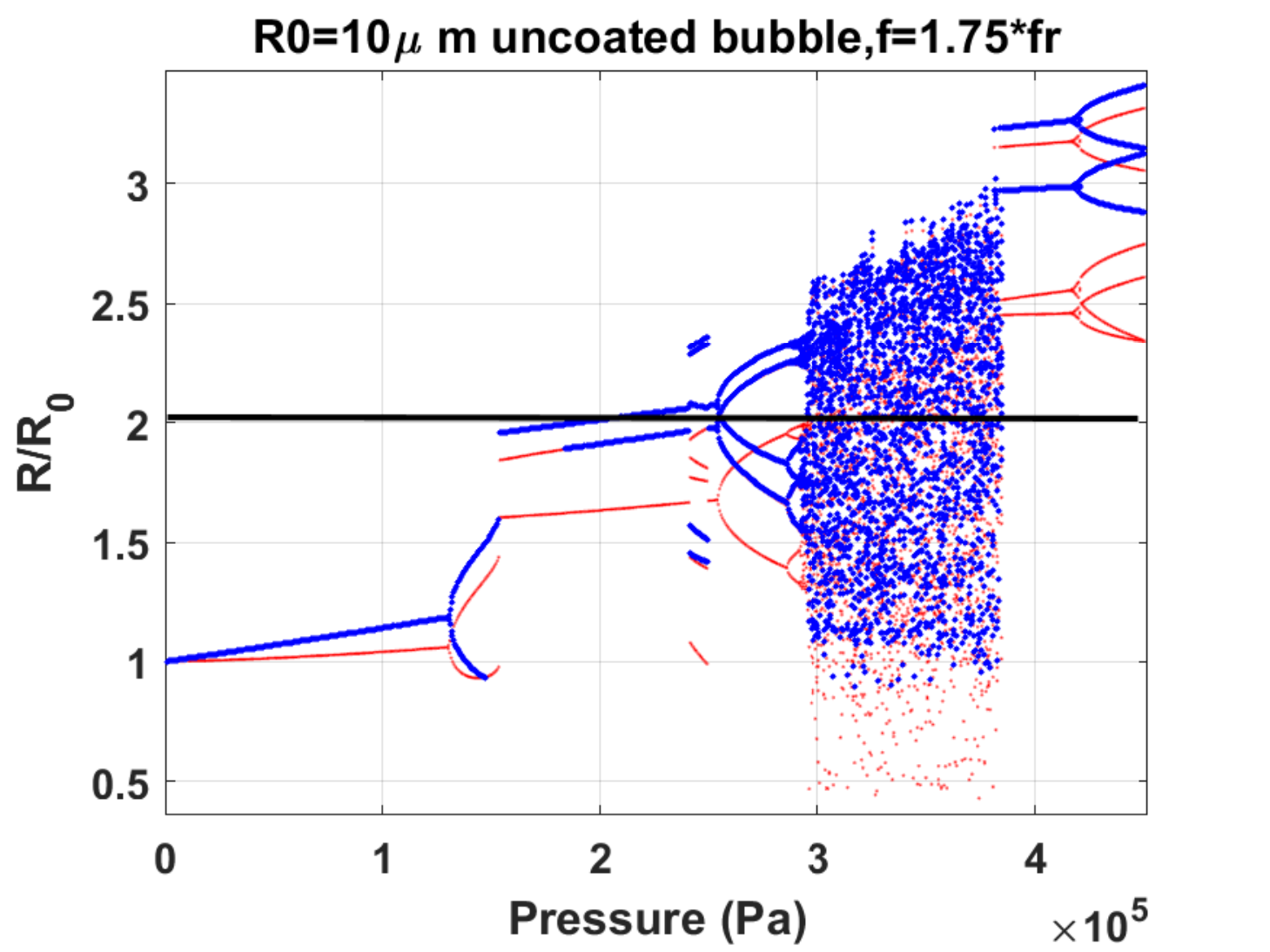}} \scalebox{0.43}{\includegraphics{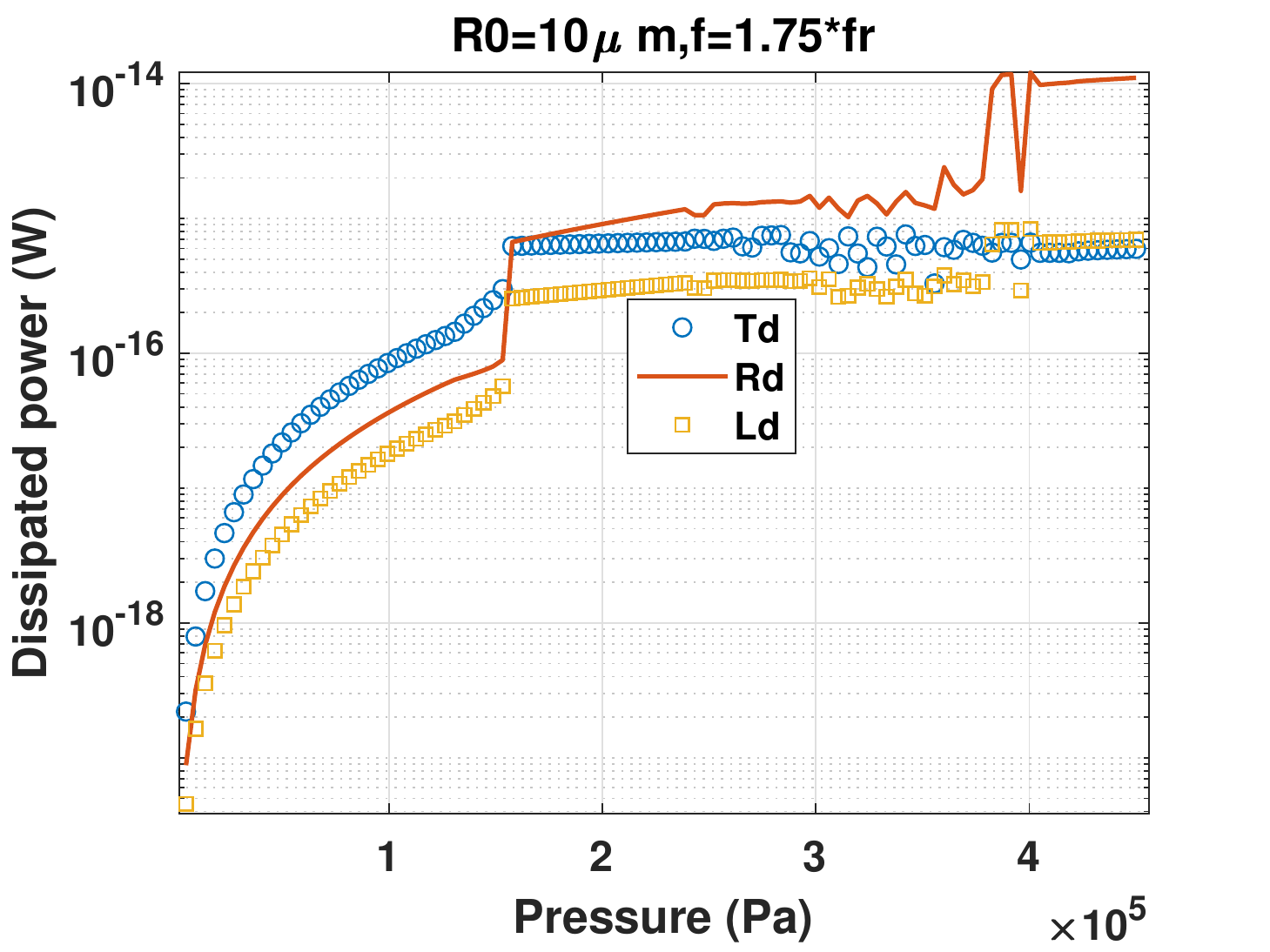}}\\
			\hspace{0.5cm} (c) \hspace{6cm} (d)\\
			\scalebox{0.43}{\includegraphics{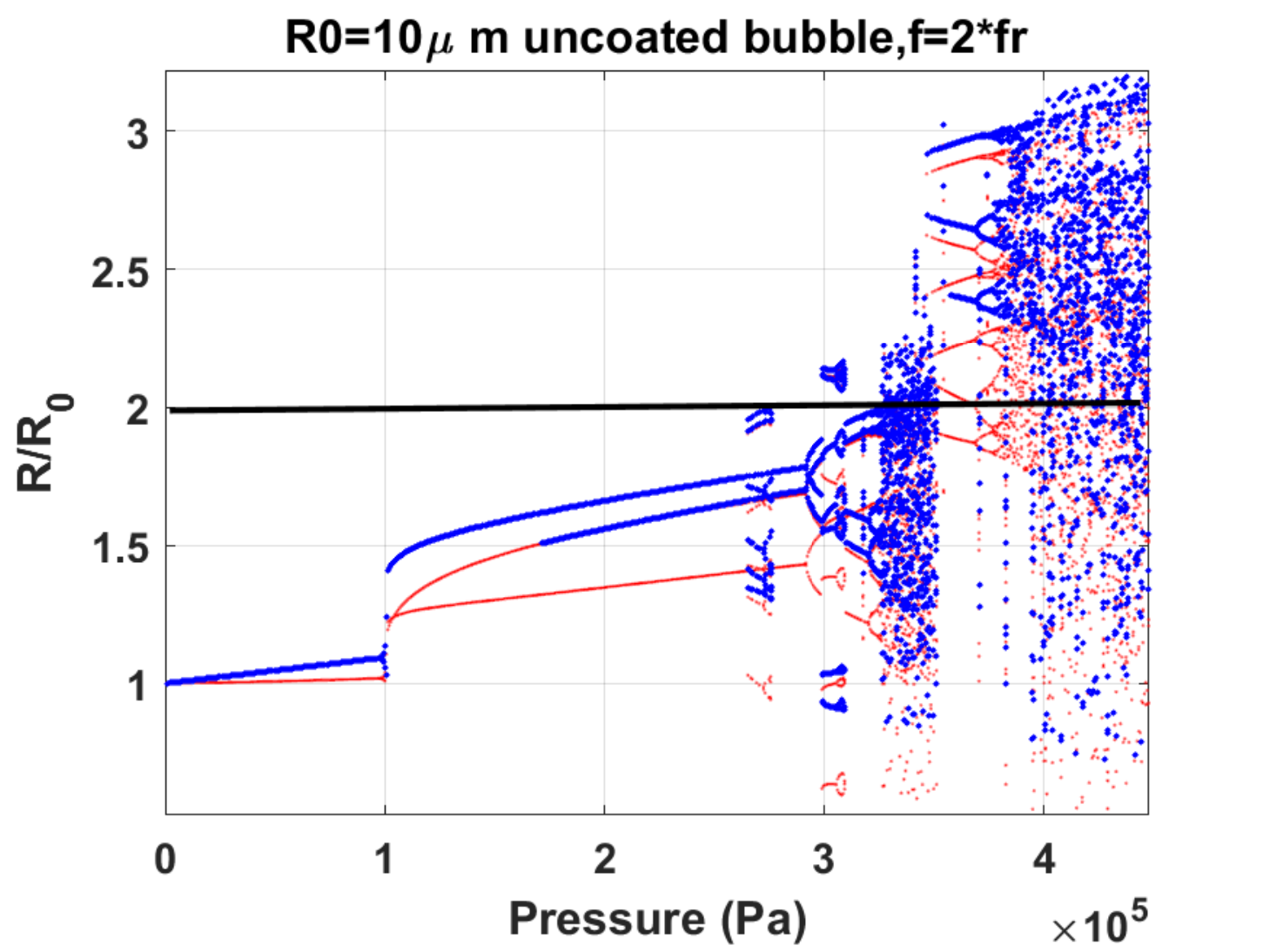}} \scalebox{0.43}{\includegraphics{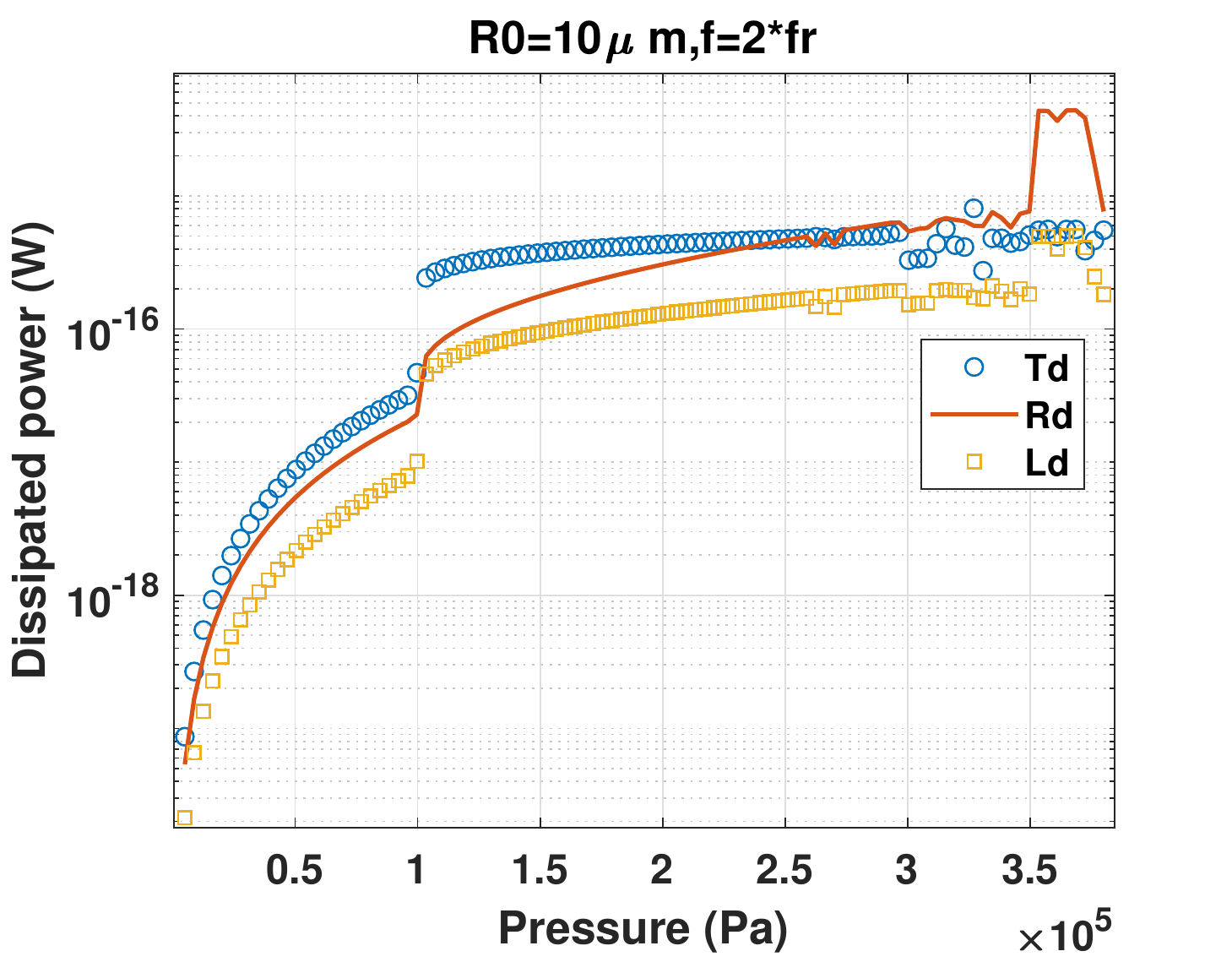}}\\
			\hspace{0.5cm} (e) \hspace{6cm} (f)\\
			\scalebox{0.43}{\includegraphics{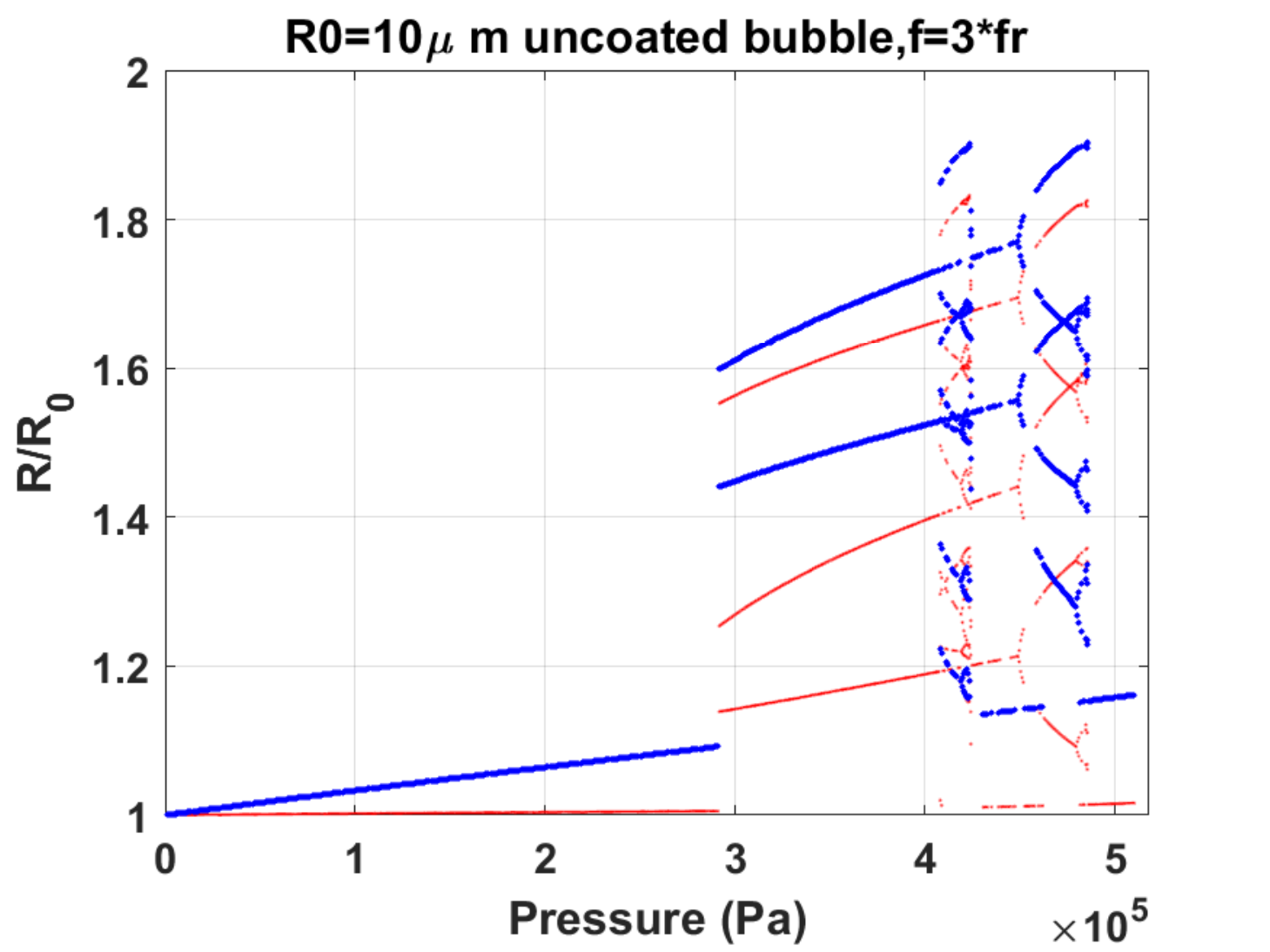}} \scalebox{0.43}{\includegraphics{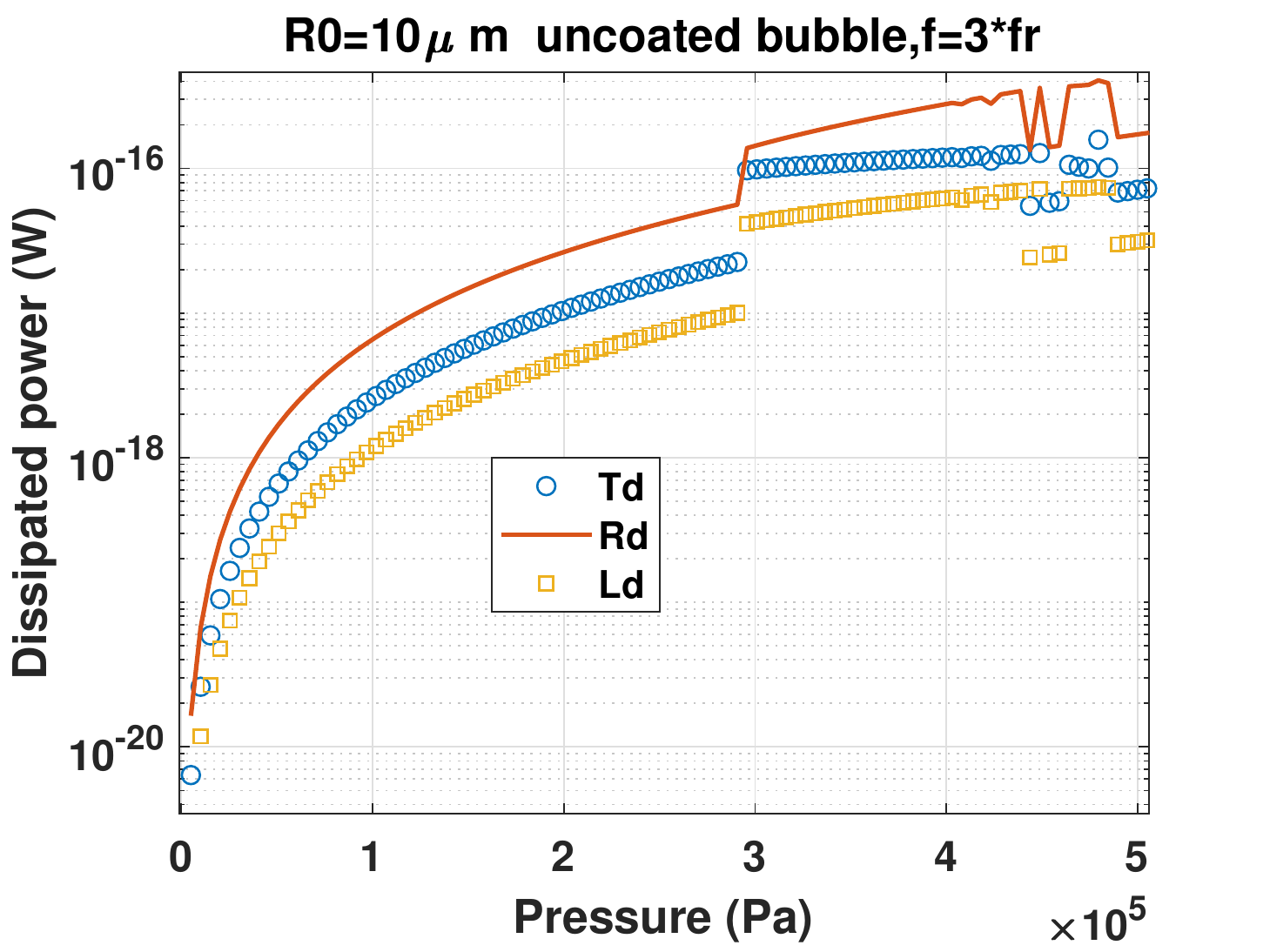}}\\
			\hspace{0.5cm} (g) \hspace{6cm} (h)\\
			\caption{Bifurcation structure (left column) and the dissipated powers (right column) of the oscillations of an uncoated air bubble with $R_0=10\mu m$ for  $f=1.2f_r$ (a-b)-$f=1.75f_r$ (c-d)- $f=2f_r$ (e-f) \& $f=3f_r$ (g-h).}
		\end{center}
	\end{figure*}
	\\Figure 2a shows the case of sonication with $f=1.2f_r$. We have chosen this frequency as the bubble is able to undergo non-destructive Pd ($R/R_0<2$). We have previously \cite{58} shown that for uncoated bubbles sonication with $f=f_r$ most likely results in bubble destruction before the development of any P2 oscillations. This was also seen in Fig. 1g ($f=f_r$). Fig. 2a shows that when $f=1.2f_r$, radial oscillations are initially of P1 and monotonically increase in amplitude as excitation pressure increases. At $P_a\approxeq 180 kPa$ Pd takes place; P2 oscillations then undergo a SN bifurcation to a P3 oscillations (properties of this P3 oscillation has been studied in \cite{58}) which can be concomitant with bubble destruction as $R/R_0>2$. The bubble oscillations return to P2 after a very small window of chaos. Another chaotic window appears through successive Pd. A giant P2 resonance emerges out of the chaotic window when $P_a>220 kPa$ which later undergo successive Pds to chaos. The dissipated powers are shown in Fig. 2b. Td is the strongest damping factor for pressures below 190 kPa. For $P_a<80$ kPa, $Td>Rd\approxeq Ld$. Rd becomes stronger than Ld with increasing pressure  above 80 kPa. At $P_a$=190 kPa, Rd becomes stronger than Td simultaneous with the SN bifurcation for P3 oscillations; however, as soon as P3 converts to P2, Td becomes larger than Rd. Emergence of the P2 giant resonance is simultaneous with a large increase in Rd and Ld and a subsequent decrease in Td. This can be due to the faster collapse with higher wall velocity and acceleration resulting in an increase in Rd and Ld; however, due to the fast collapse there is not enough time for temperature conduction thus Td decreases. In this region Rd is an order of magnitude larger than Ld and Td and its the only region in this pressure range where Ld is stronger than Td.\\
	Fig. 2c displays the case of sonication with $f=1.75f_r$ which is the pressure dependent SH resonance frequency of the bubble ($PDf_{sh}$ \cite{59}). This frequency is chosen as the SN bifurcation leads to non-destructive oscillations. Oscillations are of P1 initially; pressure increase results in Pd at $\approxeq 130 kPa$. P2 oscillations (with two maxima) undergo a SN bifurcation to P2 oscillations (with one maximum) of higher amplitude $\approxeq 130 kPa$.  At $\approxeq 180 kPa$ second maxima re-emerges with the same amplitude of the smaller solution in the red curve (indicating that wall velocity is in phase with the excitation pressure once every two acoustic cycles). At $\approxeq 200 kPa$, $R/R_0=2$ (black horizontal line); beyond this pressure the bubble may not sustain non-destructive oscillations. P2 oscillations undergo Pds to a P4 solution which later undergoes successive Pds to chaos at $P_a=300 kPa$. A giant P3 (with two maxima) resonance emerges out of the chaotic window at $\approxeq 390 kPa$. Fig. 2d shows that for $P_a<130 kPa$ $Td>Rd>Ld$. Occurrence of the SN bifurcation (over-saturation of SH signal \cite{59}) results in a fast increase in Rd and enhancement in the STDR. Rd grows with pressure increase during the P2 oscillations; however Td and Ld do not increase. Rd, Td and Ld undergo sporadic fluctuations during chaos. Emergence of giant resonance results in a sharp increase in Rd and Ld and a small decrease in Td. $Rd>Ld>Td$ for the P3 giant resonance oscillations regime. The decrease in Td and the faster and larger radial collapses indicate that higher temperatures are generated while the heat conduction becomes limited. The higher temperatures can have consequences in enhancing chemical reactions within the bubble.\\
	When sonicated with $f=2f_r$ ($f_{sh}$), oscillations undergo a Pd at $P_a=100 kPa$; P2 oscillations increase in amplitude and evolve in a shape of a bow-tie \cite{59}. Consistent with previous observations \cite{59} sonication with $f_{sh}$ results in the largest pressure range with stable P2 oscillations. At $P_a\approxeq 280 kPa$ a small window of P6 oscillations appear through a SN bifurcation with each solution undergoing Pds to P12 (the properties of this oscillation regime have been studied in the appendix of (Chapter 4) \cite{59}).  Oscillations return to P2 which then undergo Pd to P4 oscillations. For a small window of excitation pressure  P12 oscillations appear through a SN bifurcation; however, here because $R/R_0>2$ the bubble most likely undergoes destruction. P12 oscillations undergo P24 oscillations for a small window and then disappear as P4 oscillations emerge. At $\approxeq 320 kPa$ chaos appears. A P3 giant resonance emerges out of the chaotic window which later undergo successive Pds to chaotic oscillations. For $P_a<220 kPa$, $Td>Rd>Ld$. After the occurrence of Pd, Td remains relatively constant with increasing pressure while Rd grows faster than Ld as pressure increases.  Eventually at $P_a\approxeq 200 kPa$ Rd becomes equal to Td. The occurrence of giant resonance results in a sharp increase in Rd and Ld and Rd becomes the strongest damping factor with $Rd>Ld>Td$. Regeneration of chaos results in a decrease in Rd and Ld with $Rd \approxeq Td>Ld$.\\
	When $f=3f_r$ (Fig. 2g); radial oscillations grow very slowly and monotonically with pressure; at $P_a\approxeq 300 kPa$ a SN bifurcation takes place and oscillations become P3 (3 solutions for the red curve with 2 maxima).  Properties of these oscillations have been studied in \cite{60,61,62}. At $\approxeq 410 kPa$ oscillations undergo a SN bifurcation to P6 oscillations for small excitation pressure window which then transition to P12 and then back to P3 oscillations. P12 occurs at $\approx  450 kPa$ through Pds. P12 oscillations then switch to P1 oscillation with pressure increase. Power dissipation curve is shown in Fig. 2h. Here Rd is the strongest damping mechanism for all the studied pressure ranges with $Rd>Td>Ld$. SN bifurcation results in a sharp increase in the dissipated powers at 300 kPa with Td exhibiting the largest increase. 
	\subsection{Bifurcation structure and power dissipation of the oscillations of the coated  bubbles}   
	\subsubsection{The case of a coated C3F8 bubble with $R_0=1 \mu m$} 
	\begin{figure*}
		\begin{center}
			\scalebox{0.4}{\includegraphics{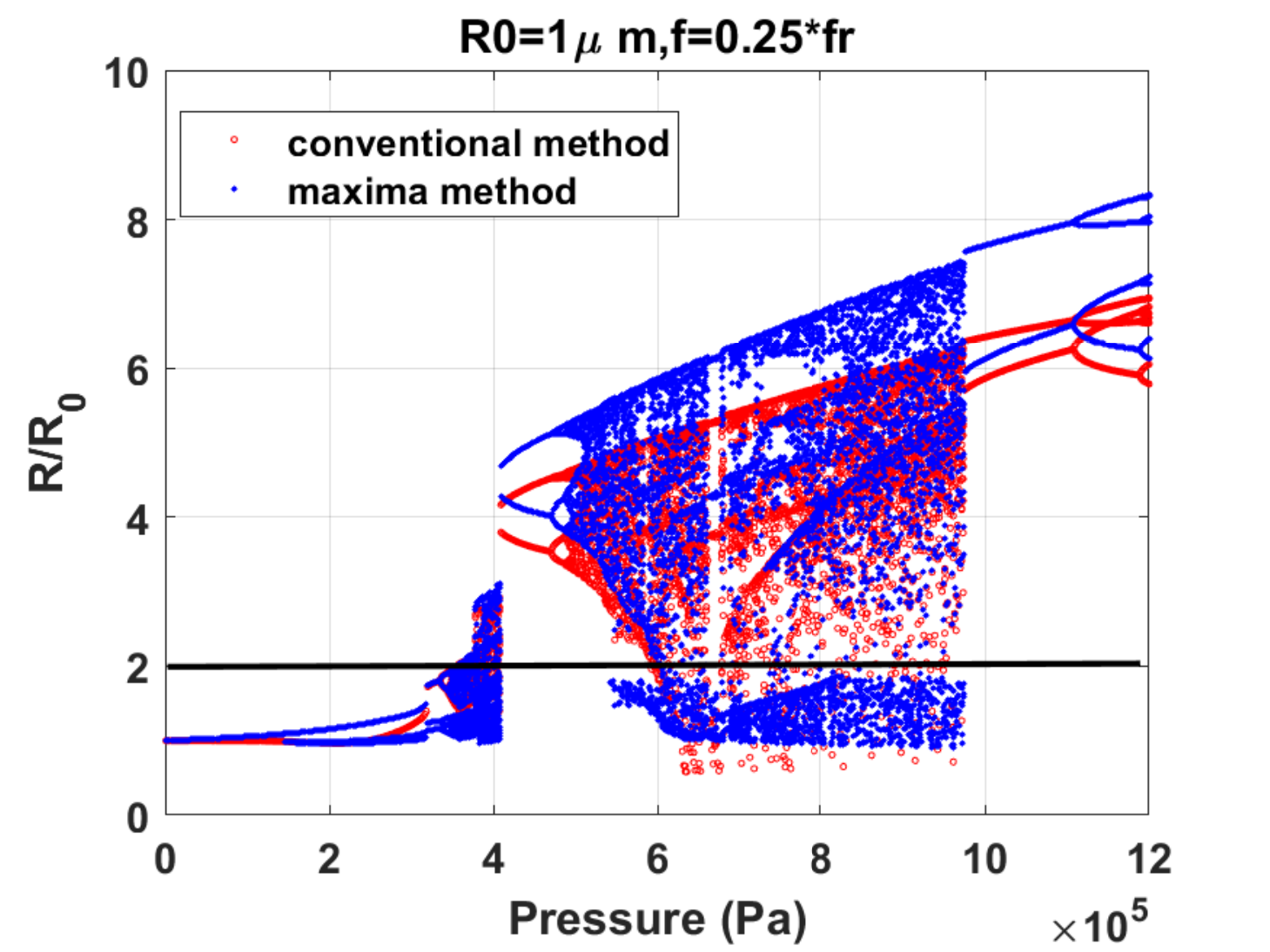}}  \scalebox{0.4}{\includegraphics{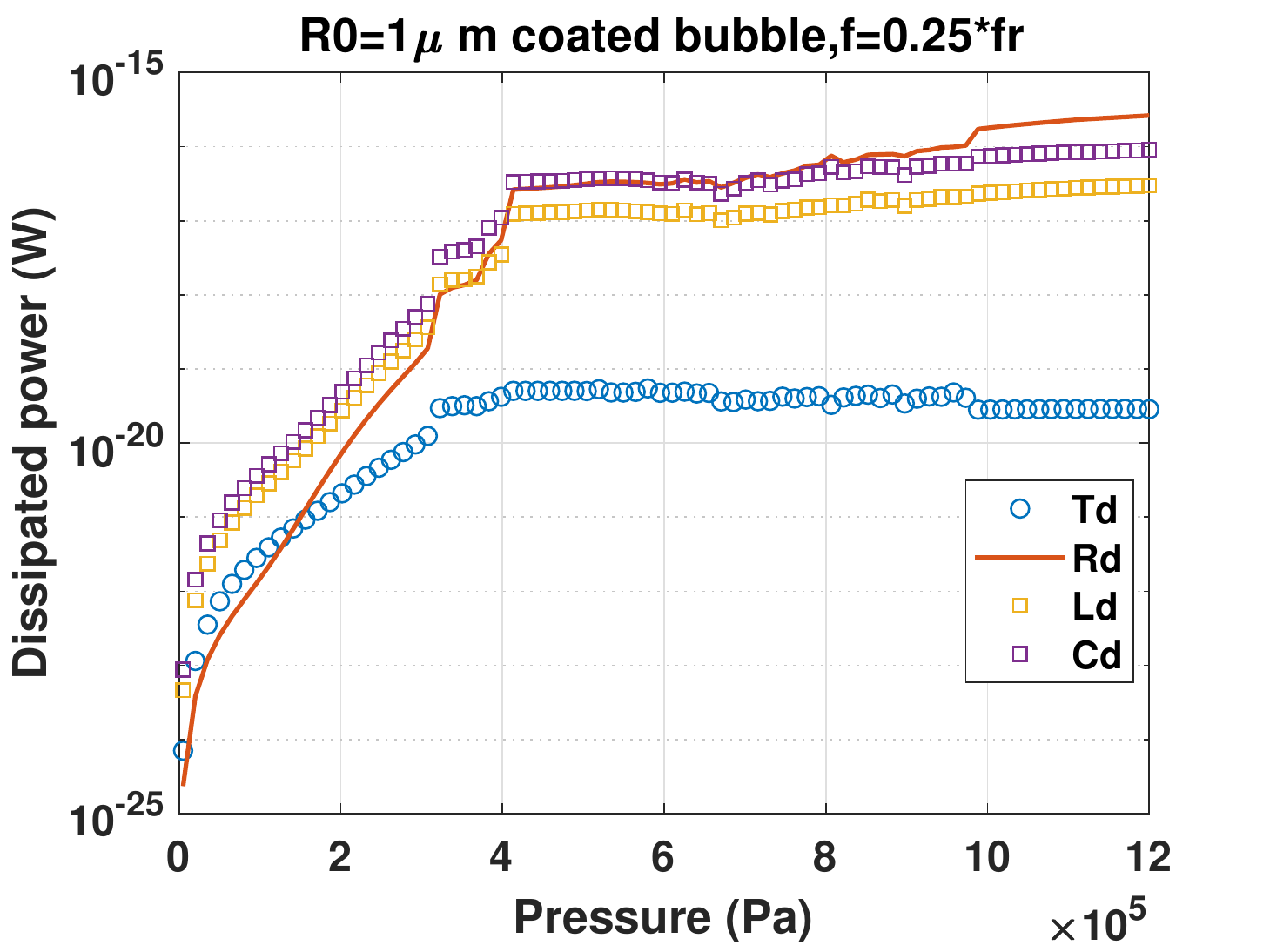}}\\
			\hspace{0.5cm} (a) \hspace{6cm} (b)\\
			\scalebox{0.43}{\includegraphics{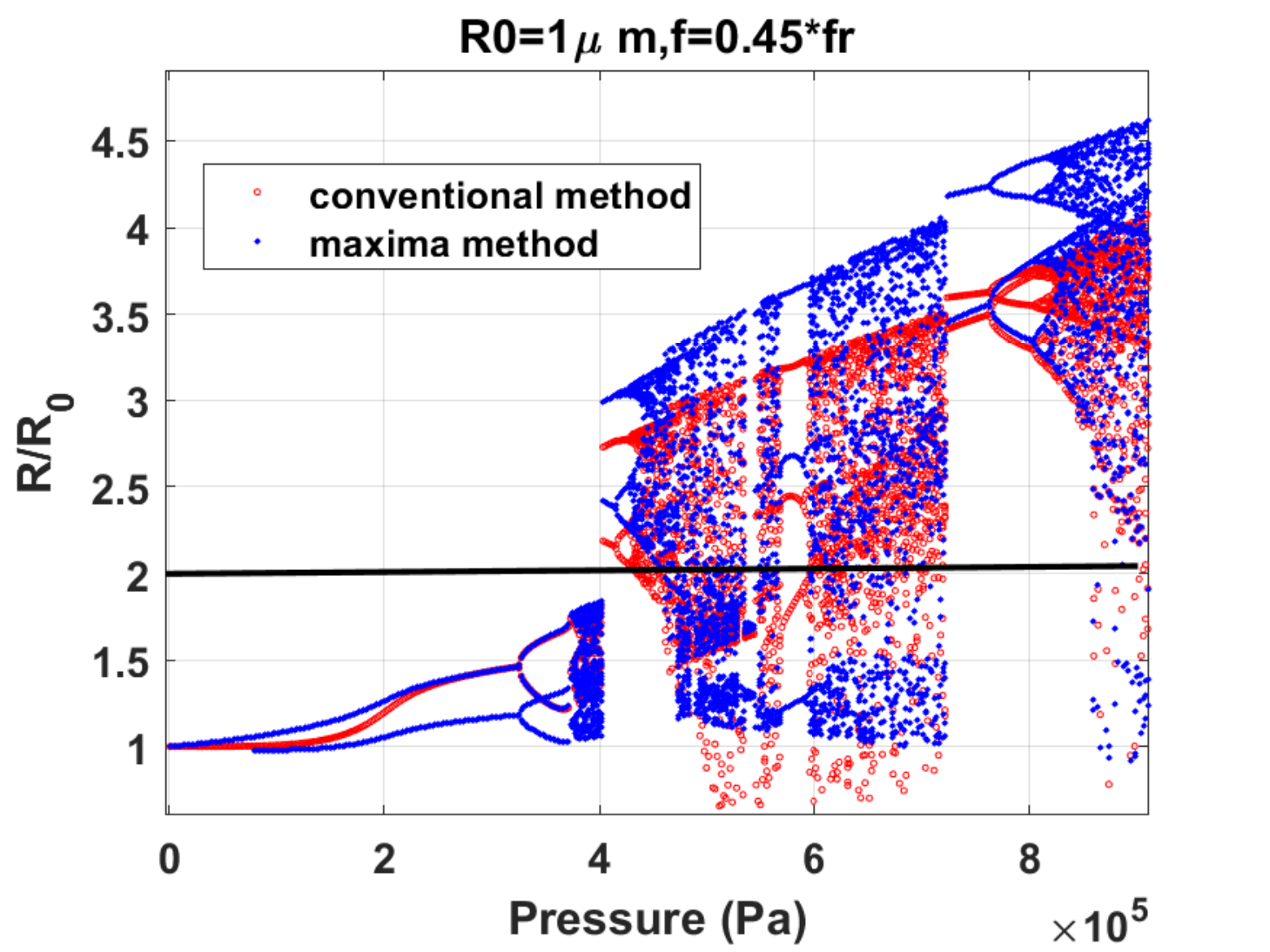}} \scalebox{0.43}{\includegraphics{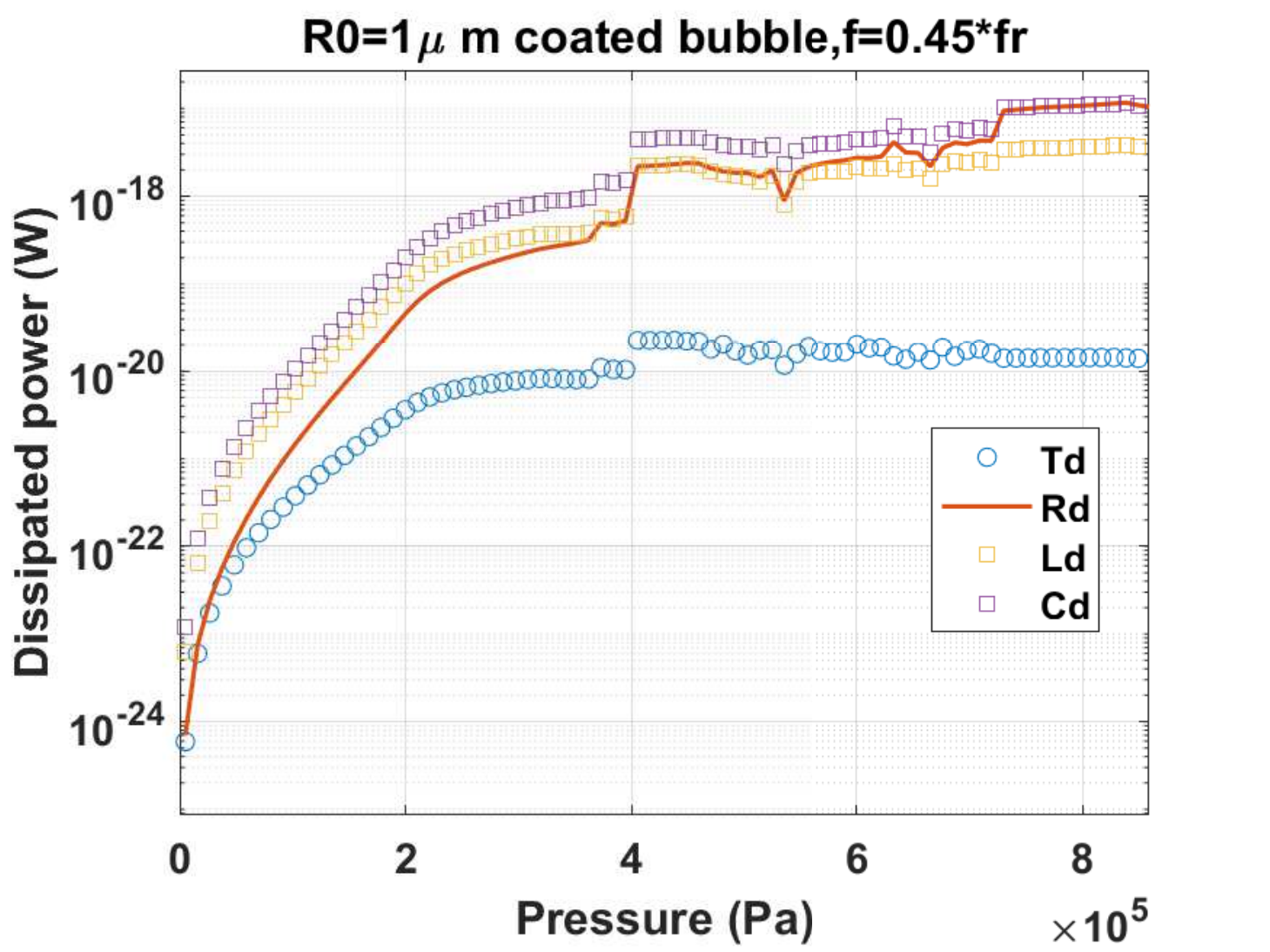}}\\
			\hspace{0.5cm} (c) \hspace{6cm} (d)\\
			\scalebox{0.43}{\includegraphics{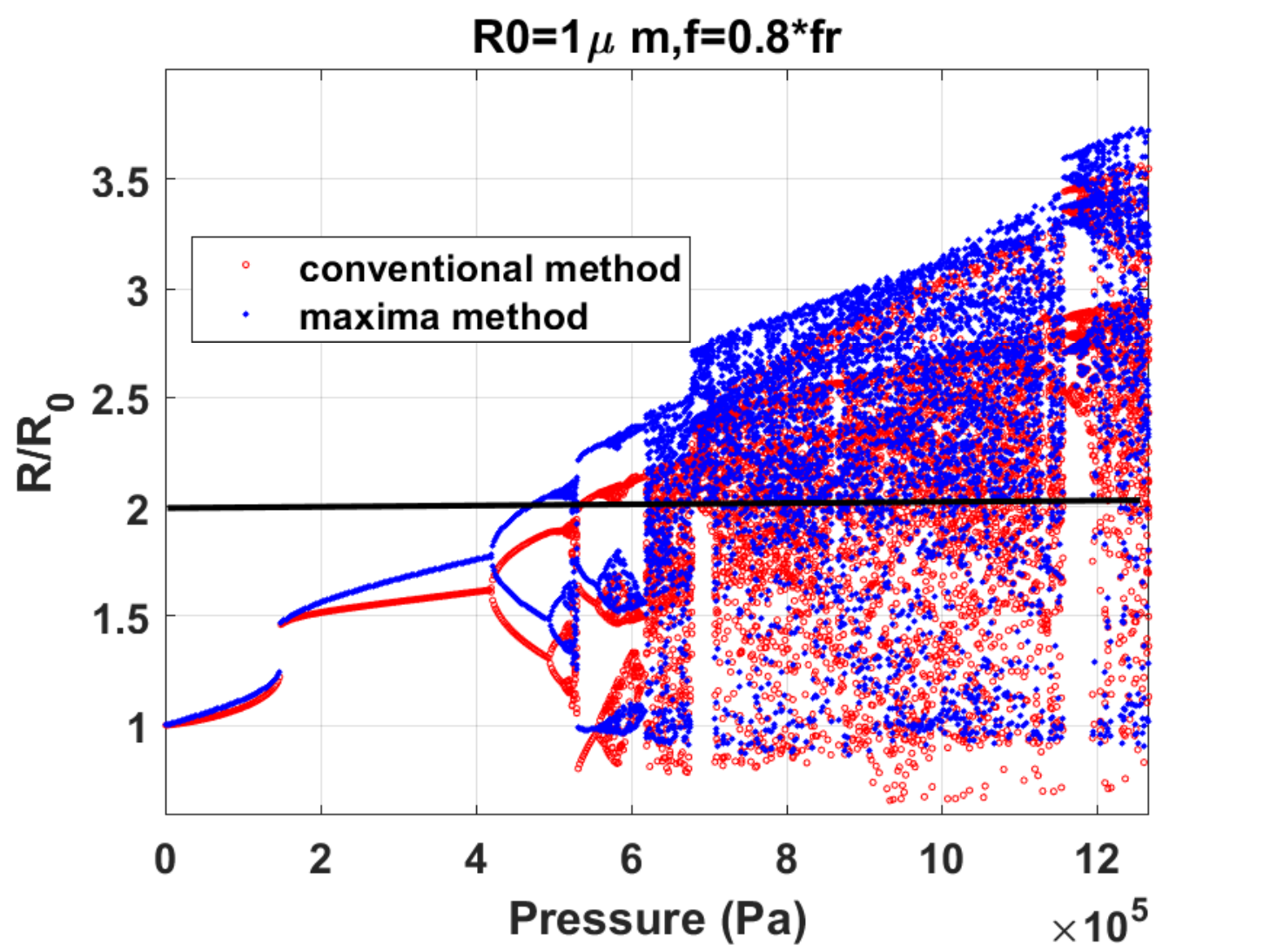}} \scalebox{0.43}{\includegraphics{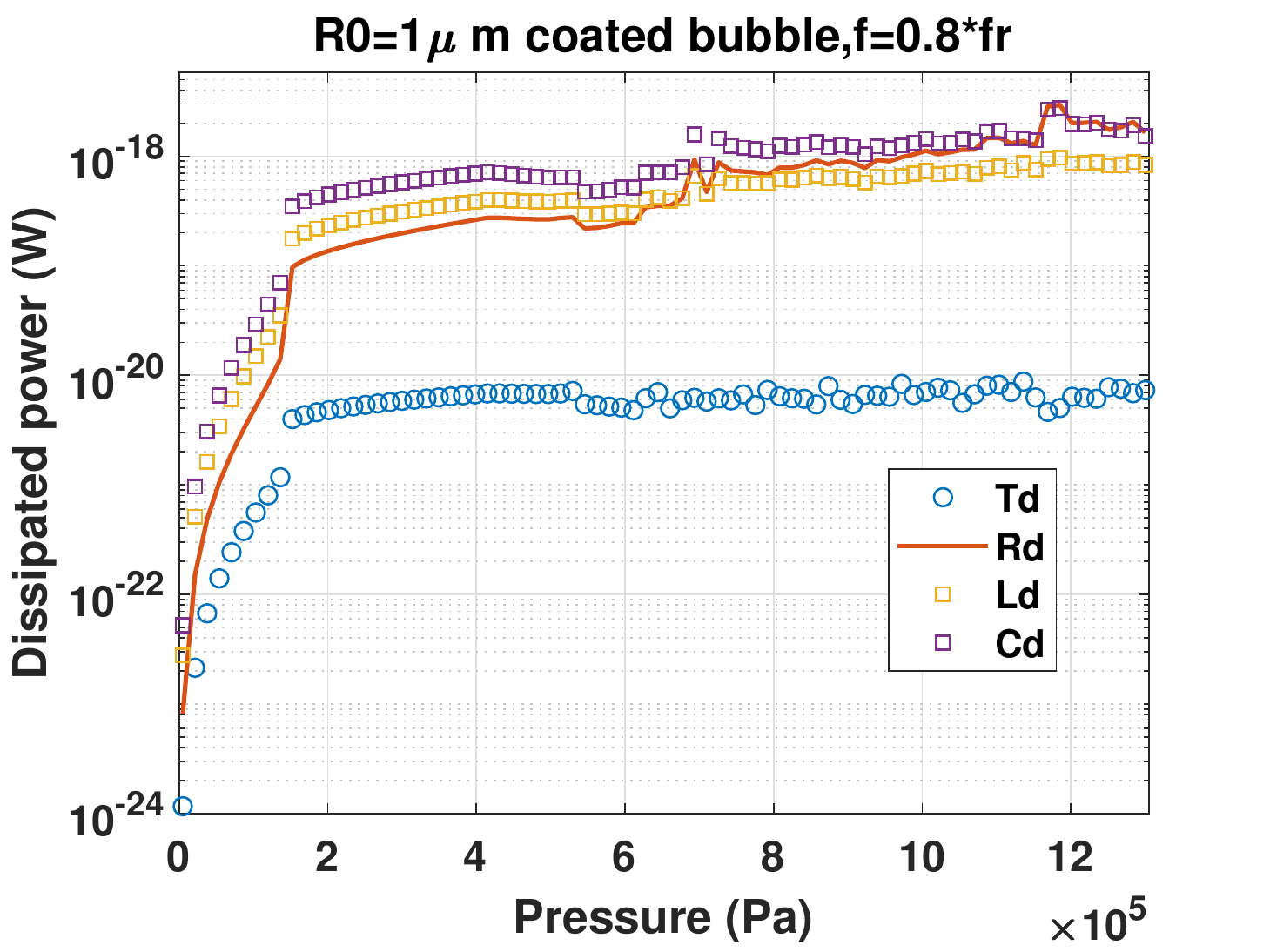}}\\
			\hspace{0.5cm} (e) \hspace{6cm} (f)\\
			\scalebox{0.43}{\includegraphics{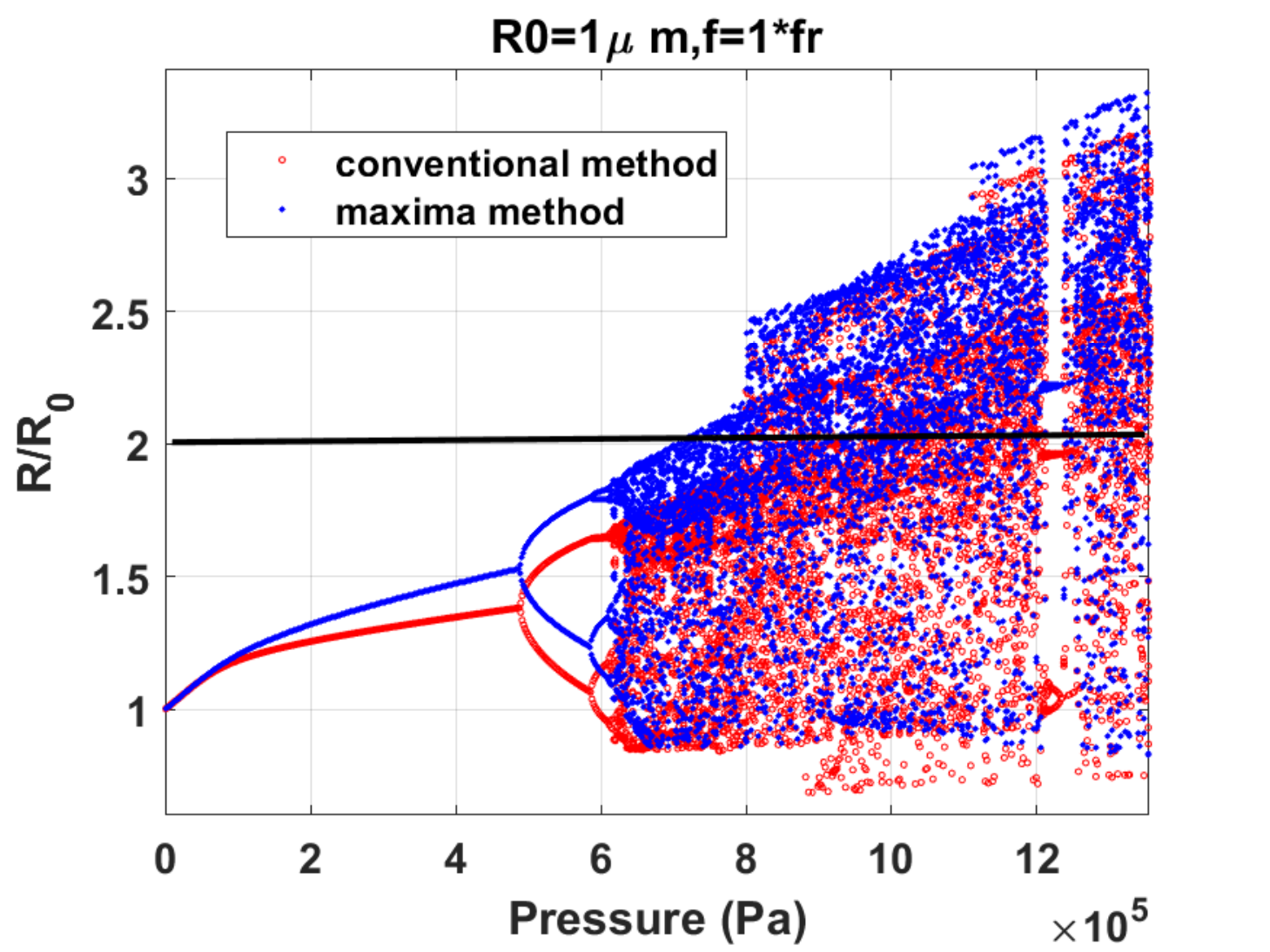}} \scalebox{0.43}{\includegraphics{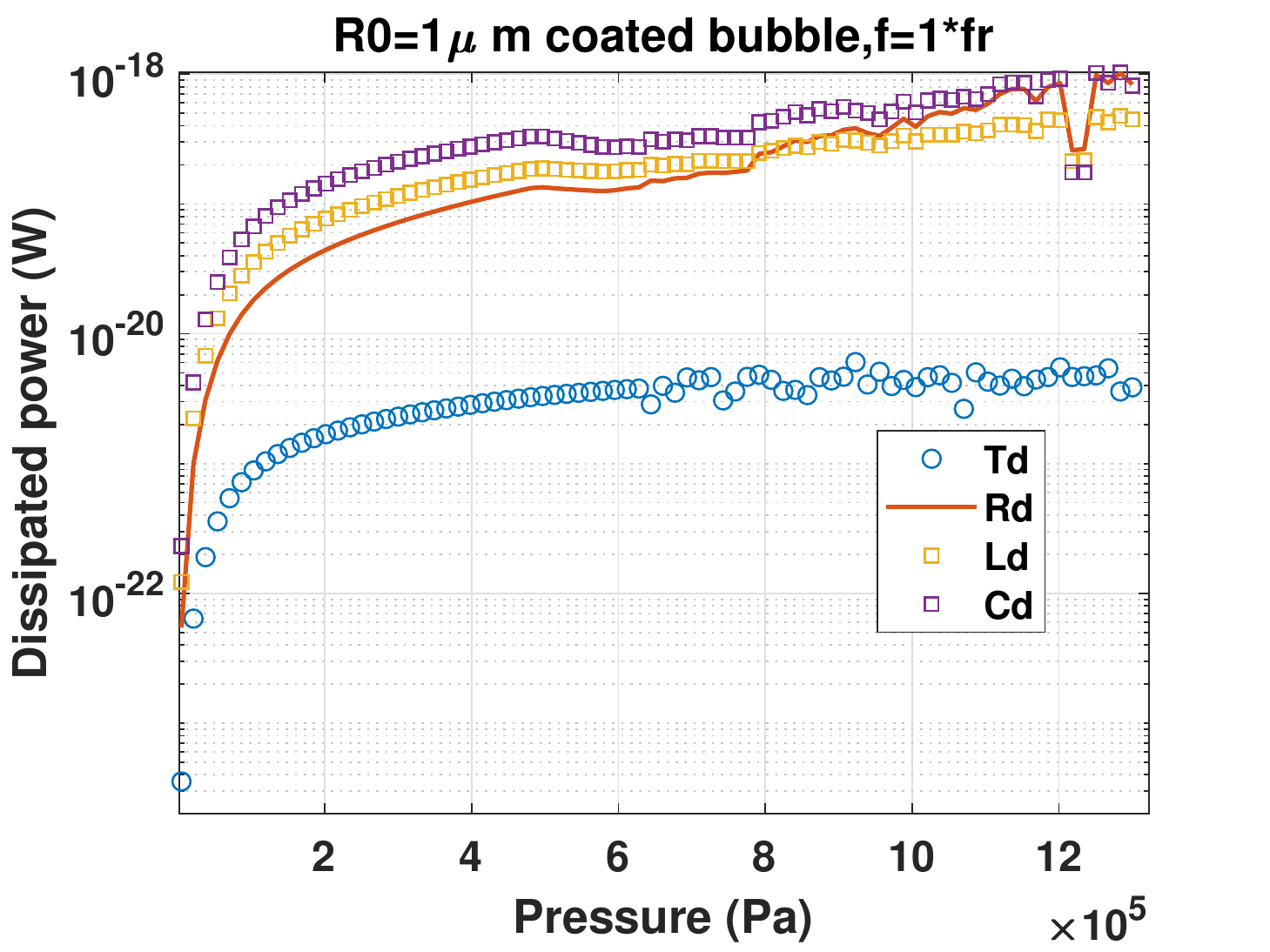}}\\
			\hspace{0.5cm} (g) \hspace{6cm} (h)\\
			\caption{Bifurcation structure (left column) and the dissipated power as a function of pressure (right column) of the oscillations of a coated C3F8 bubble with $R_0=1\mu m$ for $f= 0.25f_r$ (a-b)-$f=0.45f_r$ (c-d)- $f=0.8f_r$ (e-f) \& $f=f_r$ (g-h).}
		\end{center}
	\end{figure*}
	Fig. 3a shows the dynamics of a C3F8 coated bubble with $R_0=1 \mu m$ when $f=0.25f_r$. Oscillations are initially P1 with one maximum, later at about $P_a= 160 kPa$, 3 maxima are generated in the bubble oscillations which grow in amplitude as pressure increases, undergoing a SN bifurcation to higher amplitude oscillations at about 320 kPa. In this region (P1 with 3 maxima) the 3rd harmonic of the backscattered pressure is maximum and the bubble is in the 3rd order SuH oscillation mode.  Pd occurs at about 350 kPa, leading to a P2 signal with 6 maxima and $7/2$ UH oscillations. Oscillations become chaotic through successive Pds at $\approxeq 365 kPa$; at the same time $R/R_0$ exceeds 2 thus, the bubble may not sustain long lasting non-destructive oscillations beyond this pressure. At 405 kPa, a P2 oscillation regime with large amplitude emerges out of the chaotic window. Oscillations become chaotic again through successive Pds of the P2 signal and the chaotic window extends until $\approx 970 kPa$. Above this pressure a P2 giant resonance emerges out of chaos, and later undergoes successive Pds to chaos. Fig. 3b shows that at pressures lower than 160 kPa (generation of 3 maxima in the osculations) $Cd>Ld>Td>Rd$. Above 160 kPa, Rd increases beyond Td and $Cd>Ld>Rd>Td$ until SN bifurcation occurs at about 320 kPa. Rd grows faster and when SN occurs it undergoes a sharp increase alongside Ld and Cd. $Cd>Ld=Rd>Td$ above SN bifurcation and during UH oscillations. Rd exceeds Ld when $R/R_0 >2$ and during the chaotic oscillations. When the large amplitude P2 oscillations are generated Rd undergoes the sharpest increase and becomes equal to Cd. Thereafter $Cd=Rd>Ld>Td$ until about 900 kPa where Rd slightly exceeds Cd. Emergence of the giant P2 resonance leads to a sharp increase in Rd and a decrease in Td similar to previous cases.\\
	When $f=0.45 f_r$ (Fig. 3b) P1 oscillations increase in amplitude with pressure and at about 85 kPa a second maxima appear in the bubble oscillations. Oscillations keep growing and at about 300 kPa the red curve becomes equal in amplitude to the highest amplitude maxima (indicating the wall velocity of one of the maxima becomes in phase with the driving acoustic field). At about 320 kPa, Pd occurs and oscillations become P2 with 4 maxima indicating the generation of $5/2$ UH resonance. A small chaotic window appears, and at about 400 kPa a P2 oscillation regime with higher amplitude emerges out of the chaotic window. At this point, since $R/R_0$ exceeds 2 the bubble possibly undergoes destruction. The dissipation curves are shown in Fig. 3d. For lower pressures $Cd>Ld>Rd=Td$; however, due to the faster growth of Rd compared to other dissipation mechanisms, it supersedes Td at about 40 kPa and becomes equal to Ld when Pd takes place. When the P2 oscillations with higher amplitude emerge out of the chaotic window, the dissipation powers undergo a sharp increase with $Cd>Rd=Ld>Td$.\\
	The case of sonication with $f=0.8f_r$ ($PDf_r$) is presented in Fig.7e. P1 oscillations undergo SN bifurcation to higher amplitude at $\approxeq 150 kPa$ and at the same time the value of the red curve becomes equal to the maxima in the blue curve (indicating the wall velocity is in phase with the driving signal). Oscillations grow with pressure increase and at 410 kPa, Pd takes place leading to P2 oscillations until 500 kPa. Chaos is then generated through successive period doubling bifurcations at 510 kPa. For this frequency $Cd>Ld>Rd>Td$ in the studied pressure range. There is a sharp increase in the dissipation power when SN takes place. Furthermore concomitant with Pd; Cd, Ld and Rd  decrease due to reduced wall velocities \cite{32}.\\
	Fig. 3g displays the case of sonication with $f=f_r$. Initially the value of the red curve is equal to the oscillation amplitude in the blue curve and above 100 kPa the two curves diverge (this is because $f_r$ shifts to $PDf_r$ as pressure increases \cite{32} and when $f=f_r$ oscillations are only resonant at lower pressures) and Pd takes place at $\approxeq 480 kPa$. P2 oscillations undergo successive Pds to chaos at 620 kPa. Chaos stretches beyond 1 MPa with oscillation amplitudes exceeding $R/R_0=2$ at 700 kPa. The dissipated power curves are presented in Fig. 3h. Similar to the case of $f=0.8f_r$, $Cd>Ld>Rd>Td$ and occurrence of Pd leads to a slight decrease in Cd, Rd and Ld.\\
		Fig. 4a displays the case of sonication with $f=1.2f_r$. The P1 oscillation amplitude increases with pressure and Pd occurs at about 570 kPa.  A small period bubbling window takes place for $\approxeq 760-800 kPa$ and initiation of chaos is at about 860 kPa. When chaos is initiated, $R/R_0>2$. The corresponding power curves in Fig. 4b show that similar to the case of $f=0.8 f_r$ and $f_r$, $Cd>Ld>Rd>Td$ for $P_a <760 kPa$ where period bubbling takes place. Occurrence of Pd at 570 kPa is concomitant with a decrease in Cd, Ld and Rd and when bubbling occurs $Cd\approxeq Ld \approxeq Rd>Td$. Generation of sudden chaos at $\approx 880 kPa$ is simultaneous with a sudden increase in Cd, Ld and Rd with $Cd>Ld>Rd>Td$ right after the onset of chaos. Further increases in pressure result in a faster growth in Rd making Rd $\approxeq$ Cd at $\approx$ 1.1 MPa.\\
	Figure 8c shows the dynamics of the bubble in case of sonication with $f=1.6f_r$ ($Pdf_{sh}$). P1 oscillation amplitude grow with pressure increase and at 580 kPa a SN bifurcation from P1 oscillations to P2 oscillations of higher amplitude takes place. P2 oscillations then grow with pressure increase and undergo further Pds. After a small window of P6-P12 oscillations chaos is generated. At 1.2 MPa oscillation amplitude exceeds 2 and possible bubble destruction may take place. The corresponding power graphs are depicted in Fig. 4d. $Cd>Ld \approxeq Rd>Td$ for $P_a<800 kPa$. Occurrence of the SN results in a sharp increase in Cd, Rd, Ld and Td with Td exhibiting the highest increase.\\
	When $f=2f_r$ (Fig. 4c) P1 oscillations undergo Pd at 400 kPa; P2 oscillations later evolve in a form of a bow-tie \cite{58} (red curve) undergoing successive Pd to chaos. For $P_a$ less than the pressure threshold of Pd, $Rd\approxeq Cd>Ld>Td$. When Pd occurs, Cd grows larger than Rd and Rd becomes equal to Ld. Td exhibits the largest growth when Pd occurs. During P2 oscillations Rd grows faster than other damping factors exceeding Ld at $\approxeq 900 kPa$.\\
	When $f=3f_r$ (Fig. 4g) P1 oscillations undergo a SN bifurcation to P3 oscillations of higher amplitude. Pressure increase results in an increase in the amplitude of the P3 oscillations and at $\approxeq 2.7 MPa$, Pd takes place and oscillations become P12. Later through multiple Pds a small chaotic window appears which is followed by a sudden onset of P1 oscillations for the rest of the pressures studied here. The corresponding dissipated power graphs are shown in Fig. 4h. Rd is the strongest dissipated power for the studied pressure range here with $Rd>Cd>Ld>Td$. When SN takes place, similar to the case of $R_0=4 \mu m$ all dissipated powers undergo a sharp increase; however, Td exhibits the largest growth potentially due to more surface area available for heat transfer. 
\begin{figure*}
	\begin{center}
		\scalebox{0.43}{\includegraphics{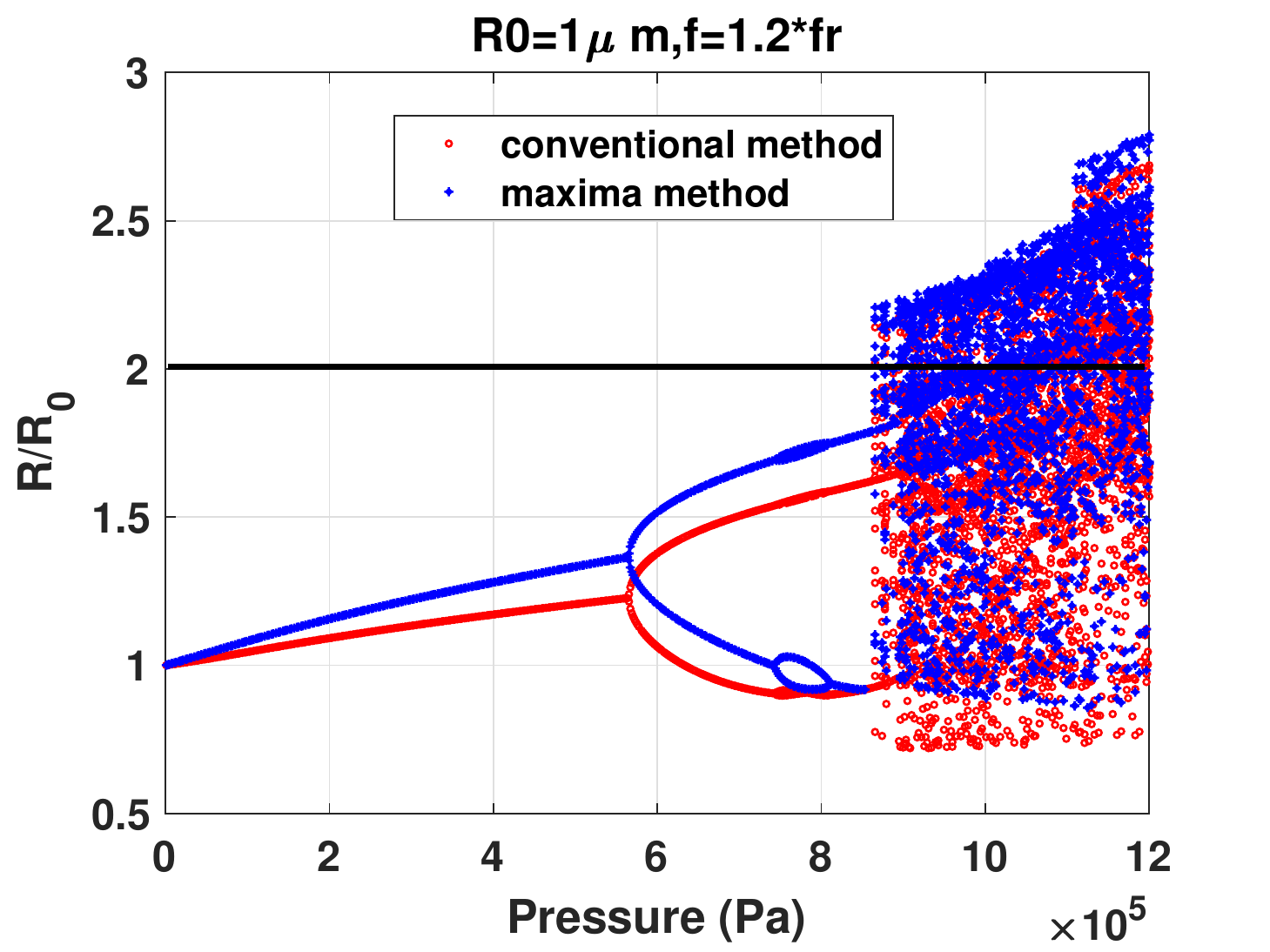}}  \scalebox{0.43}{\includegraphics{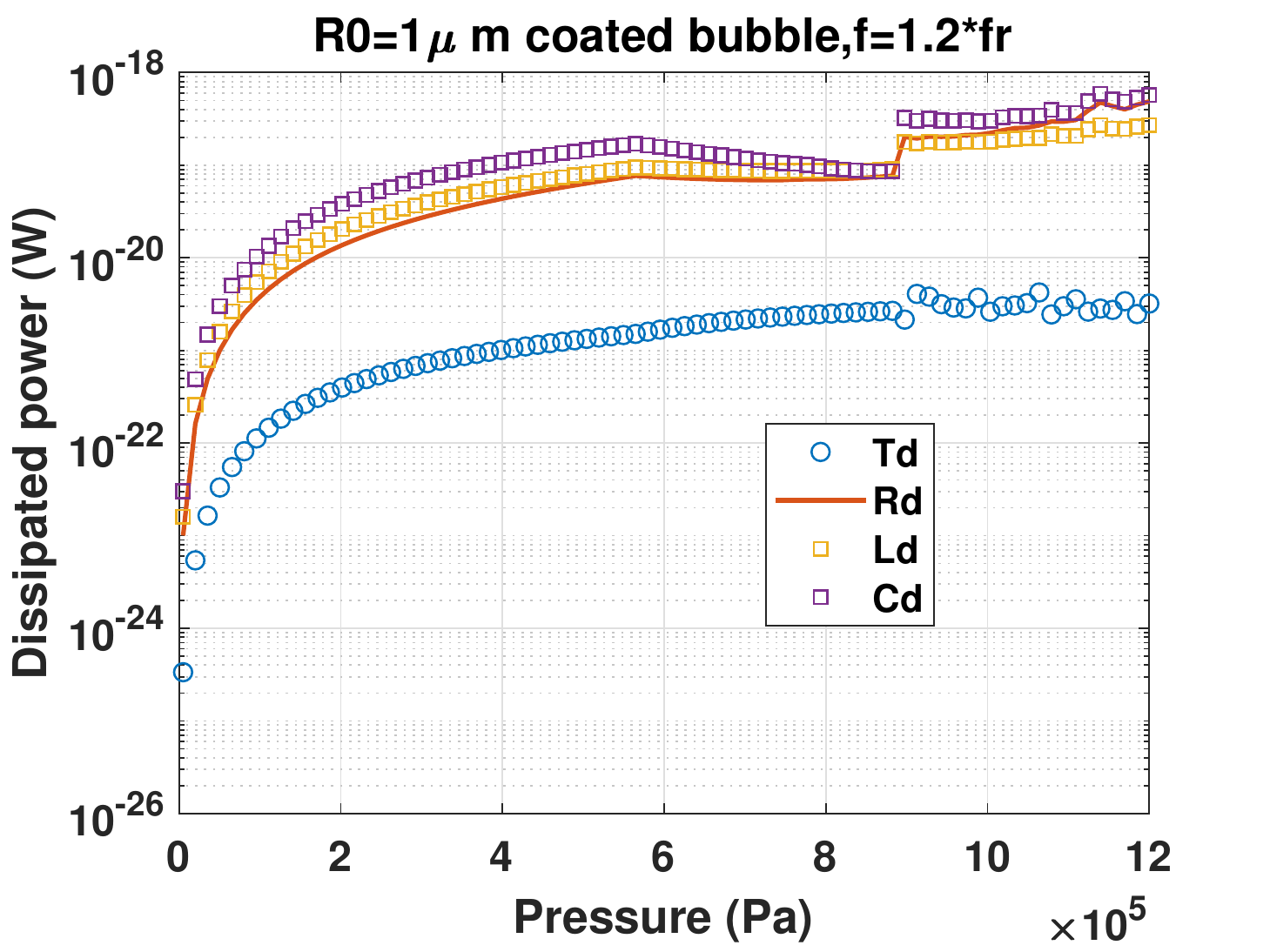}}\\
		\hspace{0.5cm} (a) \hspace{6cm} (b)\\
		\scalebox{0.43}{\includegraphics{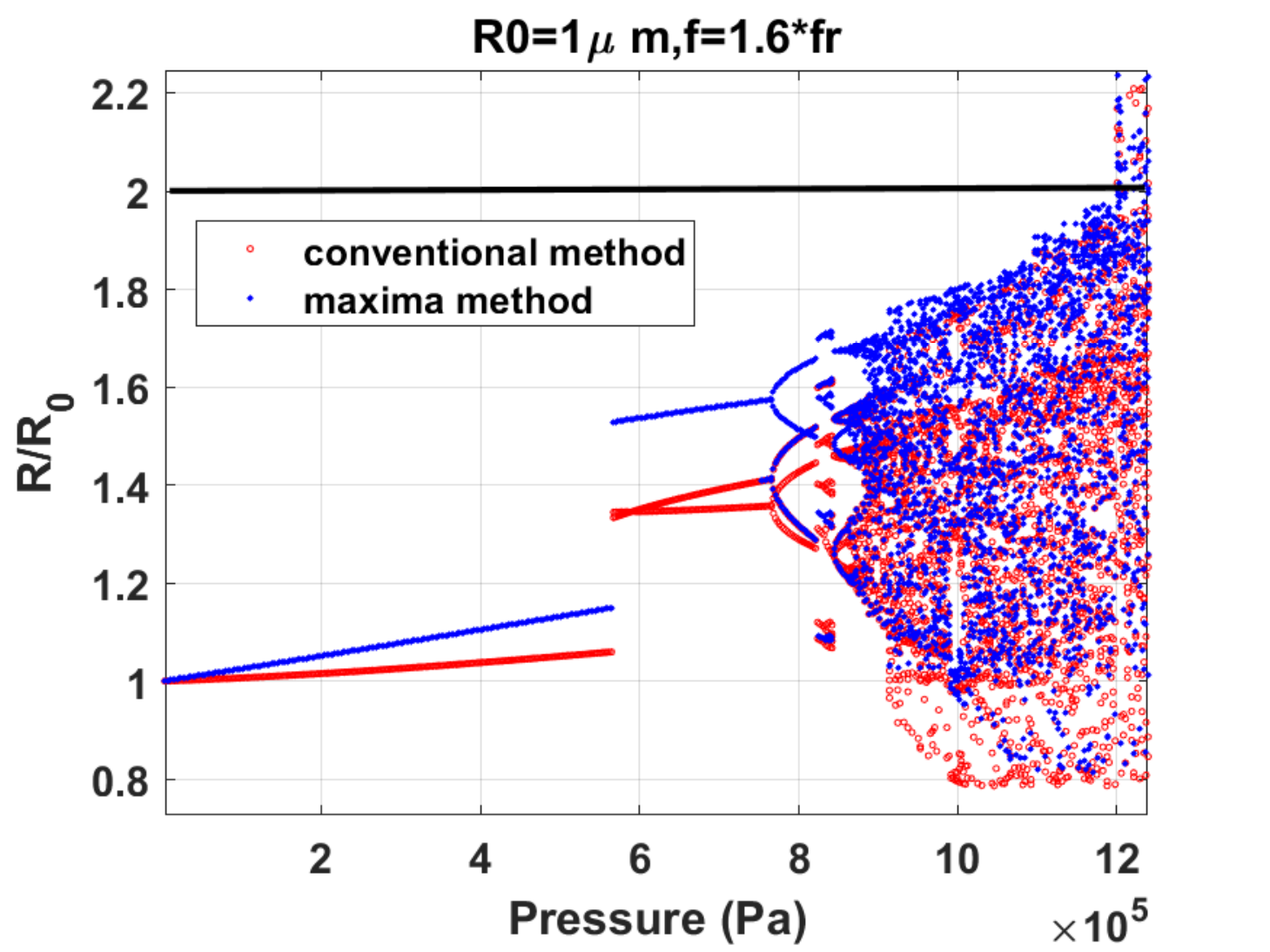}} \scalebox{0.43}{\includegraphics{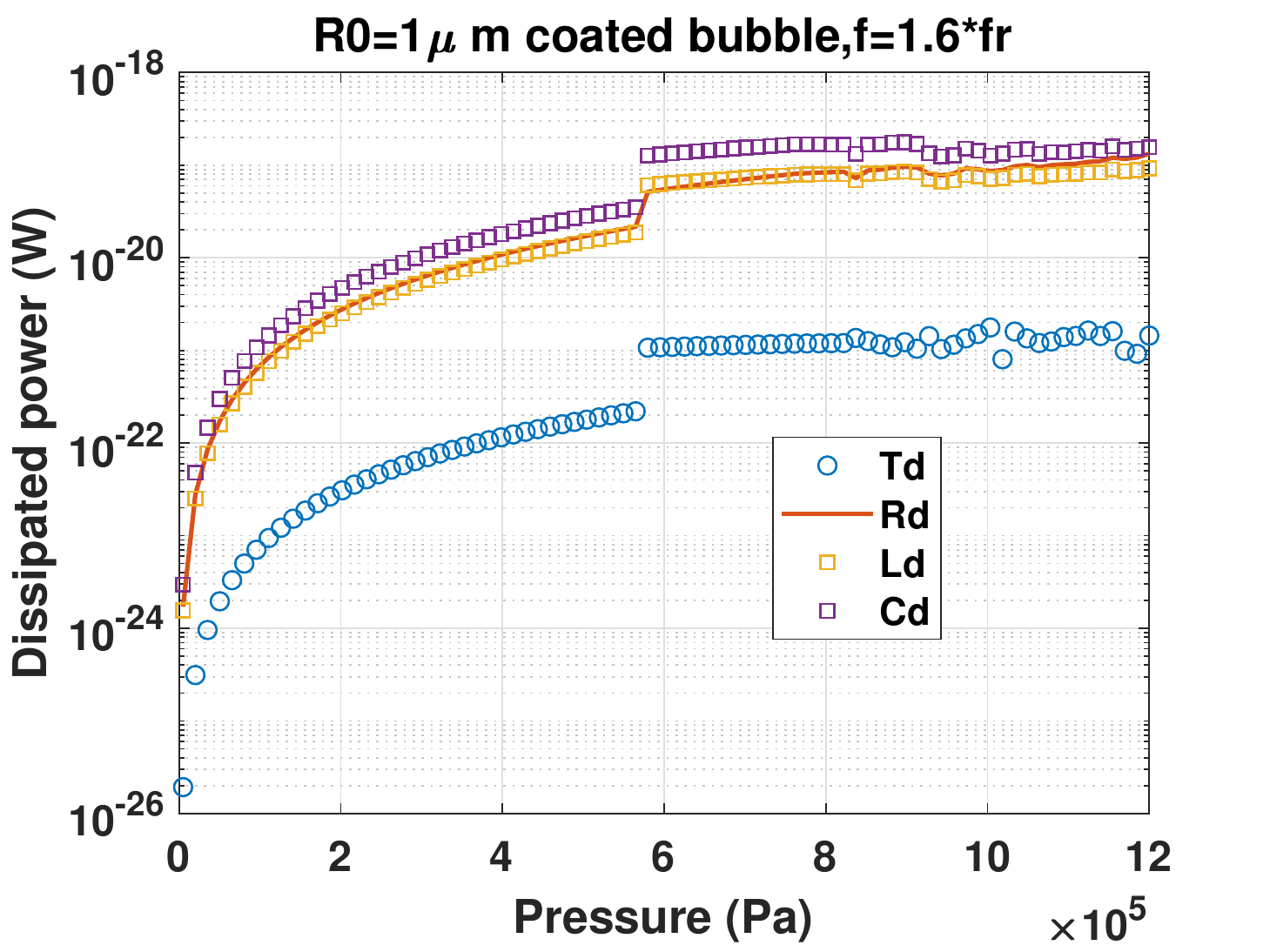}}\\
		\hspace{0.5cm} (c) \hspace{6cm} (d)\\
		\scalebox{0.43}{\includegraphics{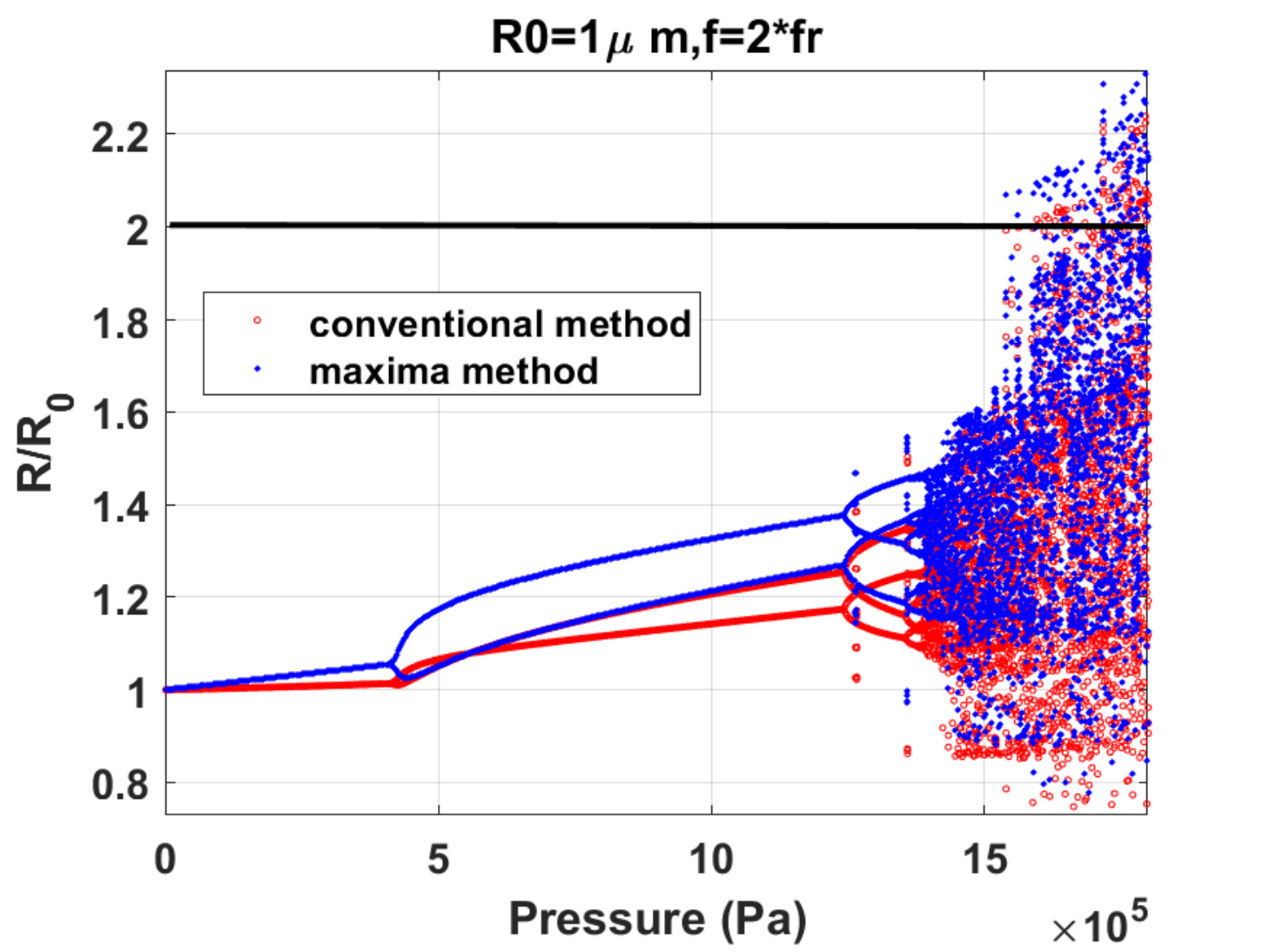}} \scalebox{0.43}{\includegraphics{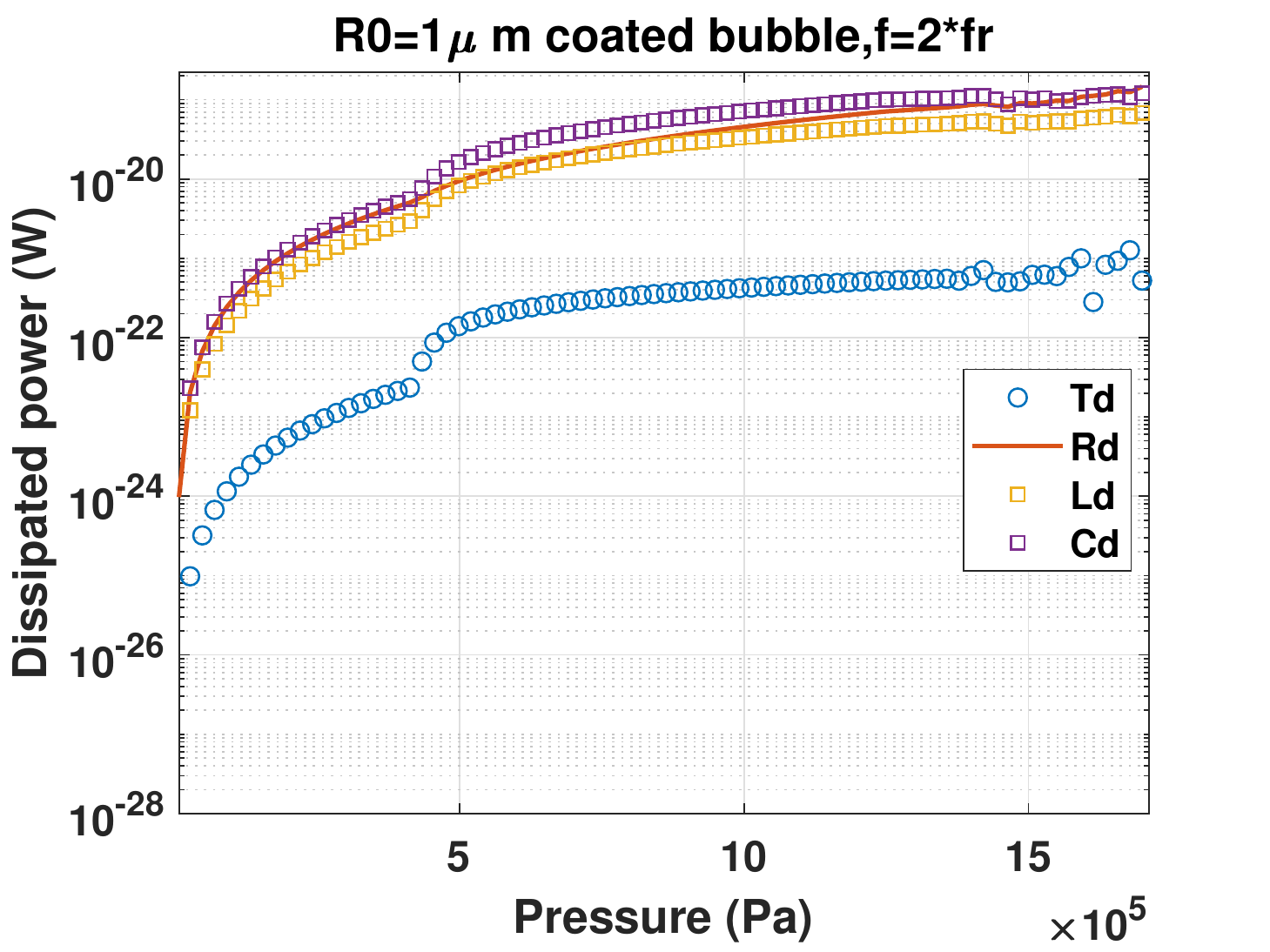}}\\
		\hspace{0.5cm} (e) \hspace{6cm} (f)\\
		\scalebox{0.43}{\includegraphics{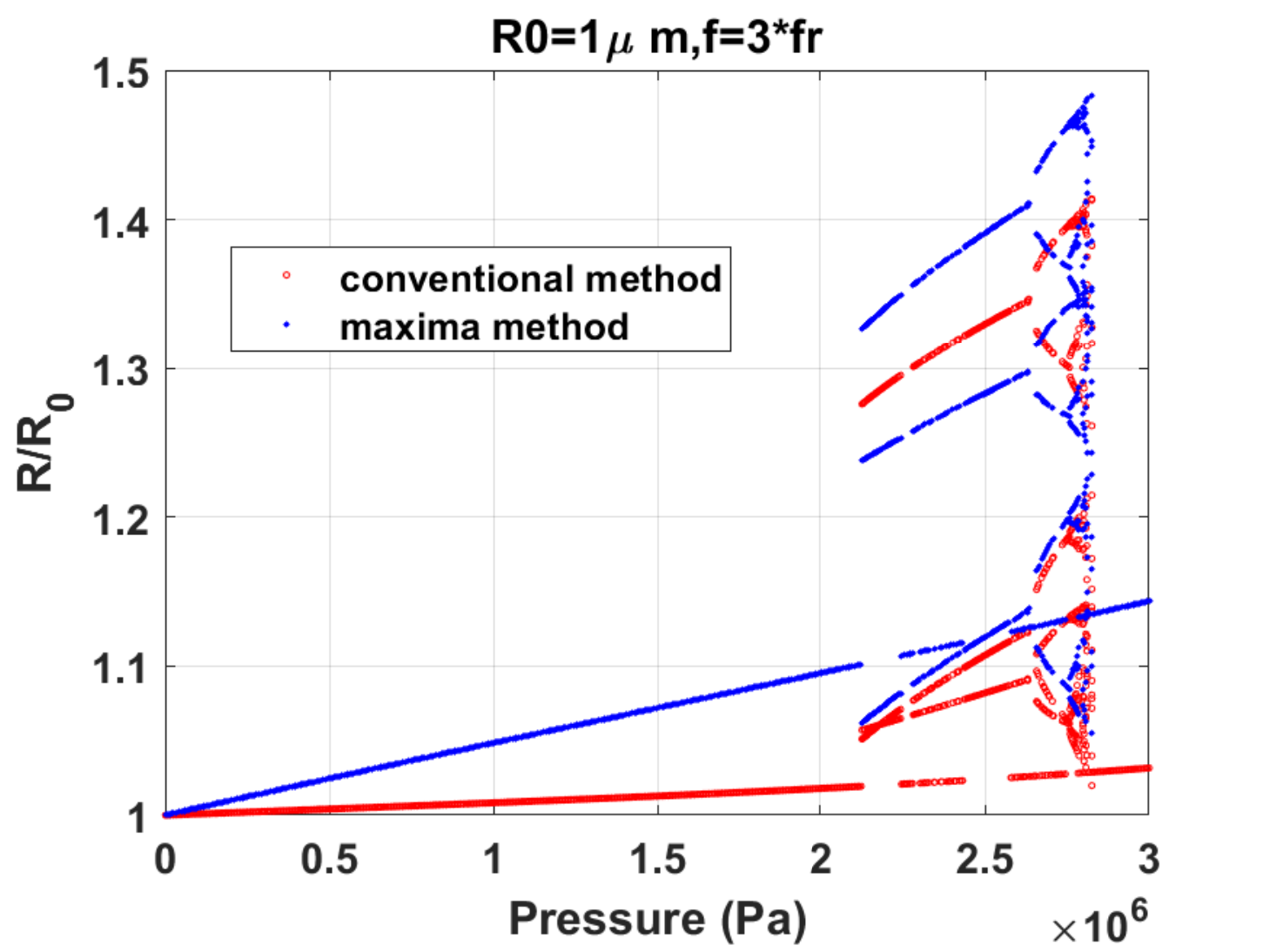}} \scalebox{0.43}{\includegraphics{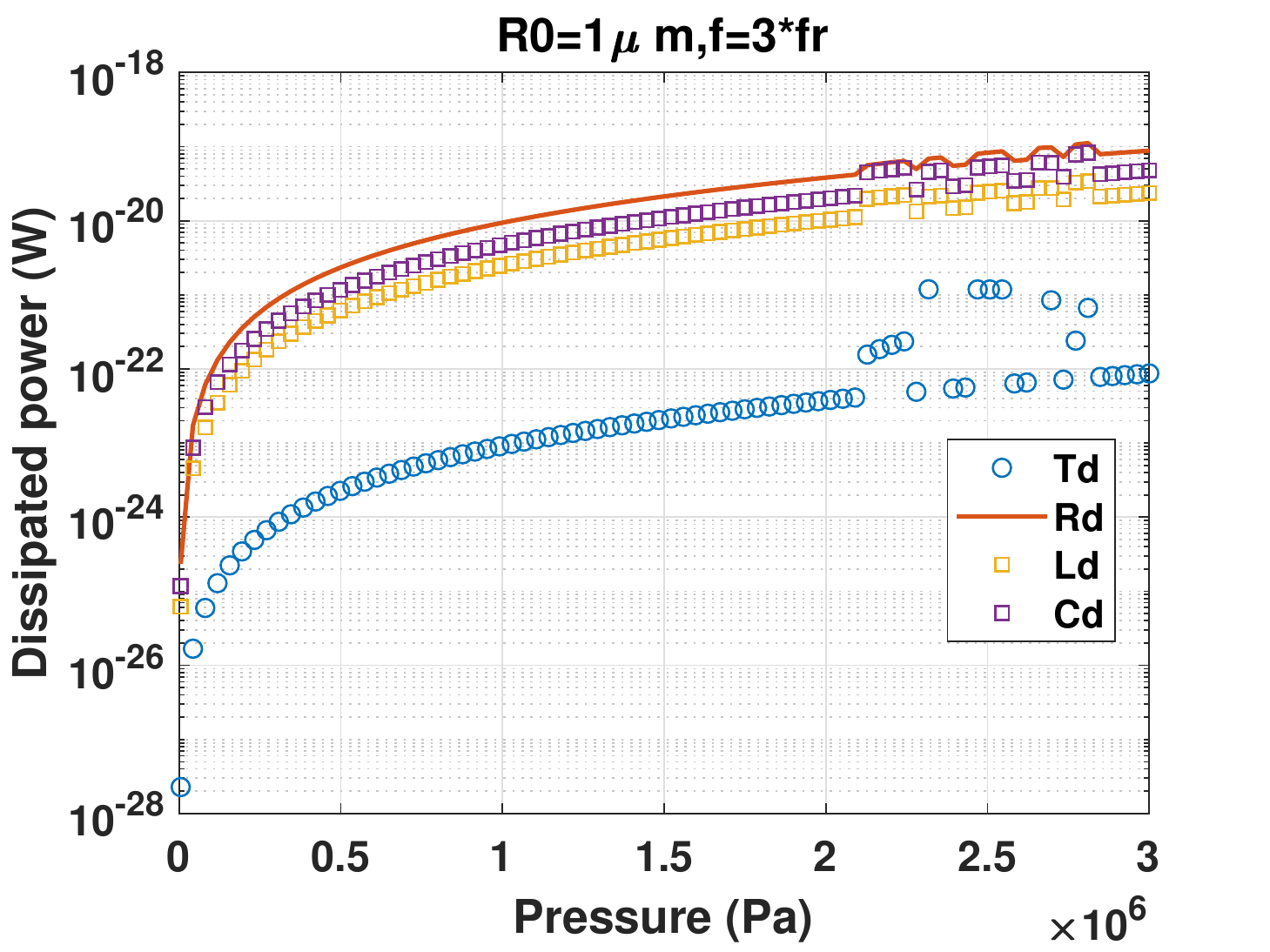}}\\
		\hspace{0.5cm} (g) \hspace{6cm} (h)\\
		\caption{Bifurcation structure (left column) and the dissipated power as a function of pressure (right column) of the oscillations of a coated C3F8 bubble with $R_0=1\mu m$ for $f= 1.2f_r$ (a-b)-$f=1.6f_r$ (c-d)- $f=2f_r$ (e-f) \& $f=3f_r$ (g-h).}
	\end{center}
\end{figure*} 
	\subsection{Concluding graphs of $|\dot{R(t)}|_{max}$, $|P_{sc}|_{max}$, total dissipated power and STDR} 
\begin{figure*}
	\begin{center}
		\scalebox{0.43}{\includegraphics{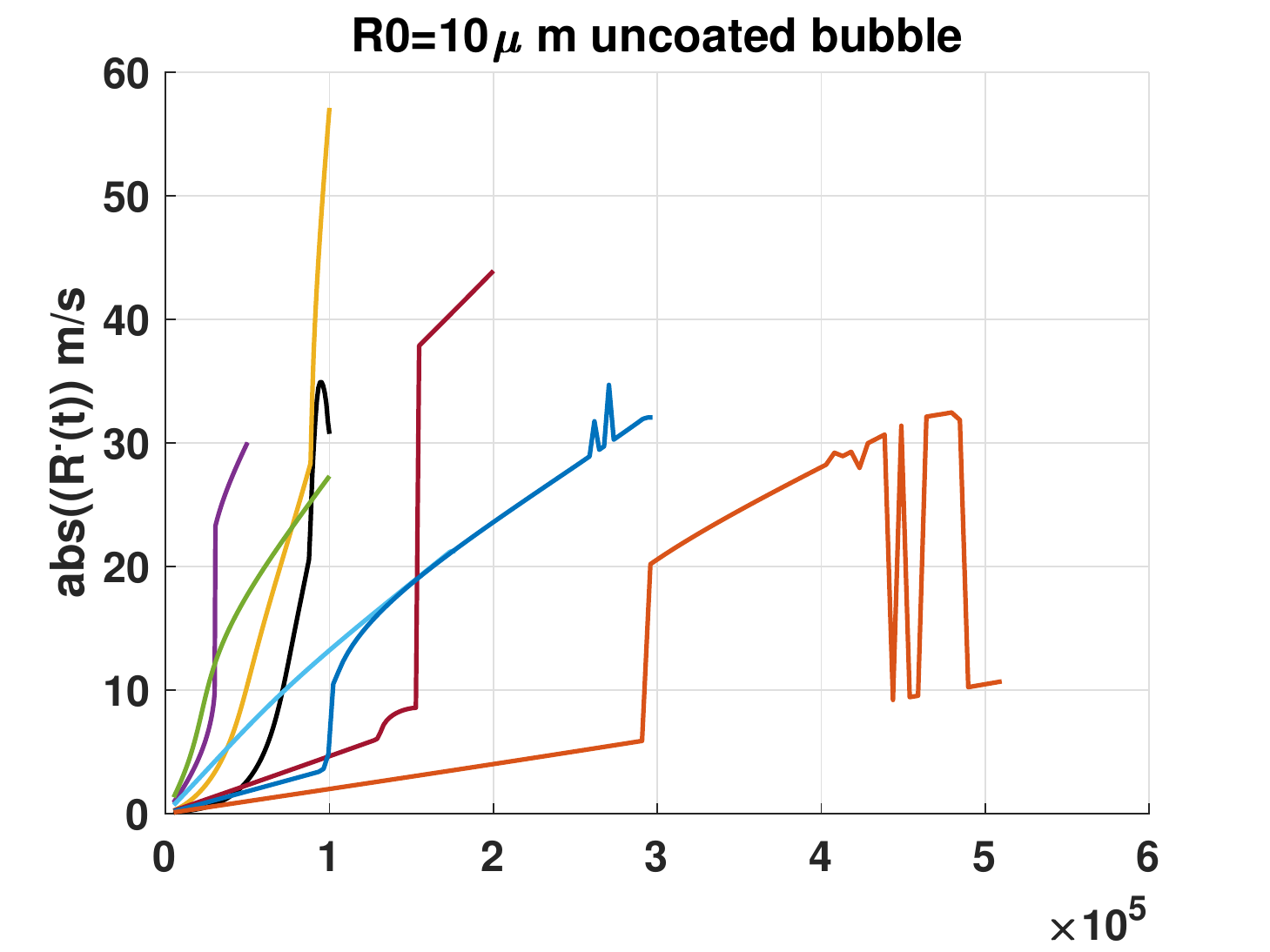}} \scalebox{0.43}{\includegraphics{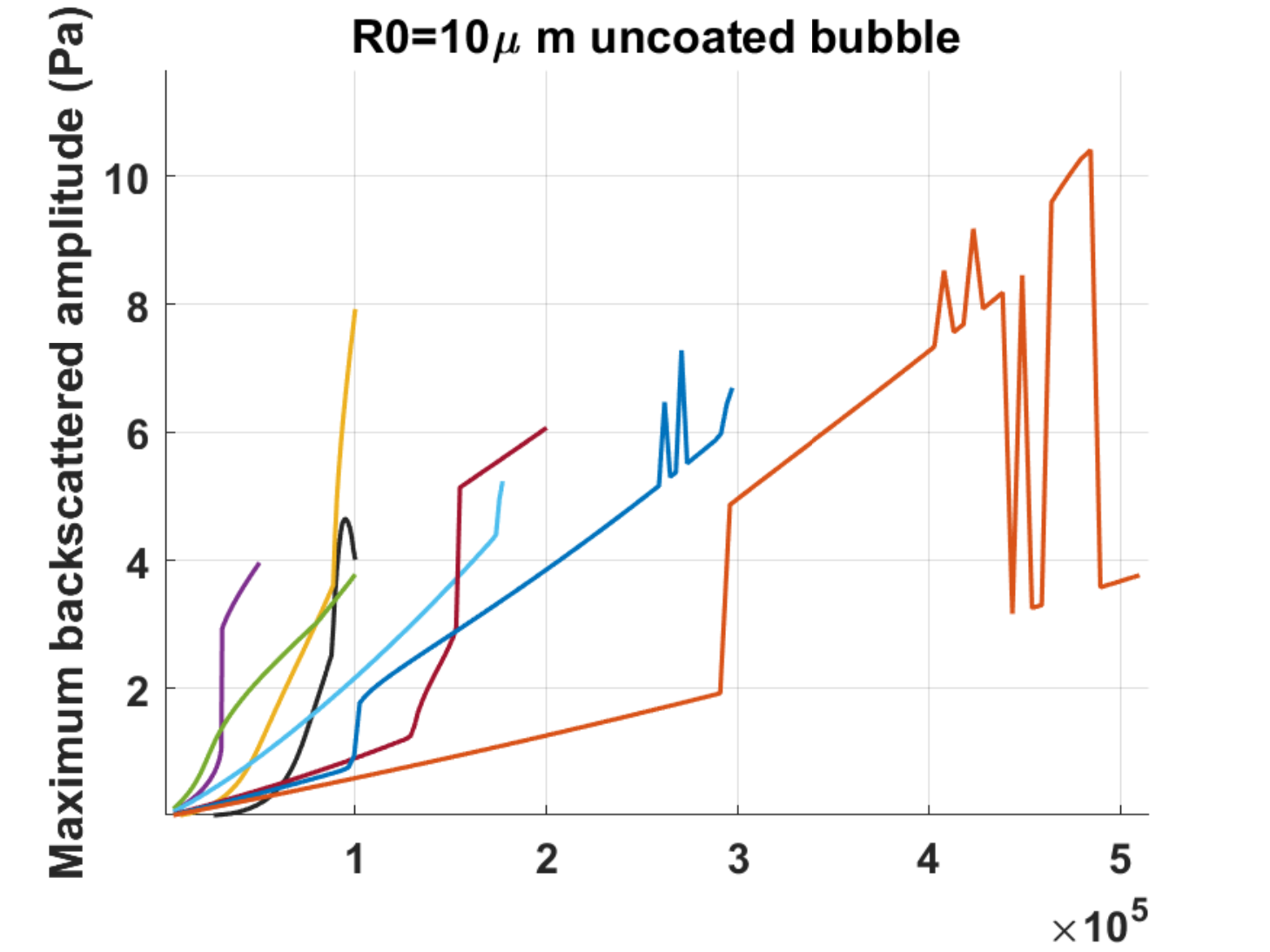}}\\
		\hspace{0.5cm} (a) \hspace{6cm} (b)\\
		\scalebox{0.43}{\includegraphics{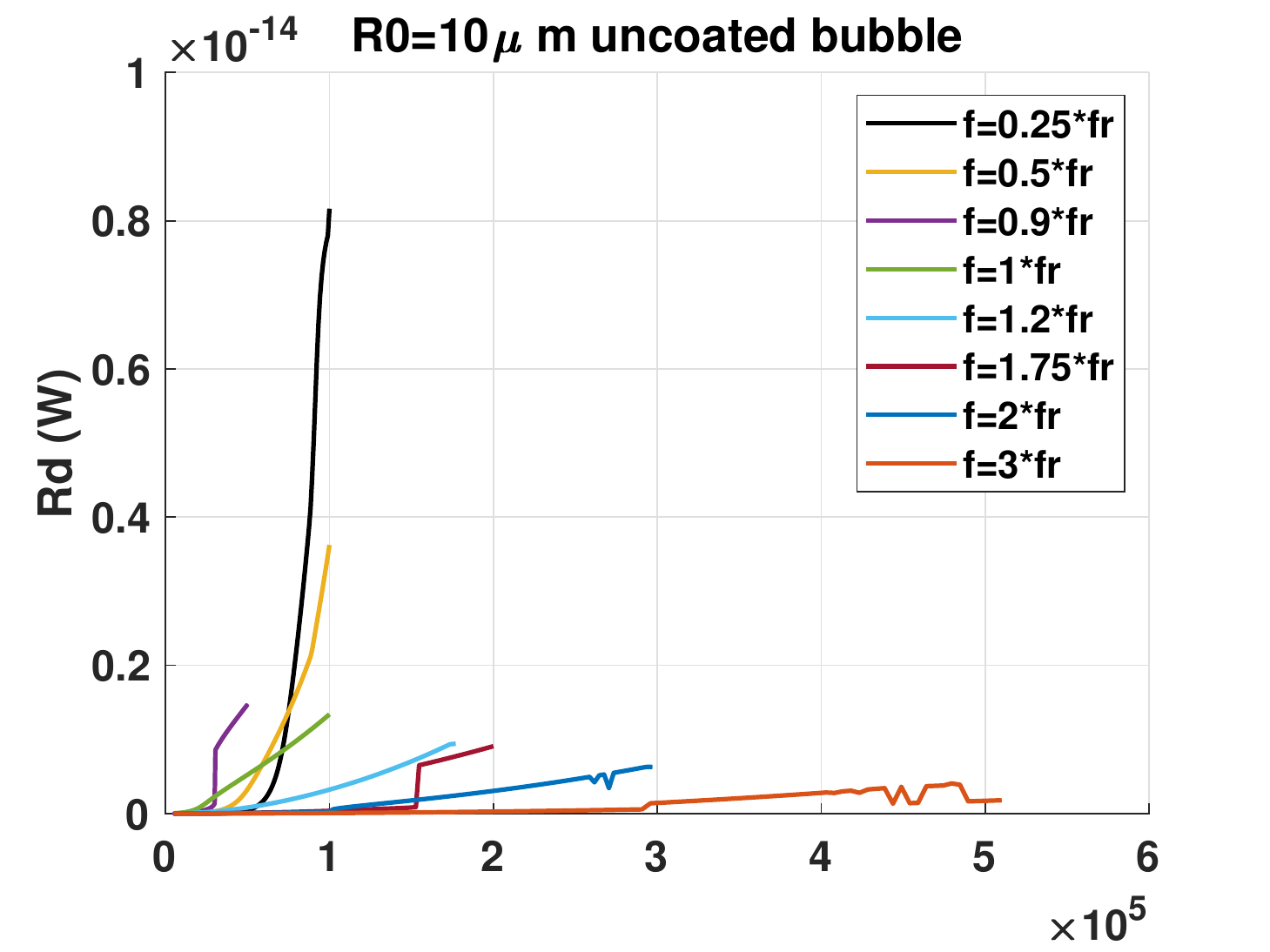}} \scalebox{0.43}{\includegraphics{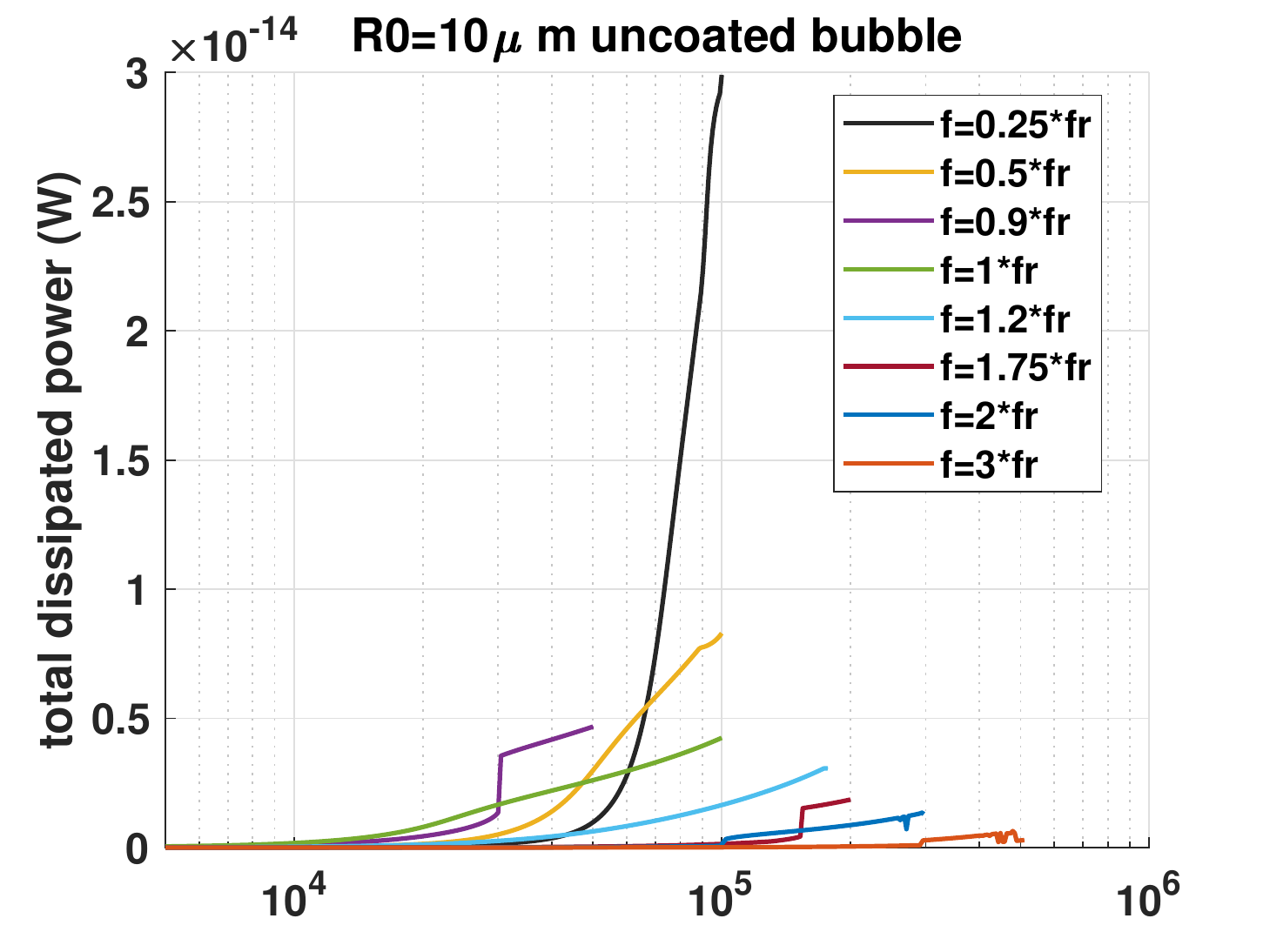}}\\
		\hspace{0.5cm} (c) \hspace{6cm} (d)\\
		\scalebox{0.43}{\includegraphics{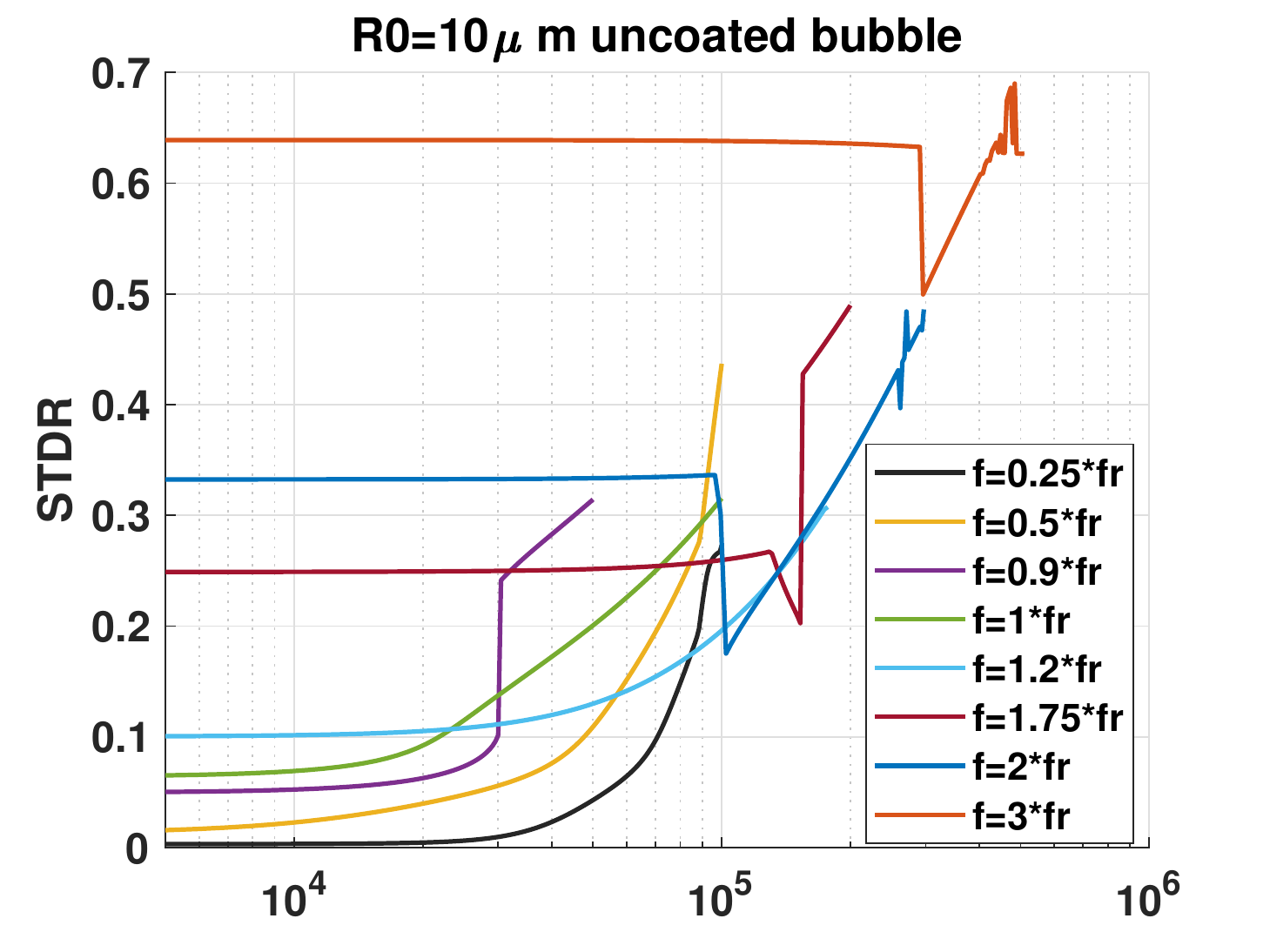}}\\
		\hspace{0.5cm} (e)\\
		\caption{Nondestructive ($R/R_0\leq 2$) values of: a) $|\dot{R(t)}|_{max}$ ($V_m$), b) Maximum backscattered pressure ($|P_{sc}|_{max}$ ($P_m$), c) Rd, d) $W_{total}$ and e) STDR as a function of pressure in the oscillations of an uncoated air bubble with $R_0=10 \mu m$.}
	\end{center}
\end{figure*}
\begin{figure*}
	\begin{center}
	\scalebox{0.43}{\includegraphics{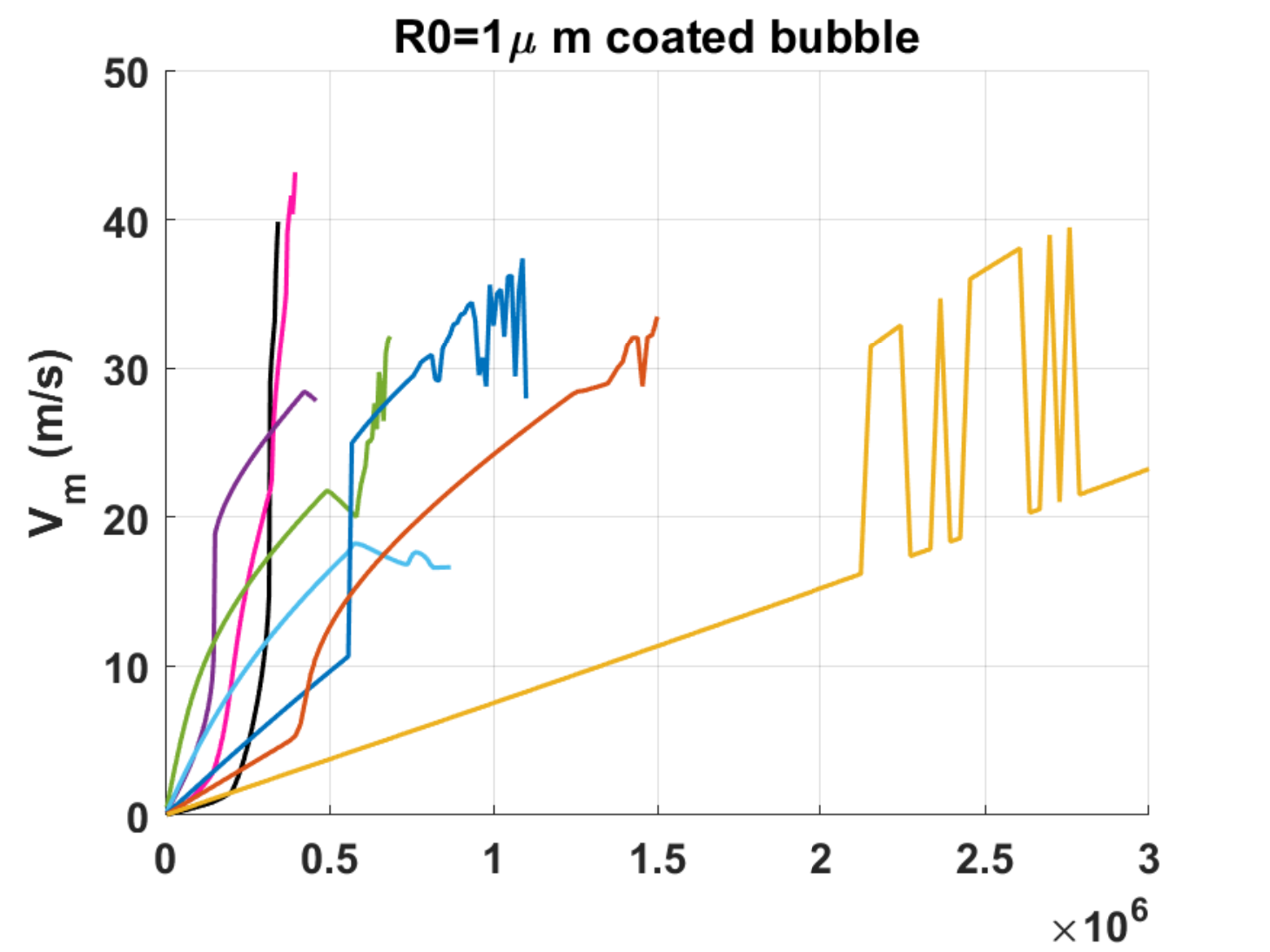}} \scalebox{0.43}{\includegraphics{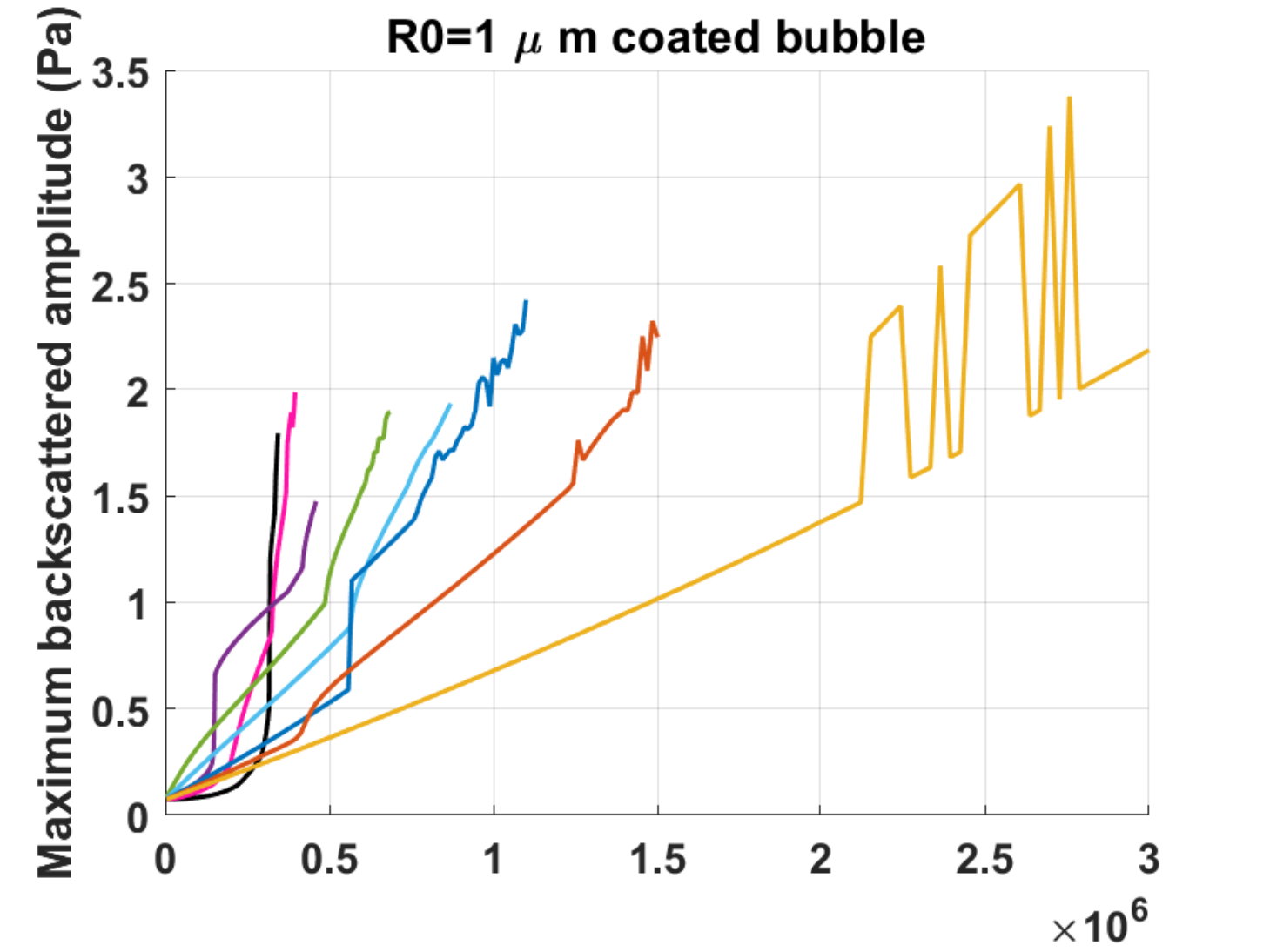}}\\
	\hspace{0.5cm} (a) \hspace{6cm} (b)\\
	\scalebox{0.43}{\includegraphics{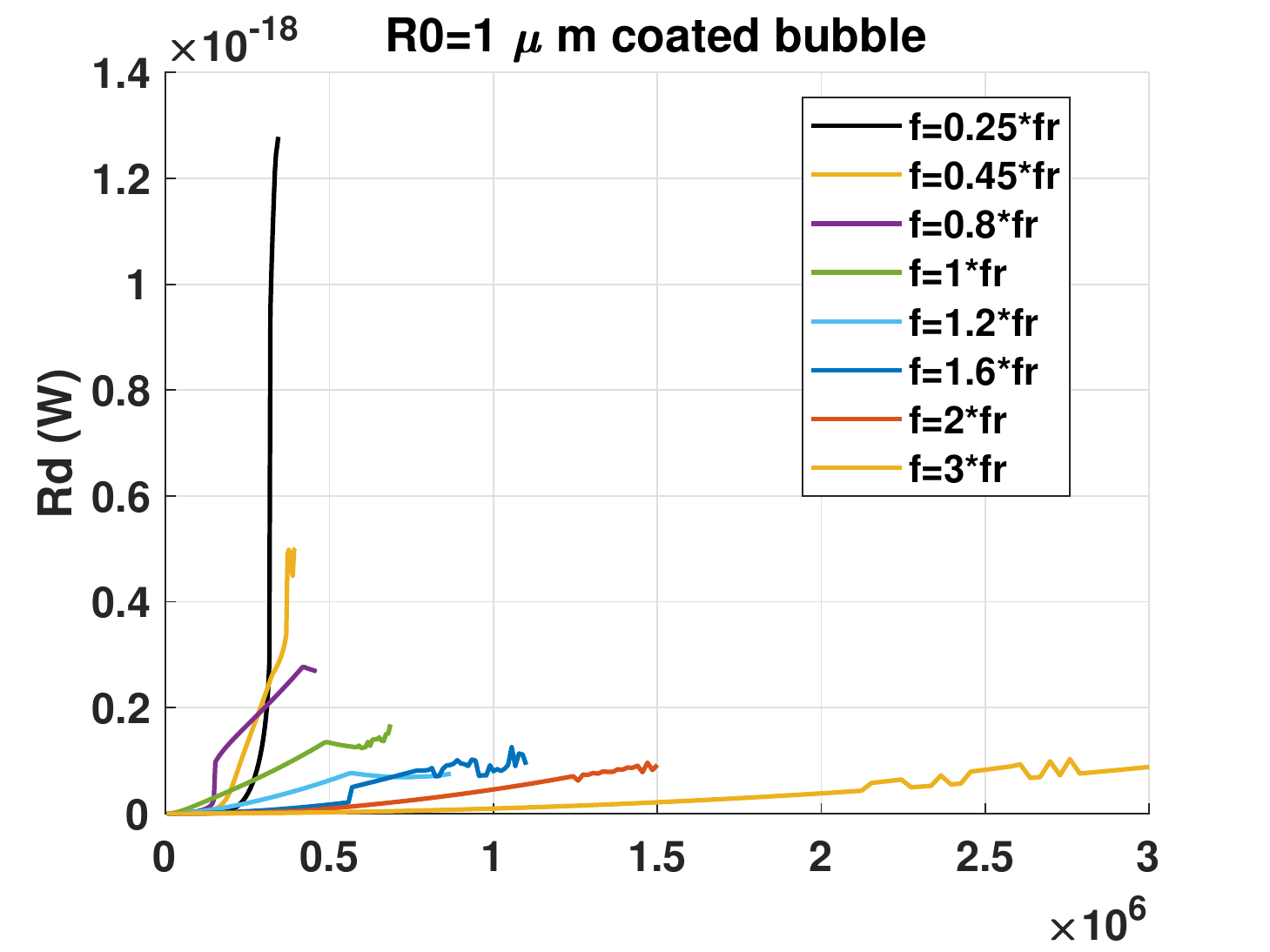}} \scalebox{0.43}{\includegraphics{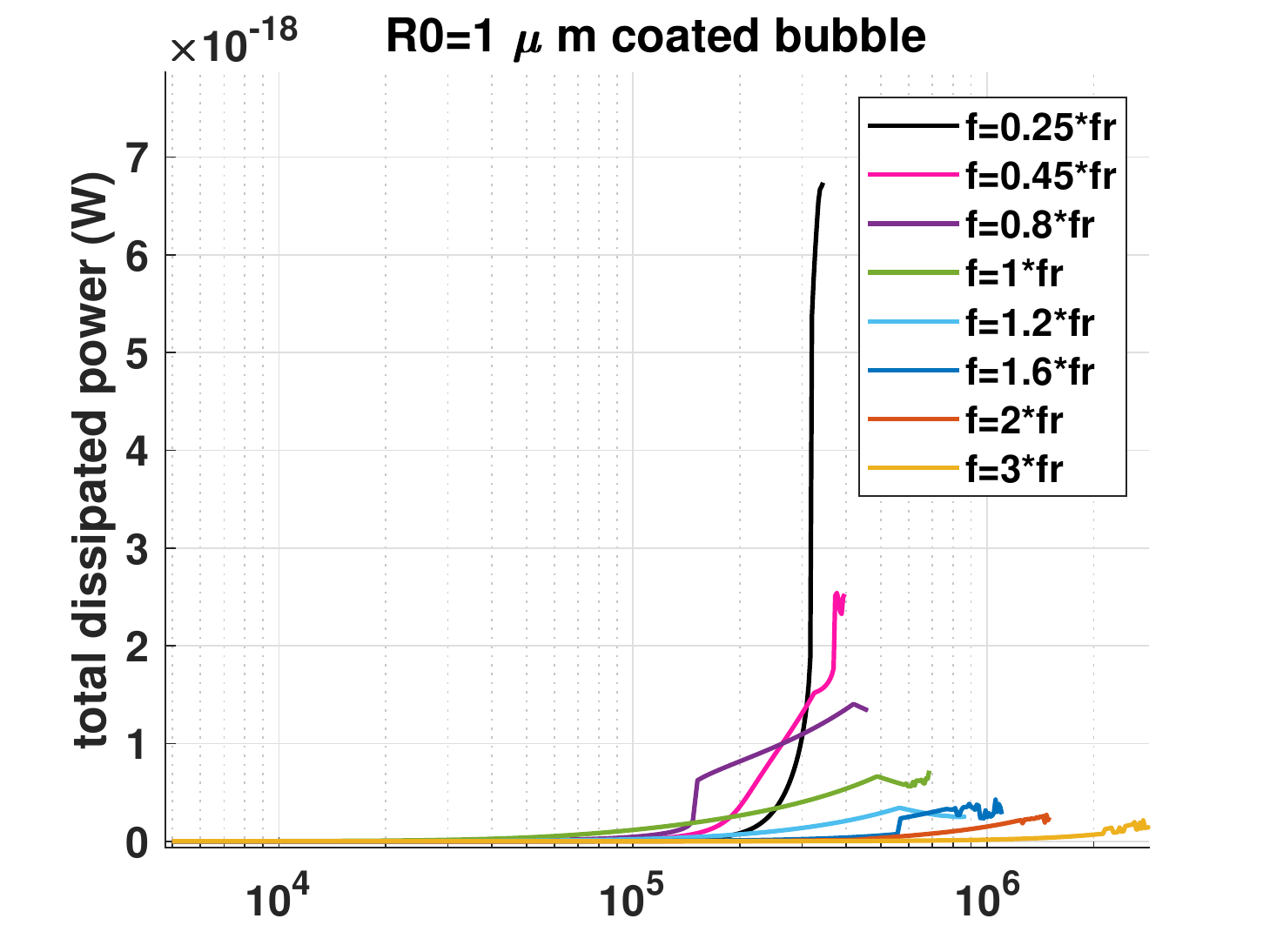}}\\
	\hspace{0.5cm} (c) \hspace{6cm} (d)\\
	\scalebox{0.43}{\includegraphics{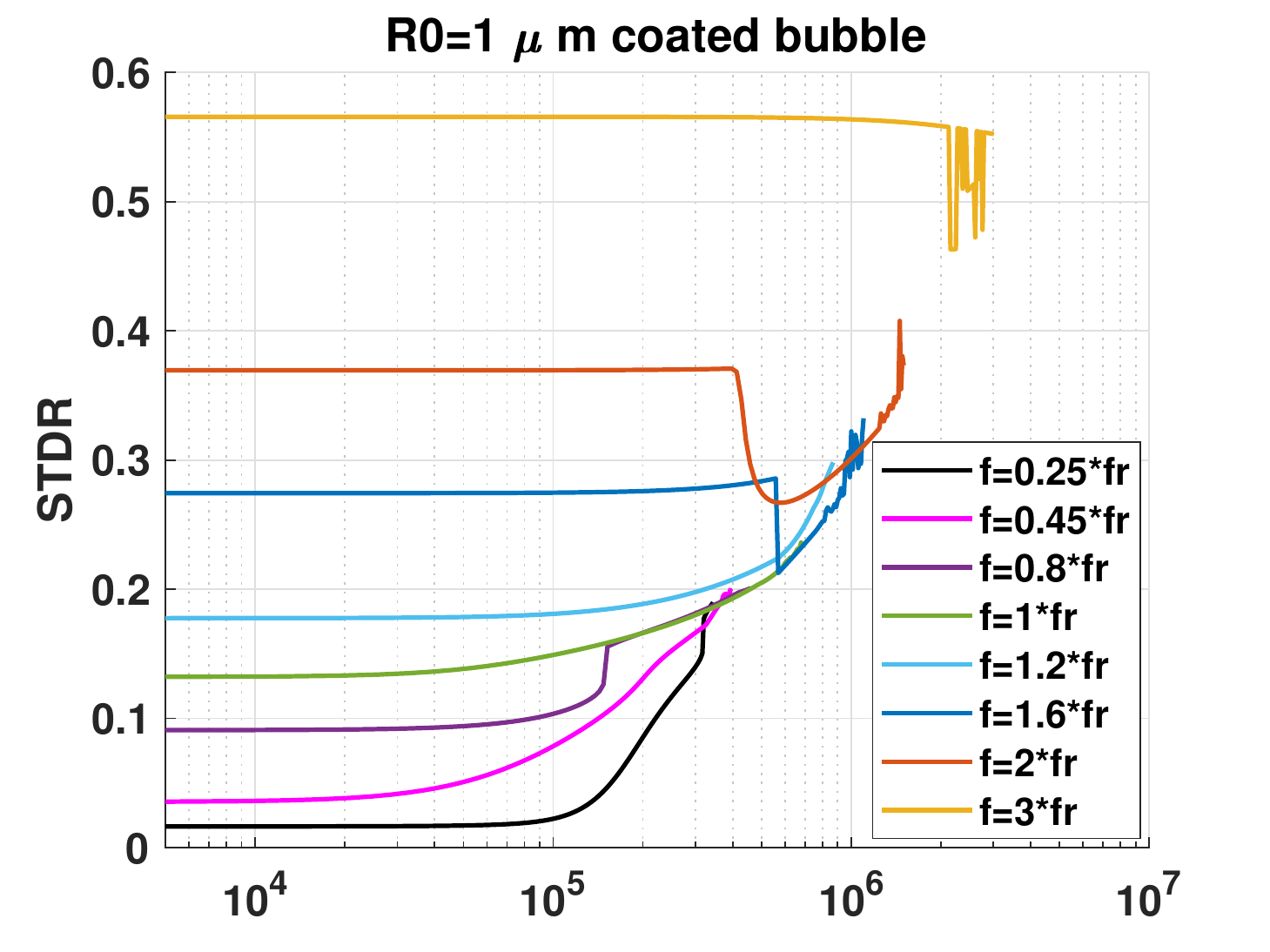}}\\
	\hspace{0.5cm} (e)\\
	\caption{Nondestructive ($R/R_0\leq 2$) values of: a) $|\dot{R(t)}|_{max}$ ($V_m$), b) Maximum backscattered pressure ($|P_{sc}|_{max}$ ($P_m$), c) Rd, d) $W_{total}$ and e) STDR as a function of pressure in the oscillations of a coated C3F8 bubble with $R_0=2 \mu m$.}
\end{center}
\end{figure*}
	In the previous subsection we investigated the evolution of the Cd, Ld, Rd and Td as a function of pressure at different frequencies and related their changes to the nonlinear behavior of the bubble. In this section we consider only stable non-destructive bubble oscillations ($\frac{R_{max}}{R_0}\leq2$ \cite{57} and for a more thorough review on bubble destruction threshold please refer to \cite{32}). Here we study maximum wall velocity amplitude ($|\dot{R(t)}|_{max}$), maximum amplitude of the re-radiated pressure ($|P_{sc}|_{max}$), total dissipated power ($W_{total}=Rd+Ld+Td+Cd$), and the scattering to dissipation ratio (STDR) when bubble is sonicated with the frequencies that are studied in Figs. 1-4 and for excitation pressures below the bubble destruction threshold. $|\dot{R(t)}|_{max}$ is abbreviated with $V_{m}$ from here on for simplicity and $|P_{sc}|_{max}$ is abbreviated with $P_{m}$. $STDR=\frac{Rd}{Rd+Ld+Td+Cd}$ where Cd is equal to zero for the uncoated bubble.\\
	Fig. 5a plots $V_{m}$ as a function of the excitation pressure for different frequencies for an uncoated air bubble with $R_0=10 \mu m$. $V_{m}$ is only presented for the oscillation regimes that most probably results in non-destructive bubble oscillations $R/R_0\leq2$ \cite{57}. Results show that when bubble is sonicated with $f=0.5 f_r$, $V_{m}$ reaches the maximum value for non-destructive oscillations. This can have advantages in drug delivery applications since higher wall velocity results in faster streaming and lower frequency of oscillations leads to smaller values for the thickness of the boundary layer \cite{63}. Since shear stress on the nearby objects is proportional to wall velocity and inversely proportional to the thickens of the boundary layer \cite{64}  sonication in this regime may result in higher shear stress values compared to other frequencies.\\ Fig. 5b shows $P_{m}$ as a function of pressure for the studied frequencies in Figs. 1-2. Sonication with $f=3f_r$ leads to the highest $P_{m}$; thus if the goal of the application is to increase the absolute amplitude of $P_{sc}$ and enhance echogenecity then sonication with $f=3f_r$ and pressures above the pressure threshold for generation of P3 oscillations will be the optimized frequency and pressure.\\
	Rd can be used as a measure for continuous bubble activity. In contrast to $P_{m}$ which denotes the maximum spontaneous back-scattered (re-radiated) pressure, Rd is a measure of sustained bubble activity as it is averaged over time. Fig. 5c shows that the maximum non-destructive Rd occurs for $f=0.25 f_r$ and frequencies below resonance display the largest non-destructive Rd.\\ The total dissipated power ($W_{total}=Rd+Ld+Td$) is shown in Fig. 5d. Maximum dissipated power occurs for $f=0.25f_r$.  Sonication below resonance have the advantage of higher dissipated powers. This is useful in applications where bubbles are used to enhance the power deposition by ultrasound to increase the generated heating. Furthermore, when compared to $f=f_r$, below  100 kPa, the dissipated power is lower when $f=0.3f_r$ and above 100 kPa the dissipated power undergoes a sharp increase and becomes approximately 4.8 times larger than the case of sonication with $f=f_r$. This has advantages in focused ultrasound heating enhancement; by taking advantage of the sharp pressure gradients of the focused ultrasound transducers, spatial heating prior to the focal point is limited and heating at the focal region can be enhanced.\\ Fig. 5e displays the STDR as a function of pressure. Frequencies above $f_r$ have larger STDR; the higher the freqeuncy the larger is the STDR which leads to the maximum STDR at $f=3f_r$. It is interesting to note that for frequencies above resonance the onset of non-linear oscillations results in a decrease in STDR; STDR then grows as pressure increases. This is mainly due to the increase in other damping parameters especially Td. As pressure increases, due to the faster growth rate of Rd, STDR raises again. Freqeuncies below resonance have lower STDRs compared to $f=f_r$. This is because of the increased Td and decreased Rd; however, the onset of SN bifurcation is concomitant with an increase in STDR for $f=0.8f_r$ (PDfr) and $f=0.5f_r$ (2nd SuH) oscillations.\\         
	The case of the C3F8 coated bubble with $R_0=1 \mu m$ is shown in Fig. 6. Same conclusions can be drawn as the case of the uncoated bubble in Figs. 5. This possibly indicates a universal behavior in the studied parameters for the cases considered in this paper.
\section{Summarizing points}
Acoustic waves are highly dissipated when they pass through bubbly media. Dissipation by bubbles takes place through thermal damping (Td), radiation damping (Rd) and damping due to the friction of the liquid (Ld) and friction of the coating (Cd). Td, Rd, Ld and Cd are nonlinear and depend on the complex dynamics of the bubbles. The correct estimation of dissipation events in the bubble oscillations will help in optimizing the relevant applications by maximizing a desirable parameter.\\ 
Most previous studies were limited by linear approximations \cite{37,66,67,68,69,70,71}. These approximations lead to inaccurate estimation of the dissipation phenomenon in applications as they are only valid for low pressures and linear regimes of low amplitude oscillations. Despite the importance of understanding the nonlinear dissipation events; only a few recent studies have attempted to investigate the problem accounting for the full non-linearity of the subject \cite{38,39,40,41,46,47}.\\ At present, the pressure dependence of the dissipation events is not well understood. Thus, in this paper we attempted to classify the bubble oscillations at various excitation frequencies as a function of pressure. Using a recent comprehensive approach \cite{50,56} in studying the nonlinear bubble dynamics we have classified the nonlinear oscillations of the uncoated and coated bubbles as a function of pressure excited with frequencies that result in 3rd and 2nd SuH regimes, pressure dependent (PD) resonance ($f_r$) oscillations, bubbles sonicate with $f_r$ (linear resonance frequency), PD subharmonic (SH) resonance ($PDf_{sh}$), $1/2$ order SH resonance ($f_{sh}$) and $1/3$ order SH resonance ($f=3f_r$). We have considered the nonlinear thermal and radiation effects in modeling the oscillations of the bubbles. Dynamics of the bubble including the generation of (2nd and 3rd order) SuH and ($7/2$ and $5/2$ order) UH oscillations, P1 resonant oscillations, cases of the occurrence of different types of giant resonances as well as different SH regimes of oscillations were revealed. Moreover, nonlinear dissipated powers due to Rd, Ld, Td and Cd were calculated and results were presented in tandem with the bifurcation curves. Using this approach different regimes of the evolution of the dissipative phenomena are linked to the responsible nonlinear effect in the bubble oscillations.\\ 
Our main results can be summarized as follows:\\ 
\subsection{Classification of the main nonlinear regimes of oscillations and the corresponding dissipation powers}
1- When a bubble (coated or uncoated) is sonicated with a frequency which is approximately between $\frac{f_r}{4} $ and $\frac{f_r}{3}$, and above a pressure threshold 3 maxima occur in the P1 oscillations. In this regime, the third harmonic component of the scattred pressure ($P_{sc}$) is larger than the other frequency components. This is suitable for applications like SuH imaging \cite{72,73}. Above a second pressure threshold period doubling (Pd) takes place and oscillations become P2 with 6 maxima. At this point the $7/2$ UH component of the $P_{sc}$ is larger than the SH and other UHs components of the signal. This frequency and pressure range is suitable for high resolution UH imaging \cite{74,75,76} and passive cavitation mapping techniques \cite{77}. Further pressure increases beyond this region most probably results in bubble destruction as $R/R_0$ exceeds 2. Pressure increase also result in the generation of chaos and periodic oscillations of higher amplitudes. At higher pressures a giant resonance (2nd order SuH) emerges out of the chaotic window. The corresponding evolution of the dissipative powers at different nonlinear regimes are summarized in table 2. \\ 
\begin{savenotes}
	\begin{table}
		\begin{tabular}{ |p{3cm}||p{2cm}|p{2cm}|p{2cm}|p{2cm}| p{2cm}| }
			\hline
			\multicolumn{6}{|c|}{Dissipation mechanisms when $f=0.25-0.35f_r$  } \\
			\hline
			Oscillation shape  &Linear& P1(3 maxima)&P2(6 maxima)&chaos&giant 2nd SuH resonance\\
					\hline	
			Uncoated air bubble with $R_0=10 \mu m$&$Td>Ld> {\color{red}Rd}$ &$Td>{\color{red}Rd}=Ld$& $Td={\color{red}Rd}>Ld$&${\color{red}Rd}>Td>Ld$&${\color{red}Rd}>Ld>Td$\\
			\hline	
			Uncoated air bubble with $R_0=2\mu m$&$Td\approxeq Ld> {\color{red}Rd}$ &$Td \approxeq Ld\approxeq {\color{red}Rd}$& $Ld\approxeq {\color{red}Rd}>Td$&${\color{red}Rd}>Ld>Td$&${\color{red}Rd}>Ld>Td$\\
			\hline
			coated C3F8 bubble with $R_0=4 \mu m$&$Cd>Ld\approxeq Td>{\color{red}Rd}$ &$Cd>Ld \approxeq {\color{red}Rd}>Td$& $Cd>{\color{red}Rd}\approxeq Ld>Td$&$Cd\approxeq {\color{red}Rd}>Ld>Td$&${\color{red}Rd}> Cd>Ld>Td$\\
			\hline
			coated C3F8 bubbles with $R_0=4 \mu m$&$Cd>Ld>Td>{\color{red}Rd}$ &$Cd>Ld>{\color{red}Rd}>Td$& $Cd>{\color{red}Rd}\approxeq Ld>Td$&$Cd\approxeq {\color{red}Rd}>Ld>Td$&${\color{red}Rd}>Cd>Ld>Td$\\
			\hline	
		\end{tabular}
		\caption{Evolution of dissipation powers at different nonlinear regimes for uncoated air and coated C3F8 bubbles when $f=0.25-0.3f_r$}
		\label{table:1}
	\end{table}
\end{savenotes}
2- When $f\approxeq0.4-0.6f_r$ both of the cases of the coated and uncoated bubbles start with P1 oscillations and above a pressure threshold a second maxima occur in the P1 oscillations. This is simultaneous with a fast growth of the 2nd harmonic component of the $P_{sc}$ (2nd harmonic component becomes the strongest frequency component in the spectra). This frequency and pressure range is suitable for 2nd harmonic imaging applications of ultrasound \cite{72,73}.  Above a second pressure threshold Pd happens and oscillations become P2 with 4 maxima. At this point between the SH and UH components of the frequency spectrum of the  $P_{sc}$, the $5/2$ UH component has the highest amplitude. This can be used to enhance contrast and resolution in UH imaging applications of ultrasound \cite{74,75,76} and passive cavitation mapping techniques \cite{77} or high resolution treatment monitoring using UH emissions \cite{78}. Further pressure increase results in chaos and $R/R_0$ exceeding 2. Pressure increase beyond this point leads to emergence of a P1 giant resonance out of the chaotic window. The corresponding evolution of the dissipative powers at different nonlinear regimes are summarized in table 3.\\
\begin{savenotes}
	\begin{table}
		\begin{tabular}{ |p{3cm}||p{2cm}|p{2cm}|p{2cm}|p{2cm}| p{2cm}| }
			\hline
			\multicolumn{6}{|c|}{Dissipation mechanisms when $f=0.45-0.55f_r$  } \\
			\hline
			Oscillation shape  &Linear& P1(2 maxima)&P2(4 maxima)&chaos&P1 giant resonance\\
					\hline	
			Uncoated air bubble with $R_0= 10 \mu m$&$Td>Ld> {\color{red}Rd}$ &$Td>{\color{red}Rd}\approxeq Ld$& $Td\approxeq {\color{red}Rd}>Ld$&${\color{red}Rd}>Td>Ld$&${\color{red}Rd}>Ld>Td$\\
			\hline	
			Uncoated air bubble with $R_0=2\mu m$&$Td\approxeq Ld> {\color{red}Rd}$ &$Td \approxeq Ld> {\color{red}Rd}$& $Ld\approxeq {\color{red}Rd}>Td$&${\color{red}Rd}>Ld>Td$&${\color{red}Rd}>Ld>Td$\\
			\hline	
			coated C3F8 bubble with $R_0=4 \mu m$&$Cd>Ld\approxeq Td>{\color{red}Rd}$ &$Cd>Ld \approxeq {\color{red}Rd}>Td$& $Cd>{\color{red}Rd}\approxeq Ld>Td$&$Cd>{\color{red}Rd}>Ld>Td$&${\color{red}Rd}\approxeq Cd>Ld>Td$\\
			\hline	
			coated C3F8 bubble with $R_0=1 \mu m$&$Cd>Ld>Td>{\color{red}Rd}$ &$Cd>Ld>{\color{red}Rd}>Td$& $Cd>{\color{red}Rd}\approxeq Ld>Td$&$Cd>{\color{red}Rd}>Ld>Td$&${\color{red}Rd}\approxeq Cd>Ld>Td$\\
			\hline	
		\end{tabular}
		\caption{Evolution of dissipation powers at different nonlinear regimes for uncoated air and coated C3F8 bubbles when $f=0.45-0.55f_r$. }
		\label{table:1}
	\end{table}
\end{savenotes}
3- When $f=0.7-0.9f_r$ (the pressure dependent resonance frequency ($PDf_r$ \cite{32})), at lower pressures oscillations are P1 with 1 maximum. Above a pressure threshold, a saddle node (SN) bifurcation takes place and P1 oscillations undergo a large increase in amplitude to another P1 oscillation with 1 maximum. At SN, the maxima curve and the Poincaré curve has the same value (the wall velocity is in phase with the excitation force). The Poincaré and maxima curves diverge from each other as pressure increases beyond the SN. The occurrence of the SN can have significant advantages in imaging techniques based on amplitude modulation \cite{32,74,79,80}.  Beyond the SN, the increase in excitation pressure leads to a monotonic increase in oscillation amplitude and above a second pressure threshold Pd takes place and oscillations become P2 with 2 maxima. Apart from the coated bubble with $R_0=1 \mu m$, other studied bubbles in this frequency range most probably are destroyed as $\frac{R_{max}}{R_0}$ exceeds 2 before any P2 is generated. Further pressure increase leads to successive Pds to chaos with a possible window (of P3/P6 oscillations with 3/6 maxima) which is located inside chaotic window. At higher pressures a P2 giant resonance may emerge out of the chaotic window. The corresponding evolution of the dissipative powers at different nonlinear regimes are summarized in table 4.\\        
\begin{savenotes}
	\begin{table}
		\begin{tabular}{ |p{3cm}||p{2cm}|p{2cm}|p{2cm}|p{2cm}| p{2cm}| }
			\hline
			\multicolumn{6}{|c|}{Dissipation mechanisms when $f=0.7-0.9f_r$  } \\
			\hline
			Oscillation shape  &Linear& SN-P1(1 maxima)&P2(2 maxima)&chaos&P2 giant resonance\\
					\hline	
			Uncoated air bubble with $R_0= 10 \mu m$&$Td>Ld> {\color{red}Rd}$ &$Td>{\color{red}Rd}\approxeq Ld$& $Td\approxeq {\color{red}Rd}>Ld$&${\color{red}Rd}>Td>Ld$&${\color{red}Rd}>Ld>Td$\\
			\hline	
			Uncoated air bubble with $R_0=2\mu m$&$Td\approxeq Ld> {\color{red}Rd}$ &$Td \approxeq Ld> {\color{red}Rd}$& $Ld\approxeq {\color{red}Rd}>Td$&${\color{red}Rd}>Ld>Td$&${\color{red}Rd}>Ld>Td$\\
			\hline	
			coated C3F8 bubble with $R_0=4 \mu m$&$Cd>Ld>{\color{red}Rd}>Td$ &$Cd>Ld \approxeq {\color{red}Rd}>Td$& $Cd>{\color{red}Rd}> Ld>Td$&$Cd>{\color{red}Rd}>Ld>Td$&${\color{red}Rd}\approxeq Cd>Ld>Td$\\
			\hline	
			coated C3F8 bubble with $R_0=1 \mu m$&$Cd>Ld>{\color{red}Rd}>Td$ &$Cd>Ld>{\color{red}Rd}>Td$& $Cd> Ld>{\color{red}Rd}>Td$&$Cd>{\color{red}Rd}>Ld>Td$&${\color{red}Rd}\approxeq Cd>Ld>Td$\\
			\hline	
		\end{tabular}
		\caption{Evolution of dissipation powers at different nonlinear regimes for uncoated air and coated C3F8 bubbles when $f=0.7-0.9f_r$. }
		\label{table:1}
	\end{table}
\end{savenotes}
4- When $f=f_r$ (linear resonance frequency of the bubbles) oscillations are P1 with 1 maxima at lower pressures. At the beginning of the bifurcation diagrams; the curve that is constructed with the method of peaks has the exact value as the Poincaré method. This indicates that the wall velocity is in phase with the driving signal. As the pressure increases the two curves start diverging as the resonance frequency changes with pressure. At higher pressures resonance frequency shifts to smaller values \cite{32} ($Pdf_r$). Above a pressure threshold, bubbles undergo Pd and oscillations become P2 with two maxima. For coated bubbles, Pd can occur when $\frac{R_{max}}{R_0}<2$ (bubbles may sustain non-destructive oscillations); however, in case of uncoated bubbles Pd only occurs when $\frac{R_{max}}{R_0}>2$ (uncoated bubbles may not sustain non-destructive P2 oscillations when they are sonicated with their $f_r$). The P2 oscillations has been extensively studied in our previous work \cite{58} without the inclusion of thermal damping effects. P2 oscillations undergo successive Pd to chaos. At higher pressures a P2 (with two maxima) giant resonance may emerge out of the chaotic window. The giant resonance oscillations undergo successive Pds to chaotic oscillations of higher amplitude. The corresponding evolution of Cd, Rd, Ld and Td are summarized in table 5.\\
\begin{savenotes}
	\begin{table}
		\begin{tabular}{ |p{3cm}||p{2cm}|p{2cm}|p{2cm}|p{2cm}| p{2cm}| }
			\hline
			\multicolumn{6}{|c|}{Dissipation mechanisms when $f=f_r$  } \\
			\hline
			Oscillation shape  &Linear resonance& Linear&P2(2 maxima)&chaos& P2 giant resonance\\
			\hline	
			Uncoated air bubble with $R_0= 10 \mu m$&$Td>Ld> {\color{red}Rd}$ &$Td>{\color{red}Rd}\approxeq Ld$& $Td\approxeq {\color{red}Rd}>Ld$&${\color{red}Rd}>Td>Ld$&${\color{red}Rd}>Ld>Td$\\
			\hline	
			Uncoated air bubble with $R_0=2\mu m$&$Td> Ld> {\color{red}Rd}$ &$ Ld> {\color{red}Rd}$& $Ld\approxeq {\color{red}Rd}>Td$&${\color{red}Rd}>Ld>Td$&${\color{red}Rd}>Ld>Td$\\
			\hline	
			coated C3F8 bubble with $R_0=4 \mu m$&$Cd>Ld>{\color{red}Rd}>Td$ &$Cd>Ld \approxeq {\color{red}Rd}>Td$& $Cd>{\color{red}Rd}> Ld>Td$&$Cd>{\color{red}Rd}>Ld>Td$&${\color{red}Rd}\approxeq Cd>Ld>Td$\\
			\hline	
			coated C3F8 bubble with $R_0=1 \mu m$&$Cd>Ld>{\color{red}Rd}>Td$ &$Cd>Ld>{\color{red}Rd}>Td$& $Cd> Ld>{\color{red}Rd}>Td$&$Cd>{\color{red}Rd}>Ld>Td$&${\color{red}Rd}\approxeq Cd>Ld>Td$\\
			\hline	
		\end{tabular}
		\caption{Evolution of dissipation powers at different nonlinear regimes for uncoated air and coated C3F8 bubbles when $f=f_r$.}
		\label{table:1}
	\end{table}
\end{savenotes}
5- When $f=1.2f_r$, at lower excitation pressures,  oscillations are P1 with 1 maximum. Contrary to the case of $f=f_r$, the Poincaré curve and the curve constructed by method of peaks start diverging right at the beginning of the bifurcation diagram. Increasing pressure leads to Pd and oscillations becomes P2 with two maxima. At this frequency P2 oscillations can be non-destructive (in case of both coated and uncoated bubbles) as when Pd occurs since  $\frac{R_{max}}{R_0}$ is below 2. Sonication with this frequency and in pressure ranges responsible for P2 oscillations may lead to $3/2$ UH resonant oscillations. Further pressure increase results in successive Pds to chaos. At higher pressures a P2 giant resonance may emerge out of the chaotic window, undergoing successive Pds to chaotic oscillations of higher amplitude. The lower amplitude branch of the curve that is generated by the method of peaks has the same value as the higher amplitude branch of the curve created by the Poincaré method. This suggests that wall velocity is in phase with excitation frequency once every two acoustic cycles. The corresponding evolution of the dissipative powers are summarized in table 6.\\
\begin{savenotes}
	\begin{table}
		\begin{tabular}{ |p{3cm}||p{2cm}|p{2cm}|p{2cm}|p{2cm}| p{2.2cm}| }
			\hline
			\multicolumn{6}{|c|}{Dissipation mechanisms when $f=1.2f_r$  } \\
			\hline
			Oscillation shape  &Linear & Linear&P2(2 maxima)&chaos&chaos/P2 giant resonance\\
					\hline	
			Uncoated air bubble with $R_0= 10 \mu m$&$Td>Ld \approxeq {\color{red}Rd}$ &$Td>{\color{red}Rd}> Ld$& $Td> {\color{red}Rd}>Ld$&${\color{red}Rd}\approxeq Td>Ld$&${\color{red}Rd}>Ld>Td$\\
			\hline	
			Uncoated air bubble with $R_0=2\mu m$&$Td> Ld> {\color{red}Rd}$ &$ Ld> TD> {\color{red}Rd}$& $Ld>Td>{\color{red}Rd}$&$Ld\approxeq {\color{red}Rd}>Td$&${\color{red}Rd}>Ld>Td$\\
			\hline	
			coated C3F8 bubble with $R_0=4 \mu m$&$Cd>Ld>{\color{red}Rd}>Td$ &$Cd>Ld \approxeq {\color{red}Rd}>Td$& $Cd>{\color{red}Rd}> Ld>Td$&$Cd>{\color{red}Rd}>Ld>Td$&${ Cd\approxeq \color{red}Rd} >Ld>Td$\\
			\hline	
			coated C3F8 bubble with $R_0=1 \mu m$&$Cd>Ld>{\color{red}Rd}>Td$ &$Cd>Ld>{\color{red}Rd}>Td$& $Cd> Ld>{\color{red}Rd}>Td$&$Cd>{\color{red}Rd}>Ld>Td$&${\color{red}Rd}\approxeq Cd>Ld>Td$\\
			\hline	
		\end{tabular}
		\caption{Evolution of dissipation powers at different nonlinear regimes for uncoated air and coated C3F8 bubbles when $f=1.2f_r$. }
		\label{table:1}
	\end{table}
\end{savenotes}
6- When $f=1.6-1.8f_r$ (pressure dependent subharmonic (SH) resonance frequency $PDf_{sh}$ \cite{58}) at lower pressures oscillations are P1 with 1 maxima. Pressure increase leads to one of the following scenarios: 1- generation of Pd above a pressure threshold; then, a SN bifurcation from P2 oscillations of lower amplitude to P2 oscillations (1 maximum) of higher amplitude. 2- SN bifurcation above a pressure threshold from P1 oscillations with 1 maxima to P2 oscillations with 1 maxima. This happens while $\frac{R_{max}}{R_0}$ is below 2; therefore, the bubbles may sustain non-destructive P2 oscillations when insonated with frequencies between $f=1.6-1.8f_r$. Further pressure increases lead to the generation of a second maximum (with the same amplitude of the higher branch of the Poincaré curve). Occurrence of the SN can provide significant advantages for amplitude modulation techniques \cite{78,79,80} and in this case because of the higher sonication frequency, we can expect higher resolution. Moreover, we have shown in \cite{59} that occurrence of SN leads to oversaturation of the $1/2$ SH and $3/2$ UH frequency content of the $P_{sc}$. This can provide higher contrast to tissue and signal to noise ratio in SH imaging techniques \cite{81,82,83} and possibly SH and UH monitoring of treatments \cite{77,78,84,85}. Oscillations undergo successive Pds to chaos. Further pressure increase may lead to the emergence of a P3 giant resonance which will undergo successive Pds to chaotic oscillations of higher amplitude. The dynamics of the bubble sonicated with their $PDf_{sh}$ (in the absence of thermal damping) has been extensively studied in our previous work \cite{59}. The corresponding evolution of dissipative powers are summarized in table 7.\\
\begin{savenotes}
	\begin{table}
		\begin{tabular}{ |p{3cm}||p{2cm}|p{2cm}|p{2.5cm}|p{2cm}| p{2.5cm}| }
			\hline
			\multicolumn{6}{|c|}{Dissipation mechanisms when $f=1.6-1.8f_r$} \\
			\hline
			Oscillation shape  &Linear & P2 through SN&P4(4 maxima)&chaos&chaos/P3 giant resonance\\
				\hline	
			Uncoated air bubble with $R_0= 10 \mu m$&$Td>{\color{red}Rd}>Ld$ &${\color{red}Rd}>TD>Ld$& $ {\color{red}Rd}>Td>Ld$&${\color{red}Rd}>Td>Ld$&${\color{red}Rd}>Ld>Td$($Gf_r$)\\
			\hline	
			Uncoated air bubble with $R_0=2\mu m$&$Td\approxeq Ld> {\color{red}Rd}$ &$Ld>{\color{red}Rd}>Td$& $Ld\approxeq{\color{red}Rd}$>Td&$Ld\approxeq {\color{red}Rd}>Td$&${\color{red}Rd}>Ld>Td$\\
			\hline	
			coated C3F8 bubble with $R_0=4 \mu m$&$Cd>{\color{red}Rd}>Ld>Td$ &$Cd>{\color{red}Rd}>Td$& $Cd>Ld\approxeq{\color{red}Rd}>Ld>Td$&$Cd>{\color{red}Rd}>Ld>Td$& $Cd\approxeq{\color{red}Rd}>Ld>Td$\\
			\hline	
			coated C3F8 bubble with $R_0=1 \mu m$&$Cd>Ld\approxeq{\color{red}Rd}>Td$ &$Cd>Ld\approxeq{\color{red}Rd}>Td$& $Cd>Ld\approxeq{\color{red}Rd}>Td$&$Cd>Ld\approxeq{\color{red}Rd}>Td$& $Cd>{\color{red}Rd}>Ld>Td$\\
			\hline	
		\end{tabular}
		\caption{Evolution of dissipation powers at different nonlinear regimes for uncoated air and coated C3F8 bubbles when $f=1.6-1.8f_r$. }
		\label{table:1}
	\end{table}
\end{savenotes}
7- when $f=2f_r$ (linear SH resonance frequency $f_{sh}$ \cite{58}) oscillations are P1 with 1 maximum at lower pressures. Above a pressure threshold Pd takes place and oscillations become P2 with 2 maxima. As pressure increases one of the maxima disappears with pressure increase and the P2 oscillations evolve in the form of a bowtie (the curve that is constructed using the Poincaré method). Later the second maxima re-appears with an amplitude equal to the higher branch of the Poincaré curve. Oscillations undergo successive Pds to chaos. When $f=2f_r$, P2 oscillations occur for the widest excitation pressure range and $\frac{R_{max}}{R_0}$ is below 1.5; therefore, bubbles have the highest probability of sustaining non-destructive P2 oscillations. Analytical solutions \cite{86,87,88,89,90,91} predict the generation of P2 oscillations at the lowest pressure threshold when the bubble is sonicated with $f=2f_r$. Later, it was shown in \cite{92,93} in case of smaller bubbles (e.g. $R_0=0.6\mu m$) that the lowest pressure threshold occurs when bubble is sonicated with a frequency near its $f_r$. They concluded that the increased damping is responsible for shift the lowest frequency threshold. However, none of the previous studies included both of the pressure dependent thermal and radiation damping effects. In this work, we have included both of these effects with their full non-linearity and observed that the lowest pressure threshold of P2 oscillations occurs at none of the $f=f_r$ or $f=2f_r$, but it occurs at frequencies below $PDFf_r$. As an instance, for the uncoated air bubble $R_0=10 \mu m$ pressure thresholds for P2 oscillations are 87.5, 82, 88, 170 and 96kPa respectively at f=0.25, 0.35, 0.5, 1 and $2f_r$. Thus for this bubble lowest P2 pressure threshold is $0.35f_r$. This may be explained with the increased damping effects due to thermal damping and pressure dependent non-linear coupling. The study of the  lowest pressure threshold for P2 oscillations and the reasons behind it is not within the scope of this paper and can be the subject of future studies. The corresponding evolution of dissipation powers are summarized in table 8.\\
\begin{savenotes}
	\begin{table}
		\begin{tabular}{ |p{3cm}||p{2cm}|p{2cm}|p{2.5cm}|p{2cm}| p{2.5cm}| }
			\hline
			\multicolumn{6}{|c|}{Dissipation mechanisms when $f=2f_r$  } \\
			\hline
			Oscillation shape  &Linear & P2 through SN&P4(4 maxima)&chaos&chaos/P3 giant resonance\\
			\hline
					Uncoated air bubble with $R_0= 10 \mu m$&$Td>{\color{red}Rd}>Ld$ &$TD>{\color{red}Rd}>Ld$& $ Td\approxeq{\color{red}Rd}>Ld$&$Td\approxeq{\color{red}Rd}>Ld$&${\color{red}Rd}>Ld>Td$($Gf_r$)\\
			\hline	
			Uncoated air bubble with $R_0=2\mu m$&$Td\approxeq Ld> {\color{red}Rd}$ &$Td\approxeq Ld>{\color{red}Rd}$& $Ld\approxeq{\color{red}Rd}$>Td&$Ld\approxeq {\color{red}Rd}>Td$&${\color{red}Rd}>Ld>Td$\\
			\hline	
			coated C3F8 bubble with $R_0=4 \mu m$&$Cd>{\color{red}Rd}>Ld>Td$ &$Cd>{\color{red}Rd}>Ld>Td$& $Cd>{\color{red}Rd}>Ld>Td$&$Cd\approxeq{\color{red}Rd}>Ld>Td$& ${\color{red}Rd}>Cd>Ld>Td$\\
			\hline	
			coated C3F8 bubble with $R_0=1 \mu m$&$Cd\approxeq{\color{red}Rd}>Ld>Td$ &$Cd>Ld\approxeq{\color{red}Rd}>Td$& $Cd>{\color{red}Rd}>Ld>Td$&$Cd\approxeq{\color{red}Rd}>Ld>Td$& $Cd\approxeq{\color{red}Rd}>Ld>Td$\\
			\hline	
		\end{tabular}
		\caption{Evolution of dissipation powers at different nonlinear regimes for uncoated air and coated C3F8 bubbles when $f=2f_r$.}
		\label{table:1}
	\end{table}
\end{savenotes}
8- When $f=3f_r$ oscillations are P1 with 1 maximum at lower pressures. Oscillation amplitude grow very slowly withe excitation pressure increase and above a pressure threshold P3 oscillations of higher amplitude are generated through a SN bifurcation. Later P3 oscillations undergo Pd to P12 followed by successive Pds to a small chaotic window before the oscillations convert to P1 with lower amplitude. The corresponding evolution of the dissipative powers is summarized  in table 9. The SN bifurcation is concomitant with a sharp increase in $P_{sc}$. This has advantages for amplitude modulation imaging techniques \cite{74,79,80} at higher frequencies \cite{94}. The pressure amplitudes for the pulses shoudl be chosen below and above the SN pressure.\\ 
\begin{savenotes}
	\begin{table}
		\begin{tabular}{ |p{3cm}||p{2cm}|p{2cm}|p{3cm}|p{2cm}| p{2cm}| }
			\hline
			\multicolumn{6}{|c|}{Dissipation mechanisms when $f=3f_r$  } \\
			\hline
			Oscillation shape  &Linear & P3 through SN&P6(6 maxima)&chaos&linear\\
					\hline	
			Uncoated air bubble with $R_0= 10 \mu m$&${\color{red}Rd}>Td>Ld$ &${\color{red}Rd}>Td>Ld$& ${\color{red}Rd}>Td>Ld$&${\color{red}Rd}>Td>Ld$&${\color{red}Rd}>Td>Ld$\\
			\hline	
			Uncoated air bubble with $R_0=2\mu m$&${\color{red}Rd}>Ld>Td$ &${\color{red}Rd}\approxeq Ld>Td$& ${\color{red}Rd}>Ld>Td$&${\color{red}Rd}>Ld>Td$&${\color{red}Rd}>Ld>Td$\\
			\hline	
			coated C3F8 bubble with $R_0=4 \mu m$&${\color{red}Rd}>Cd>Ld>Td$ &$Cd>{\color{red}Rd}>Ld>Td$& $Cd>{\color{red}Rd}>Ld>Td$&$Cd\approxeq{\color{red}Rd}>Ld>Td$& ${\color{red}Rd}\approxeq Cd>Ld>Td$\\
			\hline	
			coated C3F8 bubble with $R_0=1 \mu m$&${\color{red}Rd}>Cd>Ld>Td$ &${\color{red}Rd}>Cd>Ld>Td$& ${\color{red}Rd}>Cd>Ld>Td$&${\color{red}Rd}>Cd>Ld>Td$& ${\color{red}Rd}>Cd>Ld>Td$\\
			\hline	
		\end{tabular}
		\caption{Evolution of dissipation powers at different nonlinear regimes for uncoated air and coated C3F8 bubbles when $f=3f_r$. }
		\label{table:1}
	\end{table}
\end{savenotes}
9- within the pressure ranges that were investigated here, occurrence of the giant resonances were in the form of a large amplitude periodic oscillations that emerge out of the chaotic window at higher pressures. These oscillations were concomitant with a sharp increase in Rd, Ld and Cd (in case of coated bubbles) and at the same time concomitant with a decrease in Td. This implies that oscillations have larger wall velocity amplitudes and acceleration; moreover due to the larger instantaneous changes of the $R_{max}$ to $R_{min}$ higher core temperatures are expected. The faster collapses and rebound in these oscillation regimes leaves very little time for heat transfer thus Td decreases. This may advantages for sonochemical applications of ultrasound as higher core temperatures are achieved and thermal loss is decreased. 
\subsection{$|\dot{R(t)}|_{max}$, $|P_{sc}|_{max}$, Rd, $W_{total}$ and STDR during non-destructive oscillations and their possible applications }
In this section we summarize some important parameters related to the bubble behavior and link them to possible medical and sonochemical applications. These parameters are extracted for excitation pressures that leads to non-destructive regime of oscillations ($\frac{R_{max}}{R_0}\leq 2$) and for exciation freqeuncy range of $f=0.25f_r-3f_r$.\\
1- The maximum wall velocity amplitude ($|\dot{R(t)}|_{max}$) was the largest for the bubbles that were sonicated with $f=0.45-0.5f_r$. Higher wall velocities results in faster micro-streaming. The shear stress induced by bubbles on nearby objects is proportional to the micro-streaming velocity and to the thickness of the boundary layer \cite{63,64,96}. The thickness of the boundary layer is inversely proportional to frequency \cite{63,64,96}. Thus, sonication with $f=0.45-0.5f_r$ not only can produce the highest non-destructive  micro-streaming, but it also has a small boundary layer. Sonication in this frequency range may therefore enhance the shear stress and drug delivery efficacy. Moreover, non-destructive and  non-inertial high shear stresses in this frequency may enhance the surface cleaning \cite{97,98,99} while avoiding damage to delicate micro-structures (e.g. semi-conductor industry,  optical devices $\&$ precision apparatus) usually stemming from violent inertial collapse of bubbles \cite{98,99}. Quantification of the shear stress at various non-linear regimes is a complex task and is the subject of future studies. \\
2- The maximum amplitude of the back-scattred pressure ($|P_{sc}|_{max}$) from bubbles was the largest for bubbles sonicated with $3f_r$. Echogencity of the ultrasound images is directly proportional to  $|P_{sc}|_{max}$. Thus, in applications like B-Mode imaging \cite{82} using contrast agents, higher frequencies ($f=1.6-3f_r$) may be desired. However, one must also note that the higher $|P_{sc}|_{max}$ occurs at a higher pressure for $f=3f_r$; thus, the signal intensity from the background tissue can be higher. In the absence of any non-linear signal acquisition (as an instance amplitude modulation \cite{74,79,80} or phase inversion \cite{100}) that suppresses the tissue response in the final image, the effect of higher scattering from tissue at higher pressures should also be considered. On the other hand the abrupt increase in the $|P_{sc}|_{max}$ of the bubble when SN bifurcation takes place (e.g. at $f=1.6f_r$ or $f=3f_r$ (Fig.12b)) can be used to increase the residual signal in amplitude modulation techniques and increase the contrast to tissue ratio and signal to noise ratio. In amplitude modulation technique two pulses are sent to the target with different amplitudes. The received signals from the target are then scaled and subtracted; due to the linear tissue response the two signals cancel each other after subtraction. However, the nonlinear response of the bubbles with respect to increase in pressure results in a considerable residue which leads to enhanced CTR. Another application for the non-inertial higher $P_{sc}$ can be in drug delivery or surface cleaning. The increased pressure radiated by the bubbles can increase the permeability \cite{96} of the cells or objects in their vicinity and contribute to the drug delivery enhancement or cleaning.\\
3- Rd and $W_{total}$ were maximum for bubbles that were sonicated with $0.25-0.3f_r$. Higher Rd and $W_{total}$ are of great importance for applications related to bubble enhanced heating in high intensity focused ultrasound (HIFU) \cite{30,101,102} and ultrasound thermal therapies and hyperthermia. Enhanced heating is of particular interest especially in cases like liver and brain where there is strong cooling of tissue due to high blood perfusion and the presence of skull \cite{103} and rib cage limits the amount of ultrasound energy that can be delivered to target. In \cite{104}, it was shown that enhancing the deposited power by increasing the wave dissipation or enhancing the pressure amplitude can decrease the effect of blood flow cooling until full necrosis takes place. Sonication with $f=0.25-0.3 f_r$ can provide Rd and $W_{total}$ of at least 6 times greater when compared to the case of sonication with $f_r$. Moreover, the higher frequency component of the $P_{sc}$ signal (e.g. 3rd SuH, $7/2$ UHs) increases the absorption of the Rd in tissue and furthermore enhances the localized heating. Another advantage of sonication with $f=0.25-0.3 f_r$ is that Rd and $W_{total}$ are very small for low pressures; however, above a pressure threshold (concomitant with the generation of UHs and SHs in the $P_{sc}$) Rd and $W_{total}$ significantly increase. This finding is in line with experimental observation \cite{30}, where enhanced heating was concomitant with SH and UH emissions and broadband noise.  The lower dissipation of acoustic waves below the pressure threshold leads to minimum enhanced heating and wave dissipation in the pre-focal tissue \cite{32,105,106,107,108}. This allows higher energy delivery for bubbles in the target (especially in cases where delivery of higher acoustic energy is challenging) and enhances the safety of the treatment as the off-target bubble activity is minimized. Moreover, the generation of UHs at the target can be used to monitor and control the treatment using methods like passive cavitation detection \cite{77,85}. $W_{total}$ and Rd were minimum for higher frequencies $f=1.6-3f_r$. Thus, in addition to higher $P_{sc}$ which leads to higher echogenecity in ultrasonic imaging, sonication with these frequencies results in lower heating due to bubble activity. This is another reason why higher frequencies may be more suitable for contrast enhanced ultrasound imaging. Moreover, enhanced absorption ($W_{total}$) in the target can be used to shield \cite{34} structures with higher ultrasound attenuation (as an instance post-target bone \cite{104} in brain).\\
4- The STDR as a function of pressure is nonlinear. The highest STDR belongs to $f=3f_r$. In the absence of super-harmonic resonance, generation of SHs and UHs are concomitant with a decrease in STDR. As it was discussed in previous sections for $f=1.6-3f_r$, Td undergoes a large increase when SHs are generated which consequently leads to a decrease in STDR. Despite the decrease, STDR still remains higher than $f<2f_r$.  The higher STDR have great advantages for contrast enhanced imaging. Higher STDR means bubble scatters more and dissipates less. This has consequences in increasing the echogenecity of the target and the underlying tissue. However, higher STDR by itself does not imply that a set of exposure parameters are suitable for imaging applications. As an instance when $f=3f_r$ STDR is very high at lower pressures (e.g. 10 kPa); however, at the same time $P_{sc}$ is very small. This means that despite a high STDR, because of the weaker scattering by the bubble, the contrast signal may not be distinguishable from the background noise. Thus, STDR should be used in tandem with the $P_{sc}$ and Rd curves to study the suitable exposure parameters for the relevant application.\\
\section{Conclusion}
In this work we investigated the mechanisms of energy dissipation in bubble oscillations and their contribution to the total damping ($W_{total}$) at various nonlinear regimes of bubble oscillations. By using a comprehensive bifurcation analysis, we have classified the nonlinear dynamics of the bubbles and the corresponding dissipation mechanisms. The bifurcation structure of the uncoated and coated bubbles including the full thermal and radiation effects have been classified for the first time. Using our recently developed equations for energy dissipation in the oscillations of coated and uncoated bubbles\cite{40,41}, the pressure dissipation mechanisms of ultrasonic energy were analyzed in detail. Results were presented in tandem with the bifurcation diagrams and several nonlinear features of dissipation phenomenon were revealed and classified. We have shown that by choosing suitable frequency and pressure a particular bubble related effect can be enhanced (e.g.  maximum wall velocity amplitude, maximum scattered pressure, etc.). The exposure parameters by which each of these parameters are maximum seem to be universal and regardless of the bubble size and coating. For example within the exposure parameter ranges that were studied in this paper we show that, for all the bubbles maximum non-destructive wall velocity occurs when the bubble is sonicated with $0.4f_r\leq f\leq 0.5f_r$, maximum non-destructive scattered pressure and STDR are reached when $f=3f_r$, nondestructive radiation damping and total scattered power are maximized at $f=0.25rf_r$. These parameters can be used as a guideline to optimize possible related applications (e.g. Imaging, drug delivery, surface cleaning, etc.).  
\section{Acknowledgments}
The work is supported by the Natural Sciences and Engineering Research Council of Canada (Discovery Grant RGPIN-2017-06496), NSERC and the Canadian Institutes of Health Research ( Collaborative Health Research Projects ) and the Terry Fox New Frontiers Program Project Grant in Ultrasound and MRI for Cancer Therapy (project $\#$1034). A. J. Sojahrood is supported by a CIHR Vanier Scholarship.

\appendix
\newcommand{\hbAppendixPrefix}{A}
\section{Bifurcation structure of an uncoated Air bubble with $R_0=2\mu m$ and a coated bubble with $R_0=4 \mu m$}
\subsection{The case of an uncoated air bubble with $R_0=2 \mu m$}
\begin{figure*}
	\begin{center}
		\scalebox{0.4}{\includegraphics{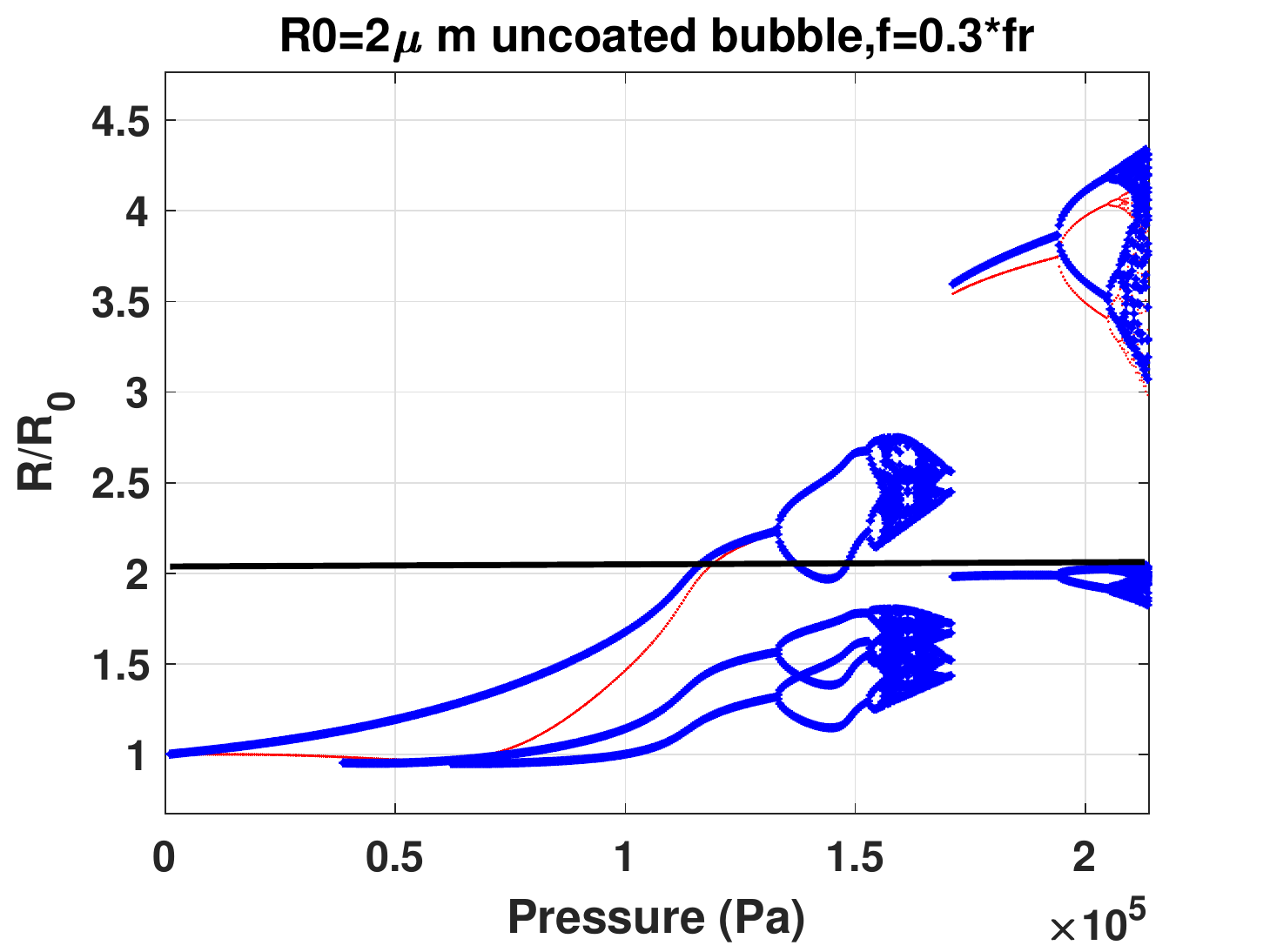}}  \scalebox{0.4}{\includegraphics{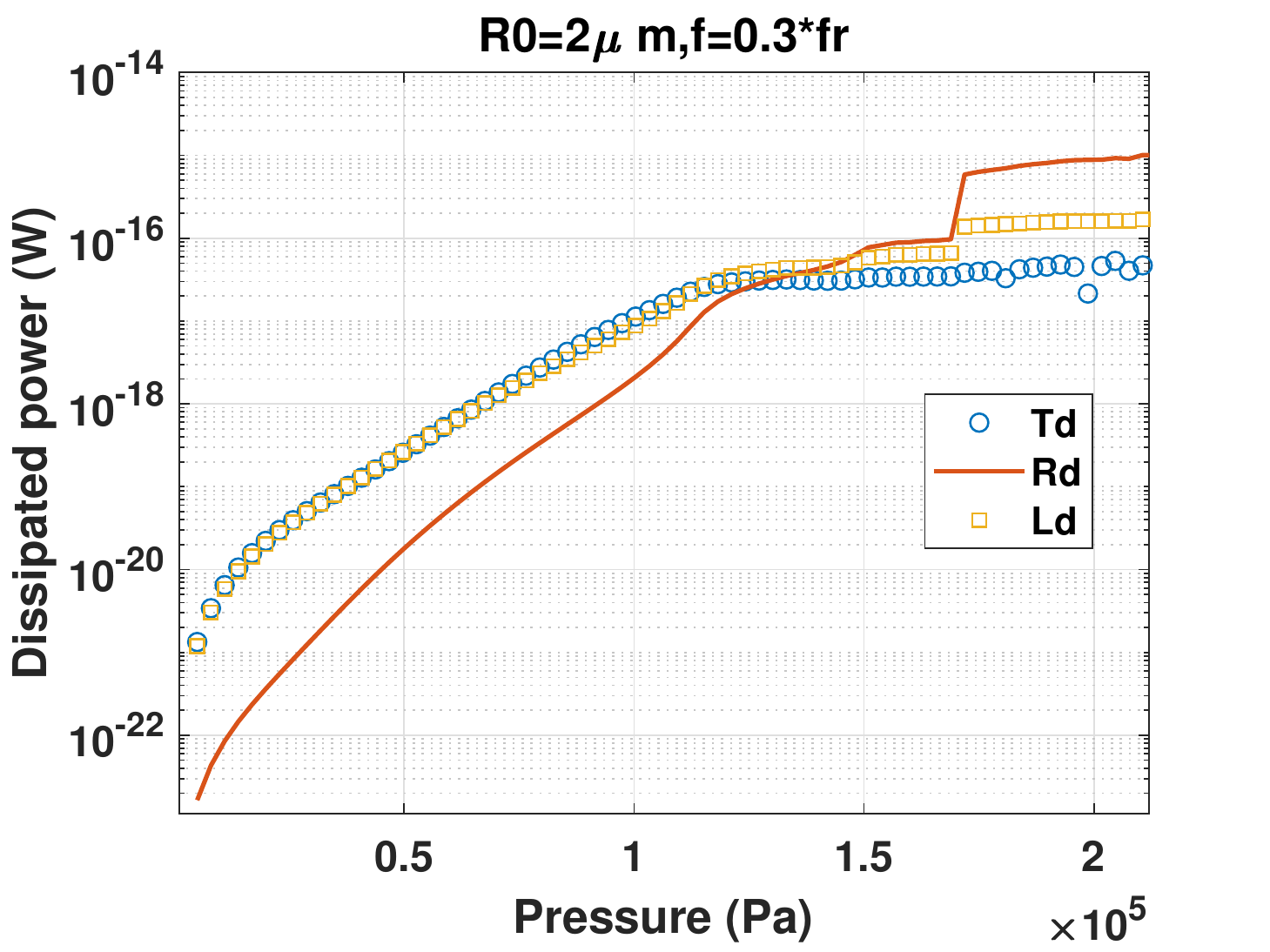}}\\
		\hspace{0.5cm} (a) \hspace{6cm} (b)\\
		\scalebox{0.43}{\includegraphics{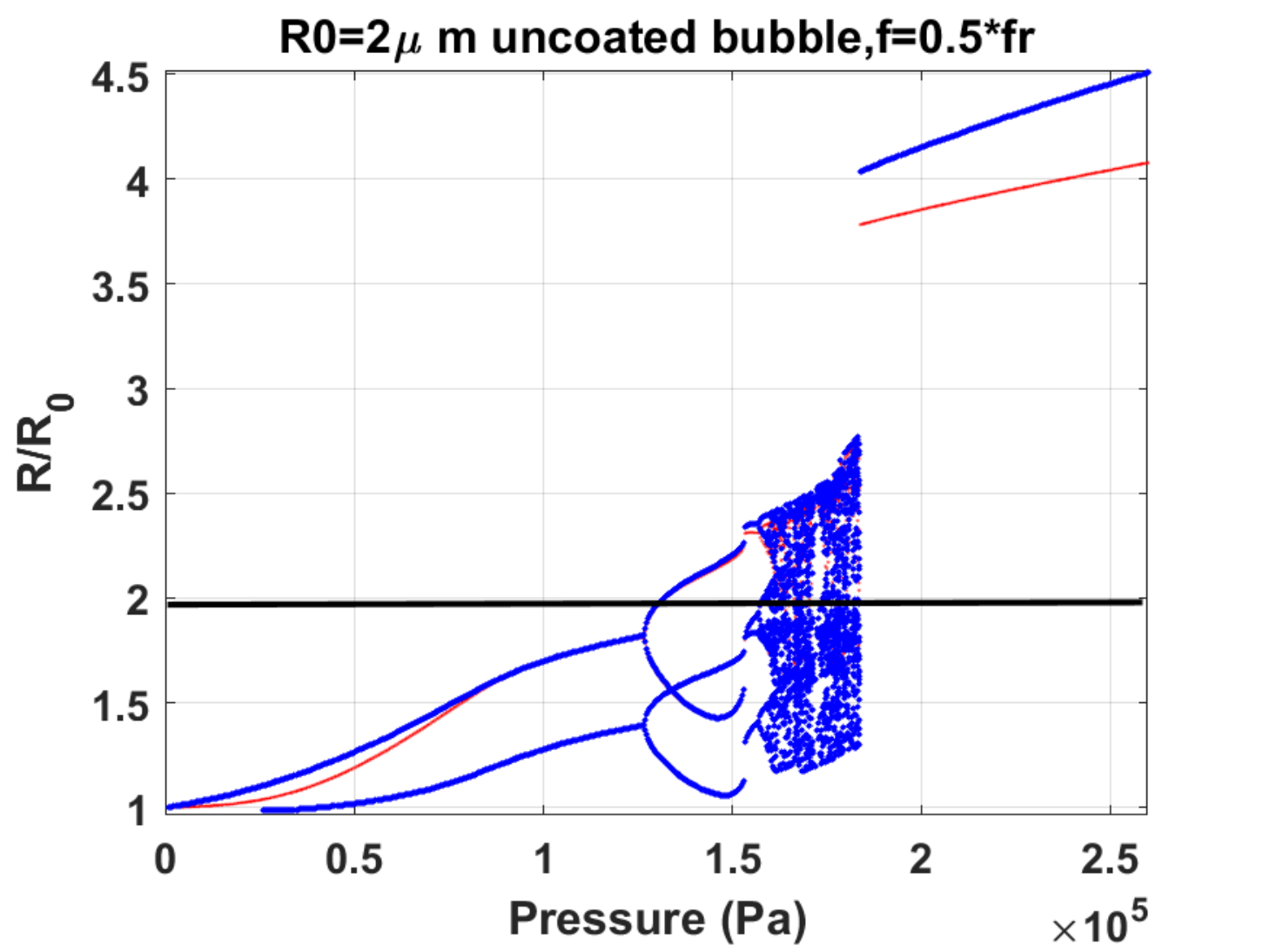}} \scalebox{0.43}{\includegraphics{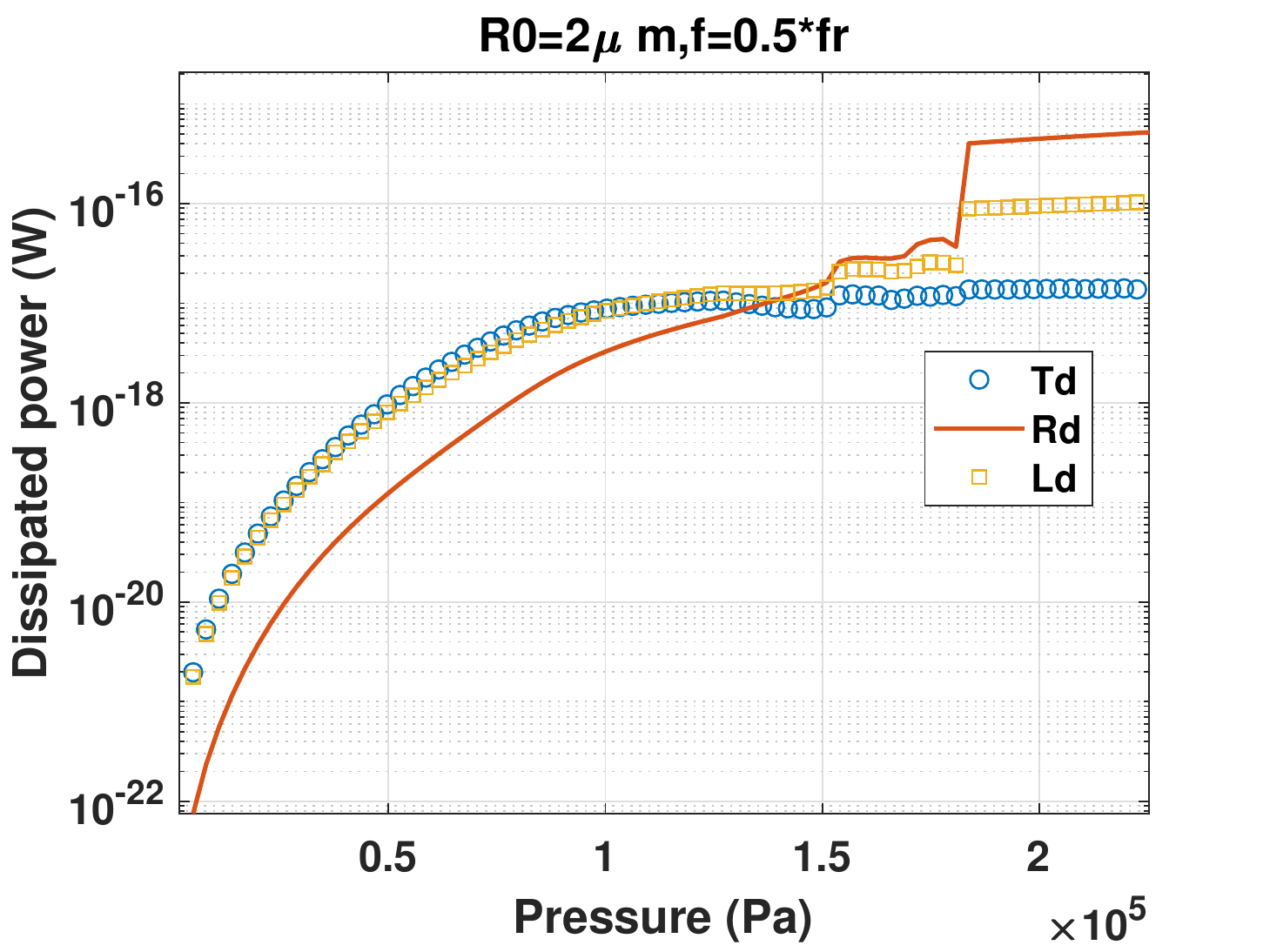}}\\
		\hspace{0.5cm} (c) \hspace{6cm} (d)\\
		\scalebox{0.43}{\includegraphics{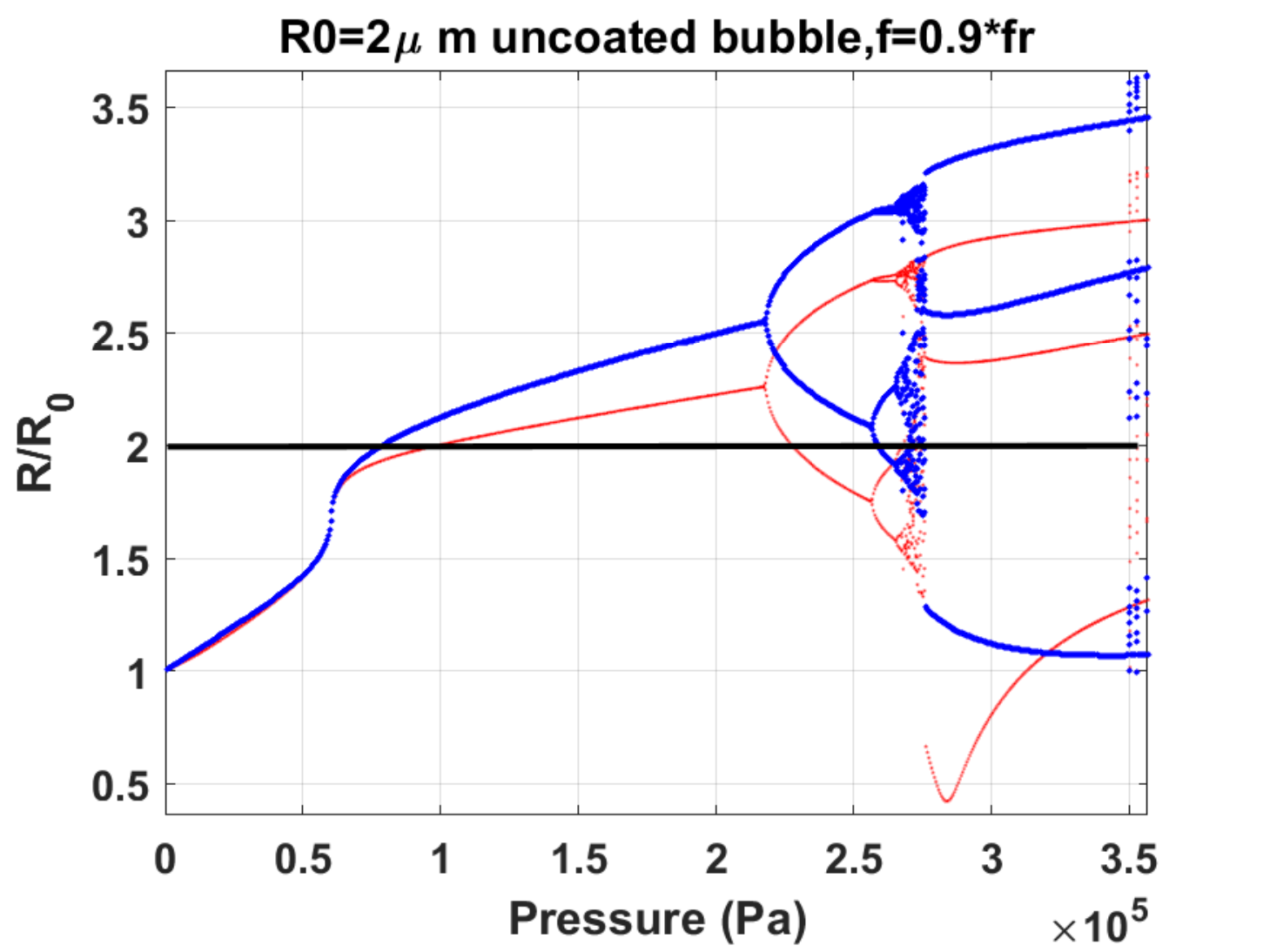}} \scalebox{0.43}{\includegraphics{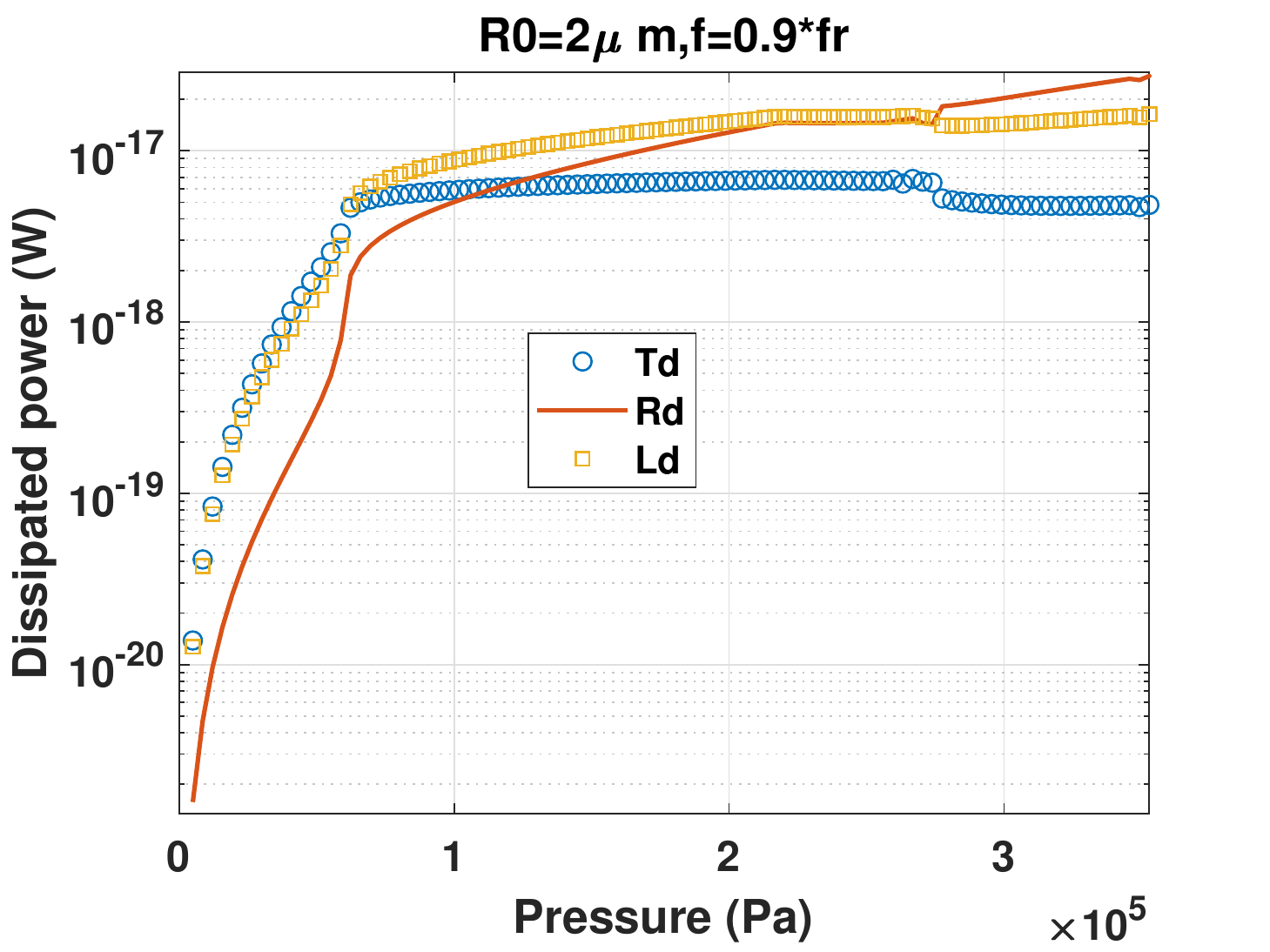}}\\
		\hspace{0.5cm} (e) \hspace{6cm} (f)\\
		\scalebox{0.43}{\includegraphics{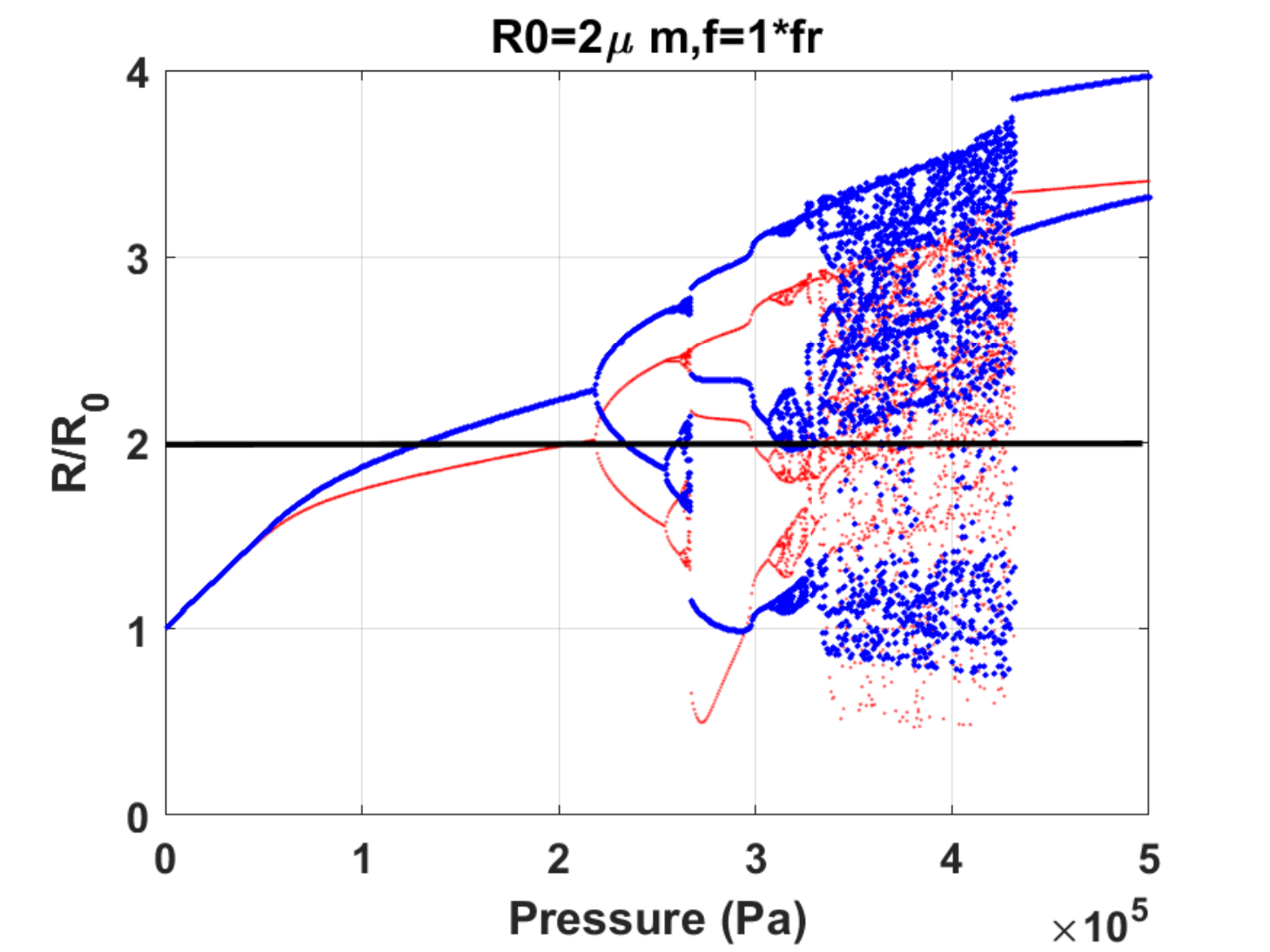}} \scalebox{0.43}{\includegraphics{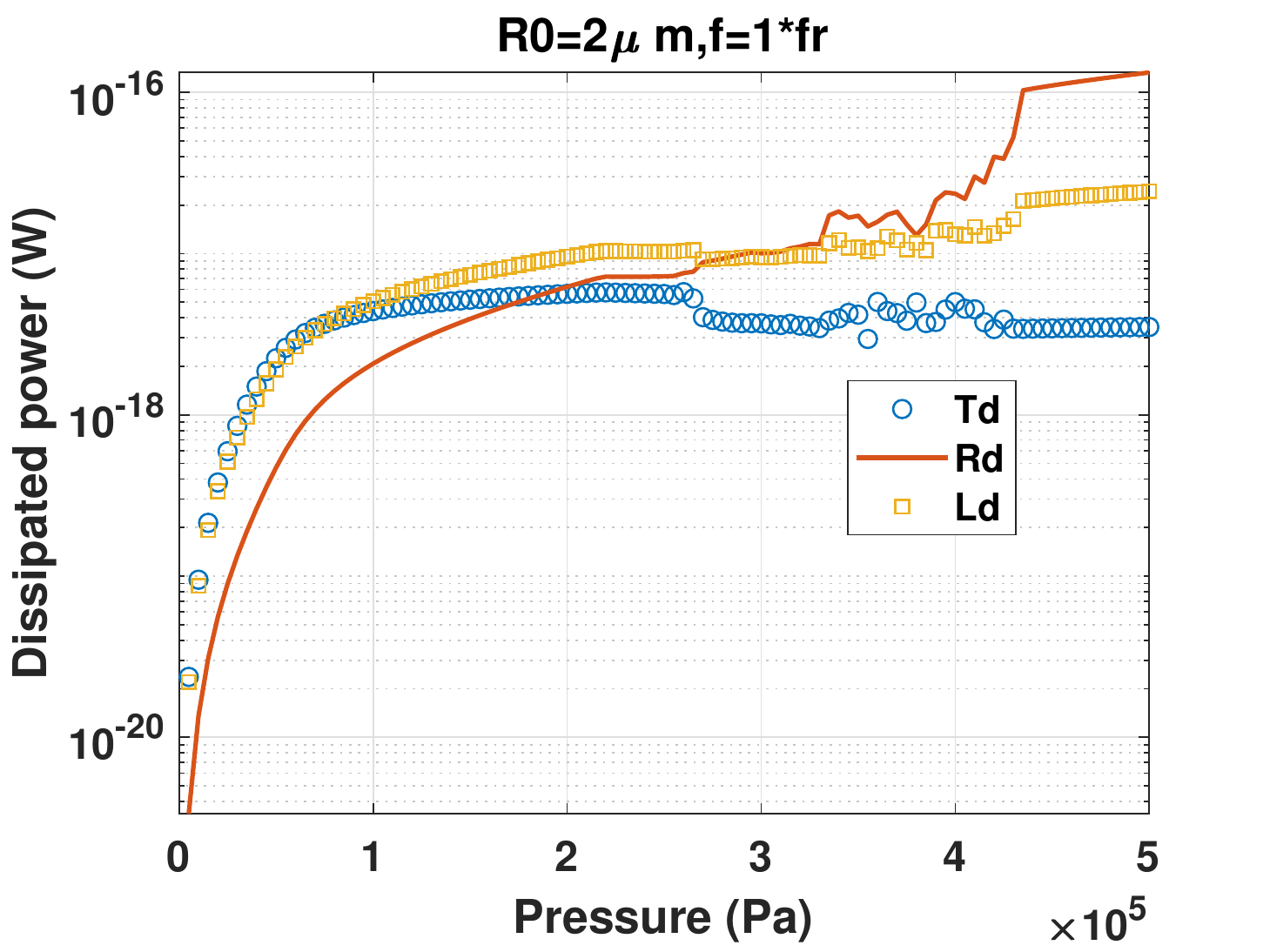}}\\
		\hspace{0.5cm} (g) \hspace{6cm} (h)\\
		\caption{Bifurcation structure (left column) and the dissipated powers (right column) of the oscillations of an uncoated air bubble with $R_0=2\mu m$ for $f= 0.3f_r$ (a-b)-$f=0.5f_r$ (c-d)- $f=0.9f_r$ (e-f) \& $f=f_r$ (g-h).}
	\end{center}
\end{figure*}
The same nonlinear oscillations regimes are also studied for an uncoated air bubble with $R_0=2 \mu m$. Due to the smaller size of the bubble, a larger contribution from viscous effects and a smaller contribution for thermal effects are expected.\\
Fig. A1.a shows the case of sonication with $f=0.3 f_r$. P1 oscillations undergo a 3rd SuH resonance at  $P_a\approxeq 60 kPa$ (P1 oscillations with 3 maxima). At $P_a\approxeq 62 kPa$, the amplitude of one of the maxima coincides with the amplitude of the red curve indicating  that the wall velocity becomes in phase with the driving pressure every acoustic cycle. At $P_a\approxeq 133 kPa$ the red curve undergoes a Pd and P1 oscillations become P2. At the same time Pd occurs in the blue curve and UH oscillations of 7/2 order develop. The red curve grows fast with increasing pressure and with the occurrence of 3rd order SuH and Pd, the amplitude of the red curve becomes the same as the highest amplitude of the blue curve right when Pd takes place, indicating wall velocity is in phase with the excitation pressure. During UH oscillations the value of the upper and lower branch of P2 oscillations in red curve are exactly the same as the two highest amplitudes of the blue curve. A small chaotic window appears at 170 kPa and then a 2nd order SuH giant (P1 with 2 maxima) resonance emerges out of the chaotic window which later undergoes successive Pds to chaos. Due to the smaller bubble size compared to the 10 $\mu m$ bubble (analyzed in Fig. 1b), Ld and Td have the same value for pressures below 120 kPa and Td is no longer the dominant power dissipation mechanism. Rd grows faster with increasing pressure and the contributions of Td, Ld and Rd become similar when Pd takes place. $Rd>Ld>Td$ for the UH regime of oscillations. The giant resonance is concomitant with a sharper increase in Rd making Rd the dominant damping mechanism during giant resonant oscillations.\\
When $f=0.5f_r$ (Fig. A.1c) oscillations are of P1 with one maximum for $P_a<25kPa$; 2nd order SuH oscillations occur at $\approxeq 25 kPa$ (2 maxima appear in the blue curve). At $P_a=90 kPa$, the amplitude of the red curve becomes equal to the highest amplitude maximum of the blue curve. This is concomitant with the saturation of 2nd order SuH frequency component of the scattered pressure ($P_{sc}$). At 127 kPa, the red curve undergoes a Pd concomitant with a Pd in the blue curve resulting in a P2 oscillation with 4 maxima ($3/2$ UH resonance \cite{50}). At $\approxeq$ 140 kPa $R/R_0=2$, and the bubble possibly can not sustain non-destructive oscillations beyond this pressure.  For $P_a>150 kPa$ a chaotic window emerges and later at $P_a\approxeq 184 kPa$ a giant period one resonance emerges out of the chaotic window. Fig. A.1d shows the corresponding  dissipated powers. Ld and Rd are approximately equal for pressures below the occurrence of Pd and UHs. Simultaneous with the $3/2$ UHs, Rd becomes stronger than Ld and Td with $Rd>Ld>Td$. Rd undergoes the sharpest increase concomitant with the generation of giant resonance, making it the strongest dissipation mechanism at higher pressures.\\ 
When  $f=0.9f_r$ (Fig. A.1e), which is a $PDf_r$ \cite{32}, the oscillations are of P1 and grow monotonically with increasing pressure and at $P_a \approxeq 60 kPa$ (the pressure of the $PDf_r$) the oscillation amplitude undergo a sharp increase with the red and blue curve coinciding with each other. Above $P_a \approxeq 75 kPa$, $R/R_0$ exceeds 2 (black horizontal curve) and beyond this point bubble destruction is likely. Oscillations undergo Pd at $\approxeq$ 220 kPa and a small chaotic window occurs at $P_a \approxeq 260 kPa$ through successive Pds. At $\approxeq$ 275 kPa a P3 oscillations with 3 maxima emerges out of the chaotic window until $P_a \approxeq$ 350 kPa where chaos is regenerated. The power dissipation graph in Fig. A.1f indicates that $Td>Ld>Rd$ before the SN bifurcation takes place. Above $P_a \approxeq 60 kPa$ 
(pressure for SN bifurcation) Ld becomes stronger and the dissipation order is $Ld>Td>Rd$. After SN, Rd grows faster with pressure increase while Td stays constant. Rd supersedes Td at  $P_a \approxeq$ 125 kPa and becomes equal to Ld when Pd takes place. Ld, Rd and Td then stay relatively constant for the P2 oscillations regimes (this can be due to the decrease in wall velocity and acceleration when Pd takes place in cases where the bubble is sonicated with a  frequency near $f=f_r$ \cite{32}). Emergence of the P3 oscillations out of the chaotic window with high amplitude is concomitant with an increase in Rd and decrease in Ld and Td; this will lead to an increase in the STDR, however, with the possible trade off the loss of stable oscillations.\\
When $f=f_r$ (Fig. A.1g); at lower pressures ($P_a<60 kPa$) oscillations are of P1 and the wall velocity is in phase with the driving acoustic force (blue and red curve are on top of each other) indicating resonant oscillations. Further pressure increase results in possible bubble destruction at $P_a=120 kPa$ (black horizontal line meets the blue line). At $P_a \approxeq 220 kPa$, occurrence of Pd results in P2 oscillations for  $220<P_a<260$ followed by the emergence of a small chaotic window through successive Pds. Similar to the previous case, a P3 emerges from the chaotic window followed by regeneration of chaos for $320kPa<P_a<425 kPa$. At $\approxeq 425 kPa$ a giant P2 resonance emerges out of the chaotic window. The corresponding dissipated powers illustrated in Fig. A.1h show that for pressures below $\approx 60 kPa$, $Td>Ld>Rd$. Above this pressure $Ld>Td>Rd$ until the excitation pressure reaches $\approxeq$ 190 kPa and Rd becomes equal to Td. When Pd occurs, Rd overcomes Td; then Rd, Ld and Td stay constant during P2 oscillations. Generation of P4 results in an increase in Rd and a subsequent decrease in Ld and Td. Emergence of P3 results in an increase in Rd making the order as $Rd>Ld>Td$ followed by a sharp increase of Rd when giant resonance takes place. Similar to previous cases Td decreases when giant resonance occurs, and Ld increases however with a smaller percentage compared to Rd. Once again, the giant resonance can lead to a significant increase in STDR; however this may lead to bubble destruction. The generation of higher temperatures due to stronger collapses and the decrease in Td may have consequences in enhancing chemical reactions within the bubble.\\ 
\begin{figure*}
	\begin{center}
		\scalebox{0.4}{\includegraphics{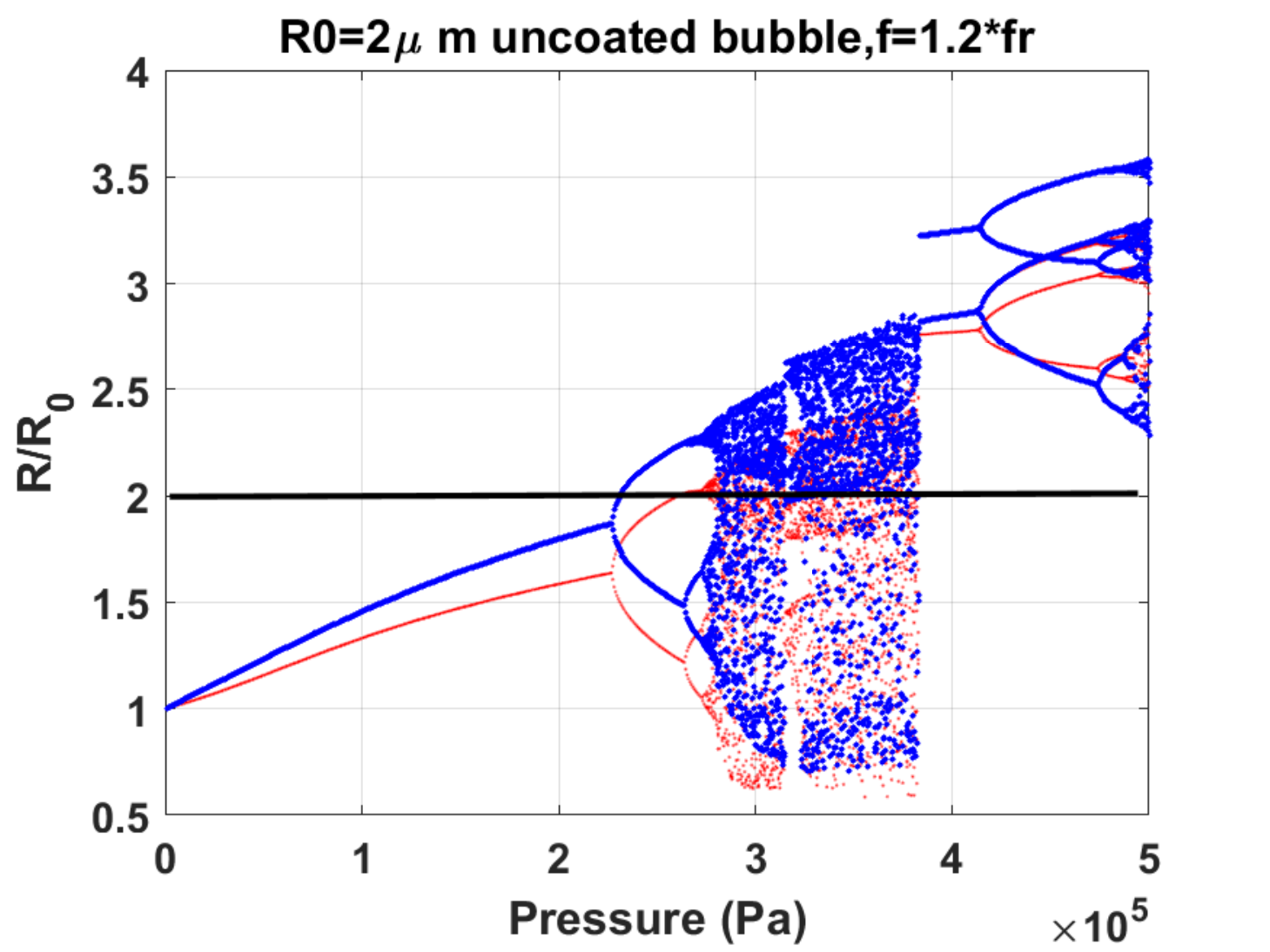}}  \scalebox{0.4}{\includegraphics{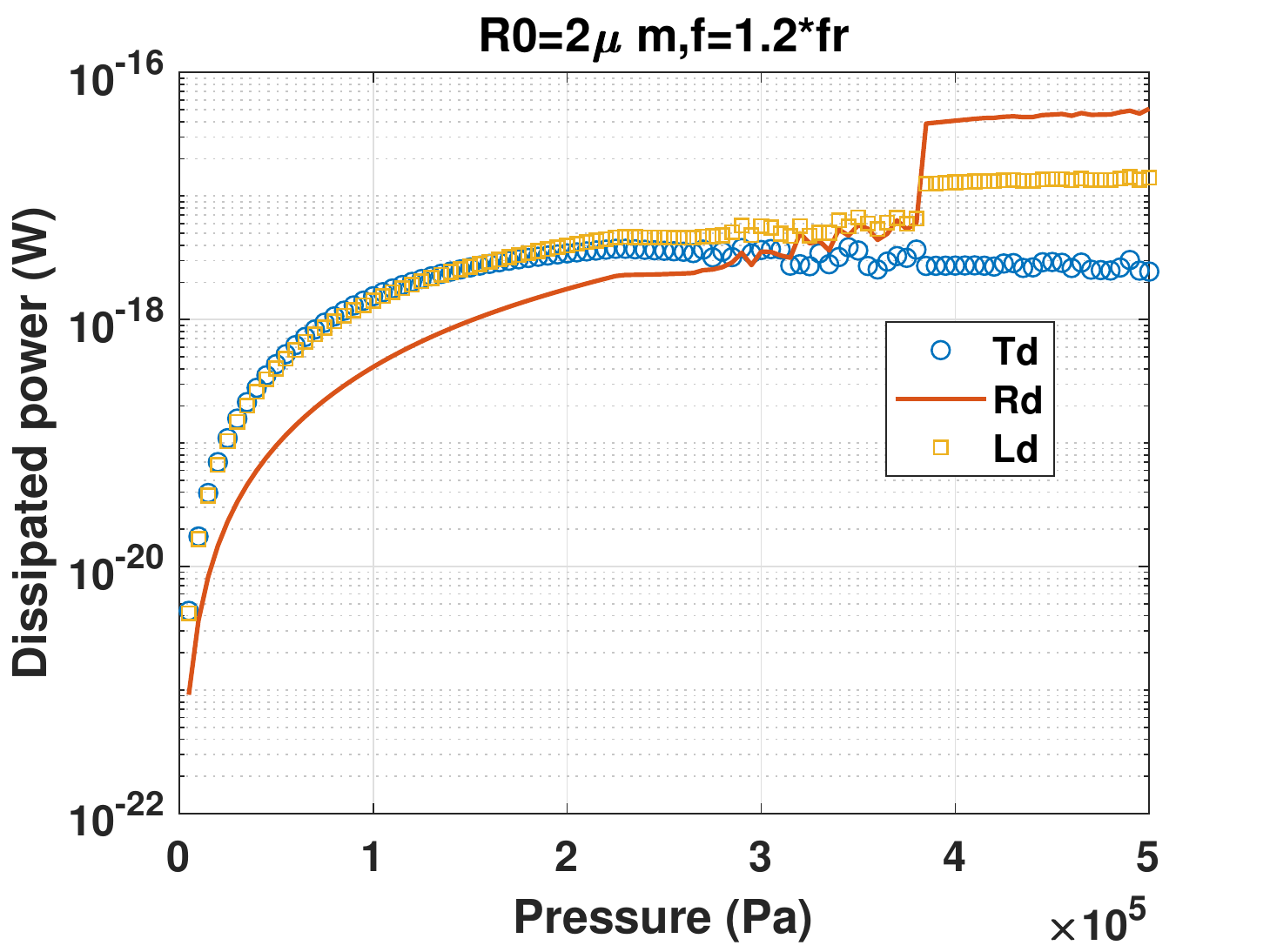}}\\
		\hspace{0.5cm} (a) \hspace{6cm} (b)\\
		\scalebox{0.43}{\includegraphics{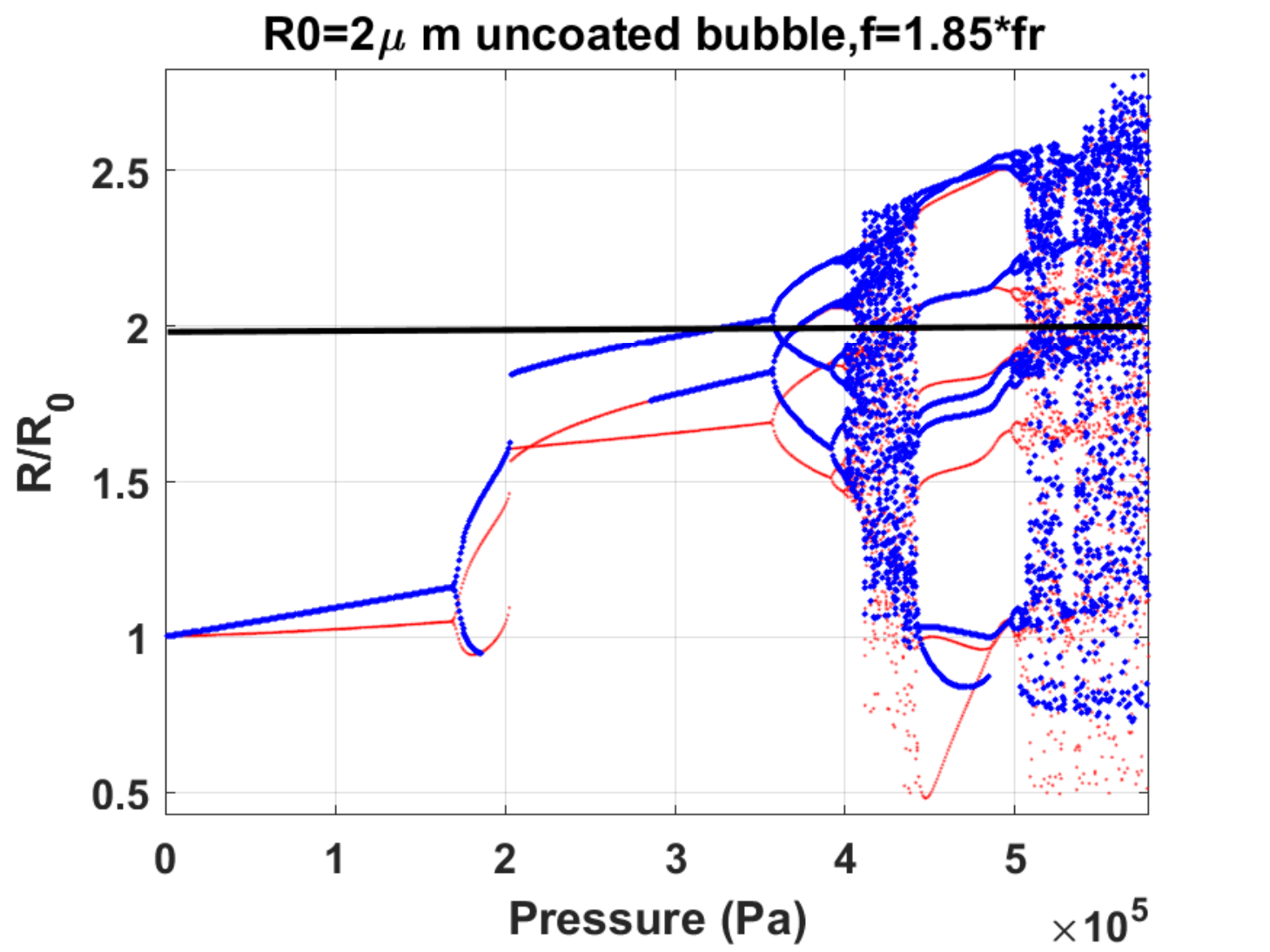}} \scalebox{0.43}{\includegraphics{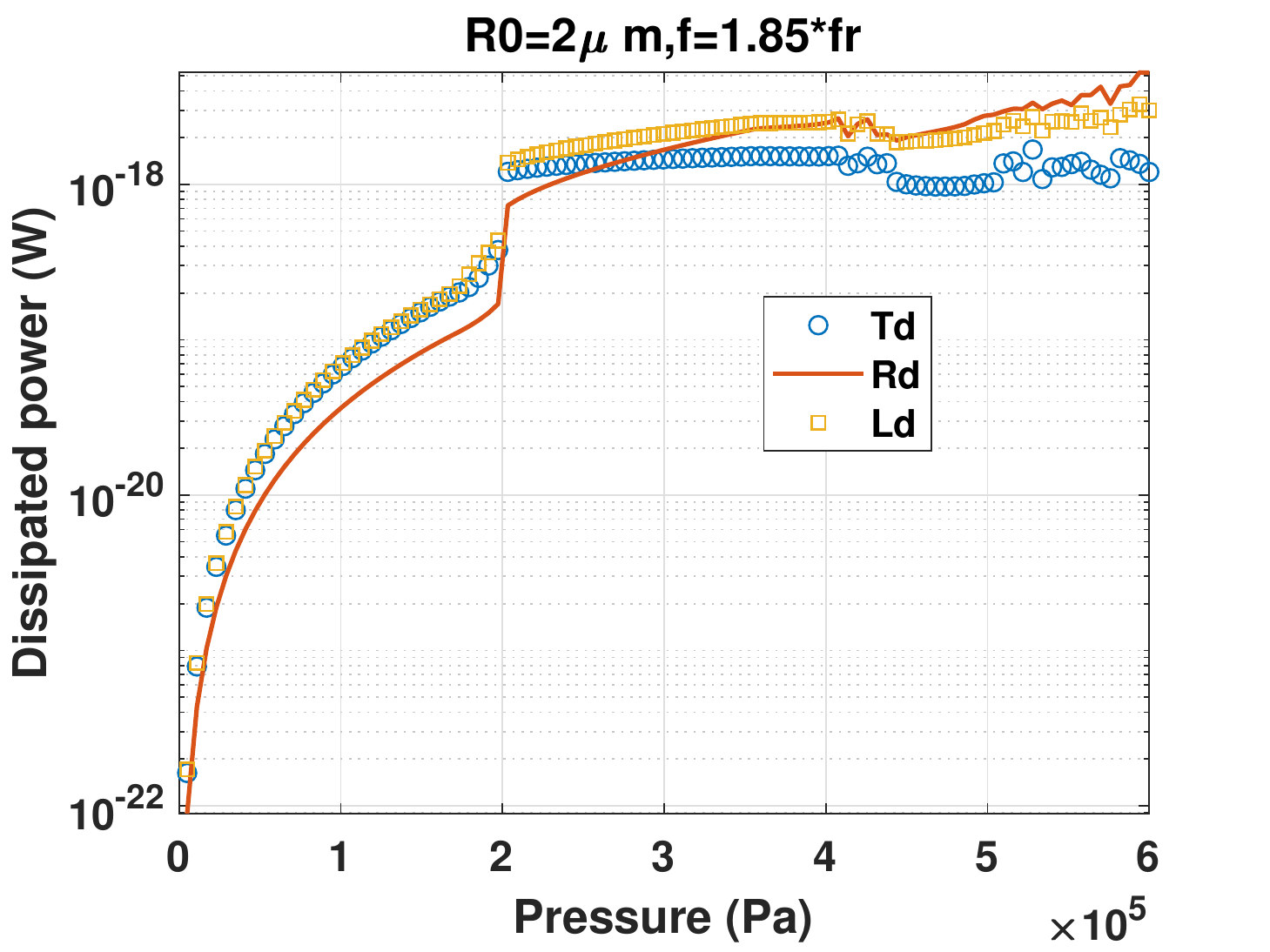}}\\
		\hspace{0.5cm} (c) \hspace{6cm} (d)\\
		\scalebox{0.43}{\includegraphics{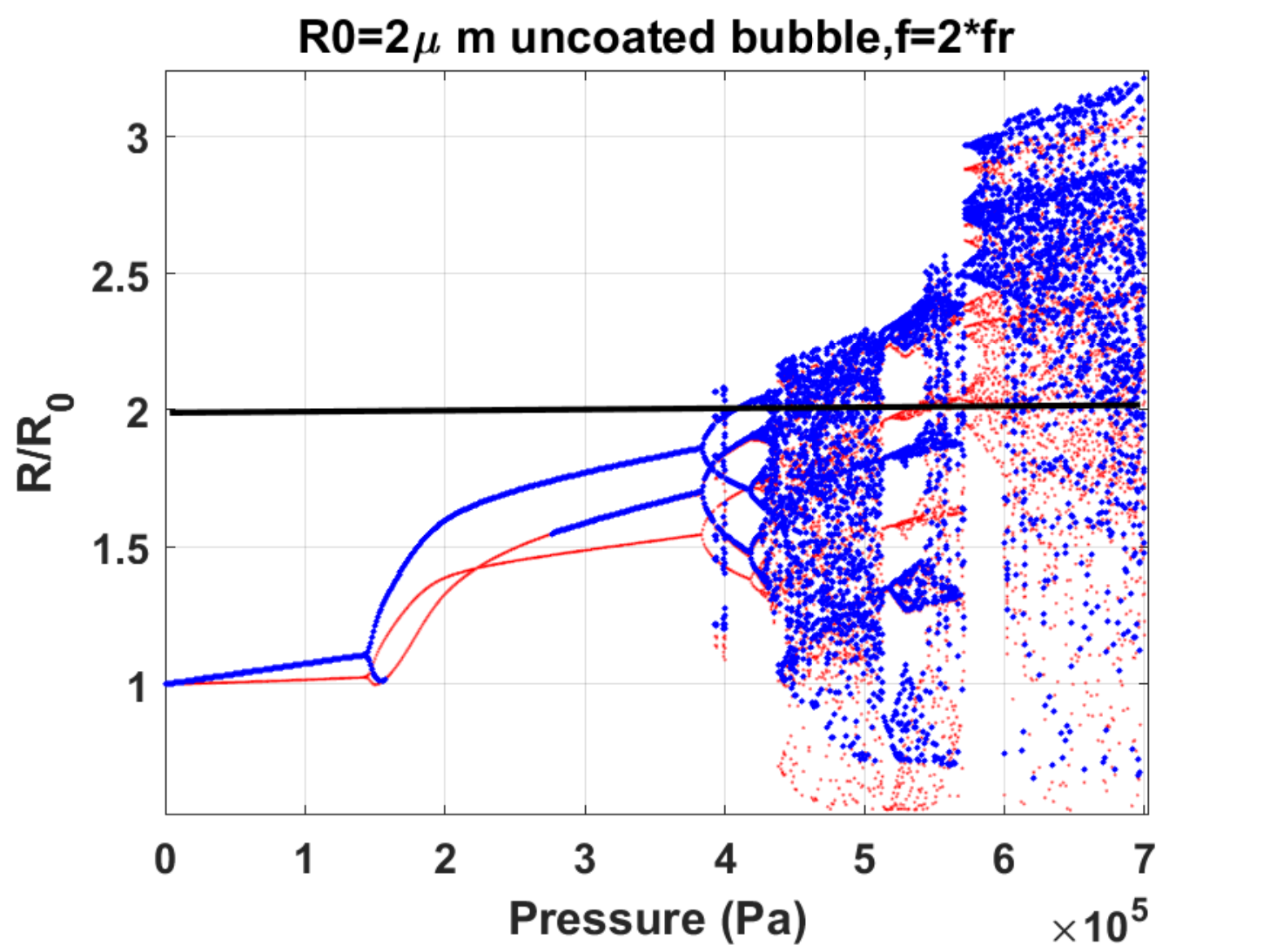}} \scalebox{0.43}{\includegraphics{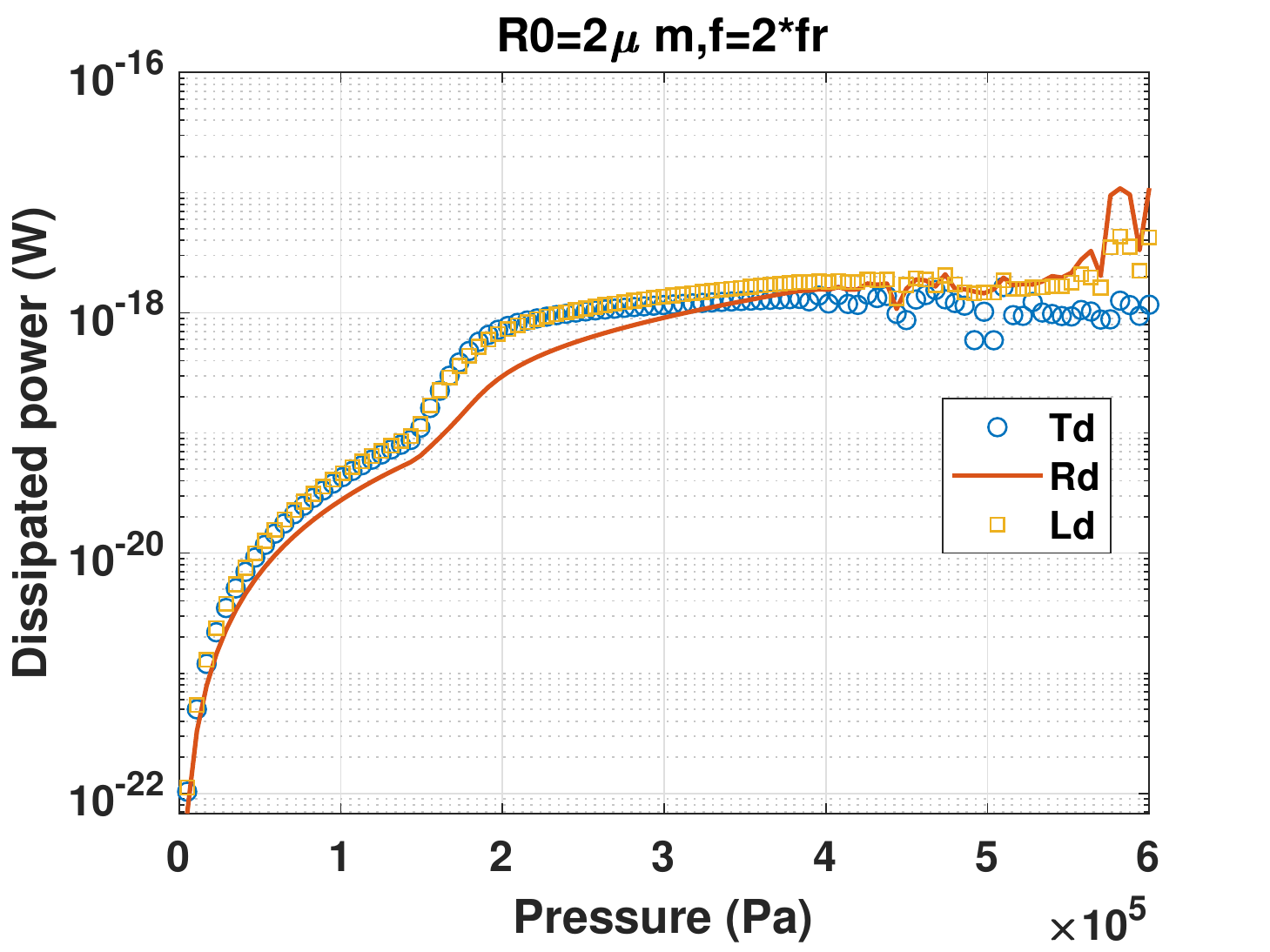}}\\
		\hspace{0.5cm} (e) \hspace{6cm} (f)\\
		\scalebox{0.43}{\includegraphics{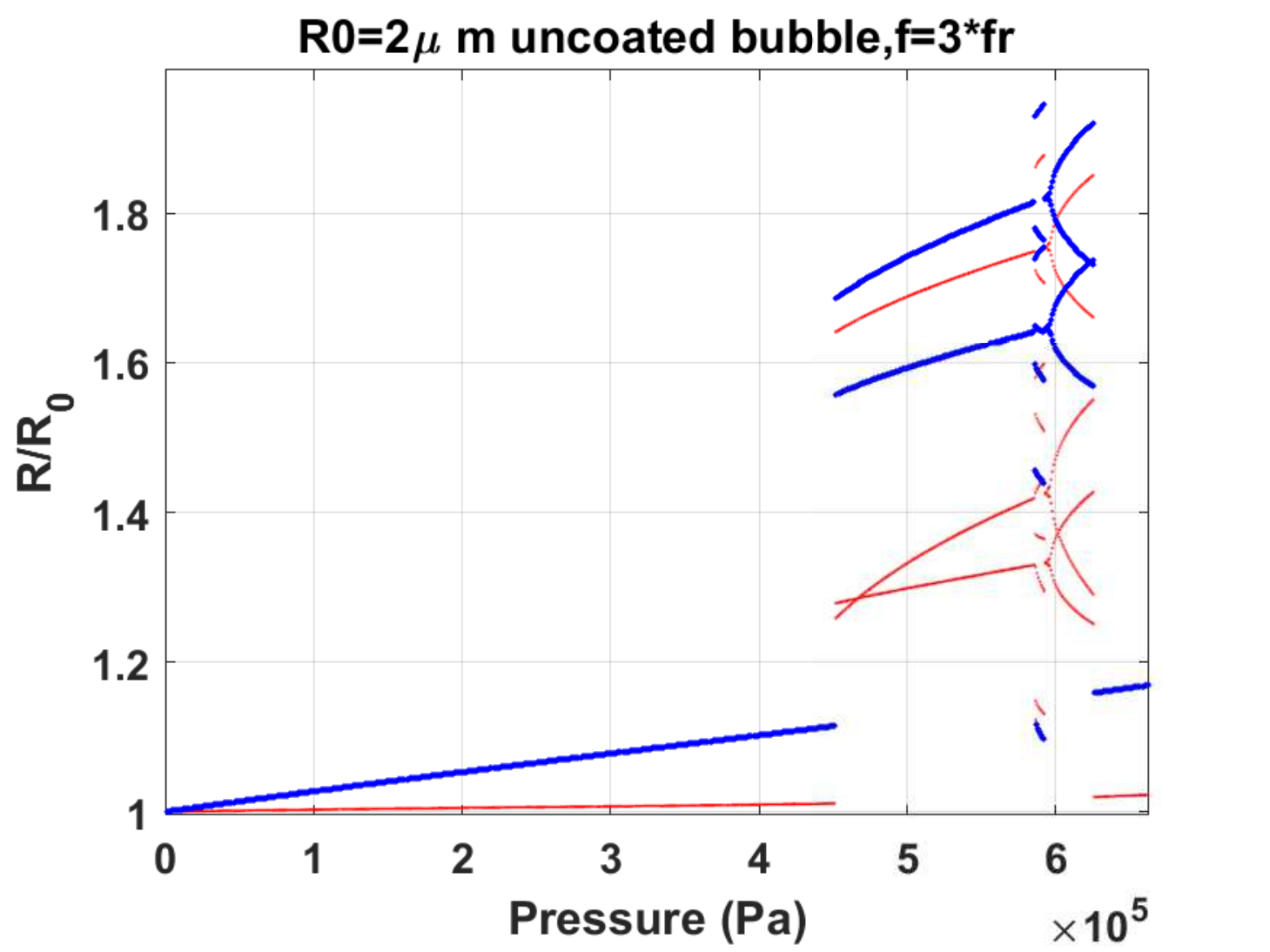}} \scalebox{0.43}{\includegraphics{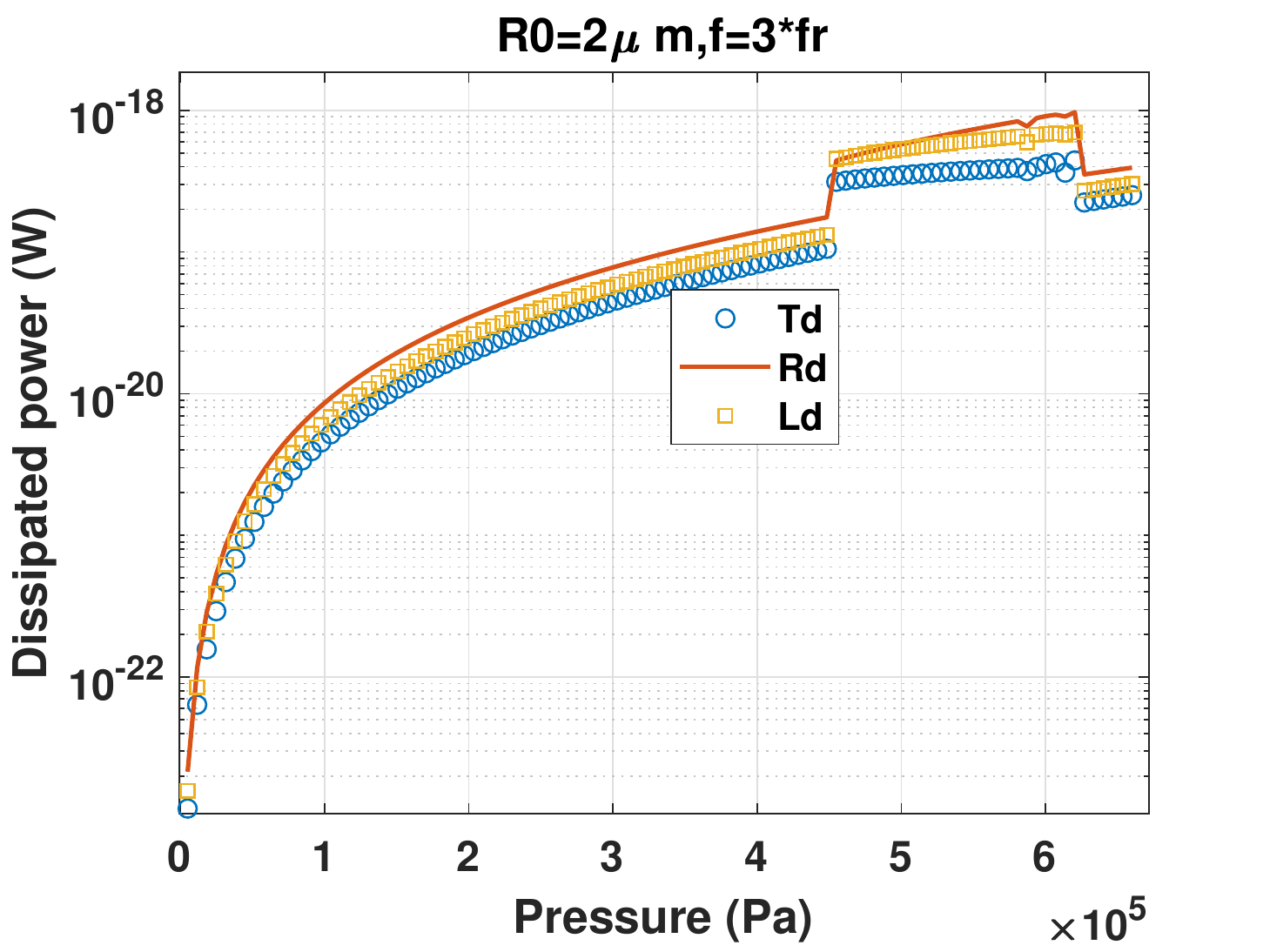}}\\
		\hspace{0.5cm} (g) \hspace{6cm} (h)\\
		\caption{Bifurcation structure (left column) and the dissipated power (right column) of the oscillations of an uncoated air bubble with $R_0=2\mu m$ for $f= 1.2f_r$ (a-b)-$f=1.85f_r$ (c-d)- $f=2f_r$ (e-f) \& $f=3f_r$ (g-h).}
	\end{center}
\end{figure*}
The case of $f=1.2f_r$ is shown in Fig. A.2a. Oscillation amplitude increases monotonically with pressure and bubble undergoes Pd at $\approxeq 225 kPa$. Unlike the case of sonication with $f=f_r$, (and similar to the uncoated air bubble with $R_0=10 \mu m$) Pd occurs when $R/R_0<2$ thus the  bubble is more likely to sustain P2 oscillations. At $\approxeq 225 kPa$ oscillations become chaotic through successive Pds and chaos stretches until $P_a \approxeq 390 kPa$. At this pressure a P2 giant resonant oscillation emerges out of the chaotic window (the solution with the higher amplitude in red curve is exactly equal to the smaller maximum indicating wall velocity becomes in phase with the driving acoustic pressure once every two acoustic cycles). The P2 giant resonance undergoes successive Pds to chaos at $\approxeq 500 kPa$. The corresponding dissipated power graphs (Fig A.2b) show that $Ld \approxeq Td >Rd$ for $P_a	\lesssim$ 225 kPa. When Pd occurs, Ld becomes stronger than Td with $Ld>Td>Rd$; once again, during P2 oscillations, Ld, Td and Rd stay relatively constant as pressure increases.  Generation of P4 and chaos is concomitant with an increase in Rd making $Ld \approxeq Rd> Td$. This is similar to the previous cases when the giant resonance emerges Rd and Ld undergo a sharp increase (Rd exhibits the sharpest increase), while Td decreases slightly.  This makes the contribution order of the dissipation mechanisms as $Rd>Ld>Td$.\\
When $f=1.85f_r$ (the $PDf_{sh}$ \cite{59}) the P1 oscillation amplitude grows slowly with increasing pressure and the bubble undergoes a Pd at $\approxeq 190 kPa$. Generation of Pd is concomitant with a sharp increase in the oscillation amplitude (oscillations are P2 and have two maxima). At 200 kPa P2 oscillations undergo a SN bifurcation to P2 oscillations of higher amplitude (here the signal looses one of its maxima \cite{59}). As pressure increases the second maxima is generated at $\approxeq 290 kPa$. At 300 kPa, $R/R_0$ becomes larger than 2 (black line). P2 oscillations undergo period doubling to P4-2 oscillations and a chaotic window appears at 405 kPa through successive Pds of the P4-2 signal. Later at 440 kPa, a P6 oscillation with 6 maxima emerges out of the chaotic window which through successive Pds translate to P12 and chaos at $\approxeq 500-505 kPa$. The corresponding power graphs (Fig. A.2d) show that $Td=Ld>Rd$ below $P_a\approxeq 190 kPa$ where Pd takes place. Generation of Pd results in a decrease in Td and Ld becomes stronger than Td. Simultaneous with the SN bifurcation at $P_a=200 kPa$, Rd, Ld and Td undergo a sharp increase (with Rd exhibiting the highest increase). This makes Rd approximately equal to Ld and for the rest of the P2 oscillations, power dissipation stays relatively constant with increasing pressure and $Rd=Ld>Td$. Generation of chaos results in some sporadic fluctuations and when P6 emerges out of chaos Ld and Td decrease resulting in $Rd>Ld>Td$.\\
When $f=f_{sh}$ ($f=2f_r$ in Fig. A.2e), P1 oscillations slowly grow with increasing pressure and at $P_a\approxeq 140kPa$, a Pd takes place, and concomitant with Pd, oscillation amplitude start growing quickly. P2 oscillations evolve in the form of a bow-tie.  Right when Pd occurs, oscillations have two maxima, one of the maxima disappears shortly after Pd but re-emerges with a value equal to the higher amplitude of the red curve at $\approxeq 290kPa$ (in \cite{58} we have shown that this may be the point where $1/2$ SH frequency component of the $P_{sc}$ gets saturated). Oscillations undergo Pds at 395 kPa to P4-2 oscillations and when $P_a\approxeq 410 kPa$, $R/R_0$ exceeds 2 (black horizontal line collides with the blue curve). A small window of P6-2 (\cite{58}) occurs right before 400 kPa. Later chaos appears at $\approxeq 440 kPa$. For $P_a<300 kPa$, $Td=Ld=Rd$ and simultaneous with Pd,  dissipation powers undergo a fast increase; but, they quickly plateau with pressure increase. Further increase in pressure results in a slight decrease in Td and a slight increase in Rd and Ld. At $P_a \approxeq 580 kPa$ where the amplitude of the chaotic oscillations sharply increases; Rd becomes stronger than Ld.\\
When $f=3f_r$ (Fig. A.2g), oscillations grow very slowly with pressure increase until at $\approxeq 420 kPa$ at which P1 oscillations undergo a SN bifurcation to P3 oscillations with 2 maxima.  Oscillation amplitude increase slowly with increasing pressure and a small P6 window appears at $\approx$ 590-598 kPa followed by a return to P3 oscillations and then to P6 oscillations.  P6 oscillations return to P1 oscillations at $P_a\approxeq 610 kPa$. The corresponding dissipated power graphs (Fig. A.2h) show that similar to the case of unacoated air bubble with $R_0= 10 \mu m$, $Rd>Ld>Td$ before the occurrence of the SN bifurcation. Occurrence of SN bifurcation is concomitant with  a sharp increase in Rd and Ld and Td with ($Rd>Ld>Td$). As pressure increases the difference between Rd, Ld and Td diverges resulting in an increase in STDR. A return to P1 oscillations is concomitant with a decrease in the dissipation. Due to the larger increase in Td when P3 occurs, the STDR decreases. The increase in Td is due to the large average surface area of the bubbles and a slower rebound during P3 oscillations.   
\subsection{The case of a coated C3F8 bubble with $R_0=4 \mu m$}  
\begin{figure*}
	\begin{center}
		\scalebox{0.4}{\includegraphics{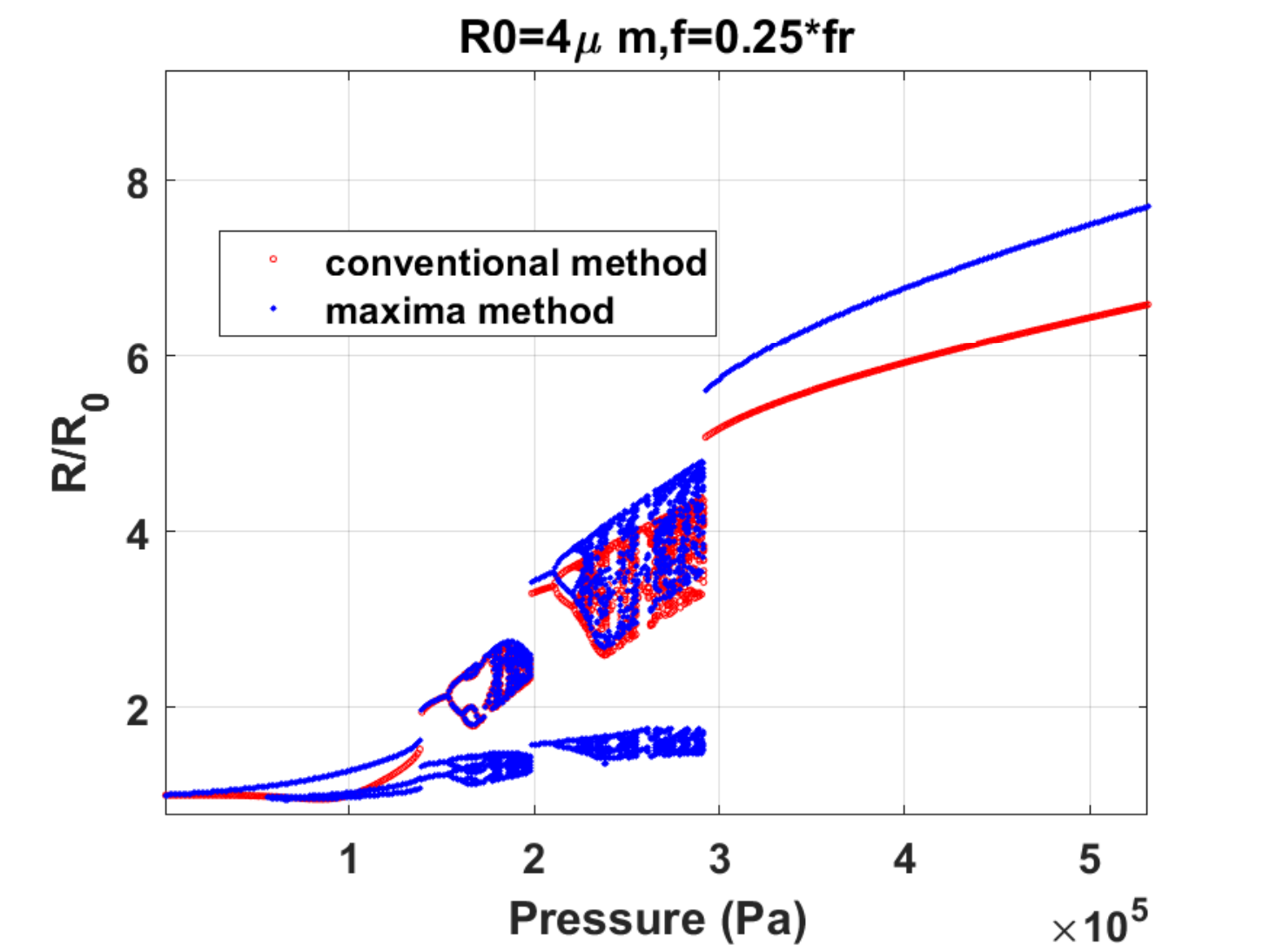}}  \scalebox{0.4}{\includegraphics{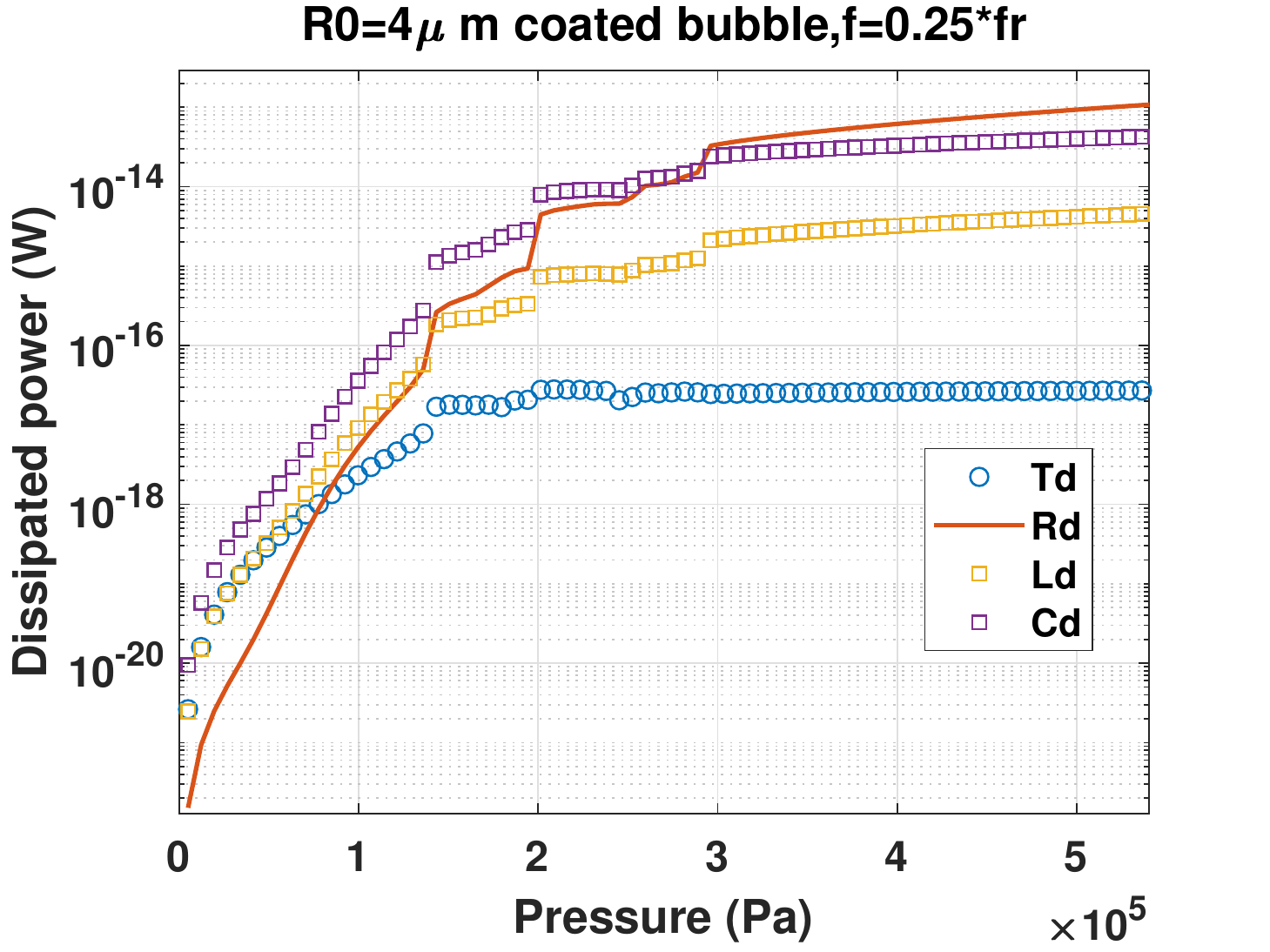}}\\
		\hspace{0.5cm} (a) \hspace{6cm} (b)\\
		\scalebox{0.43}{\includegraphics{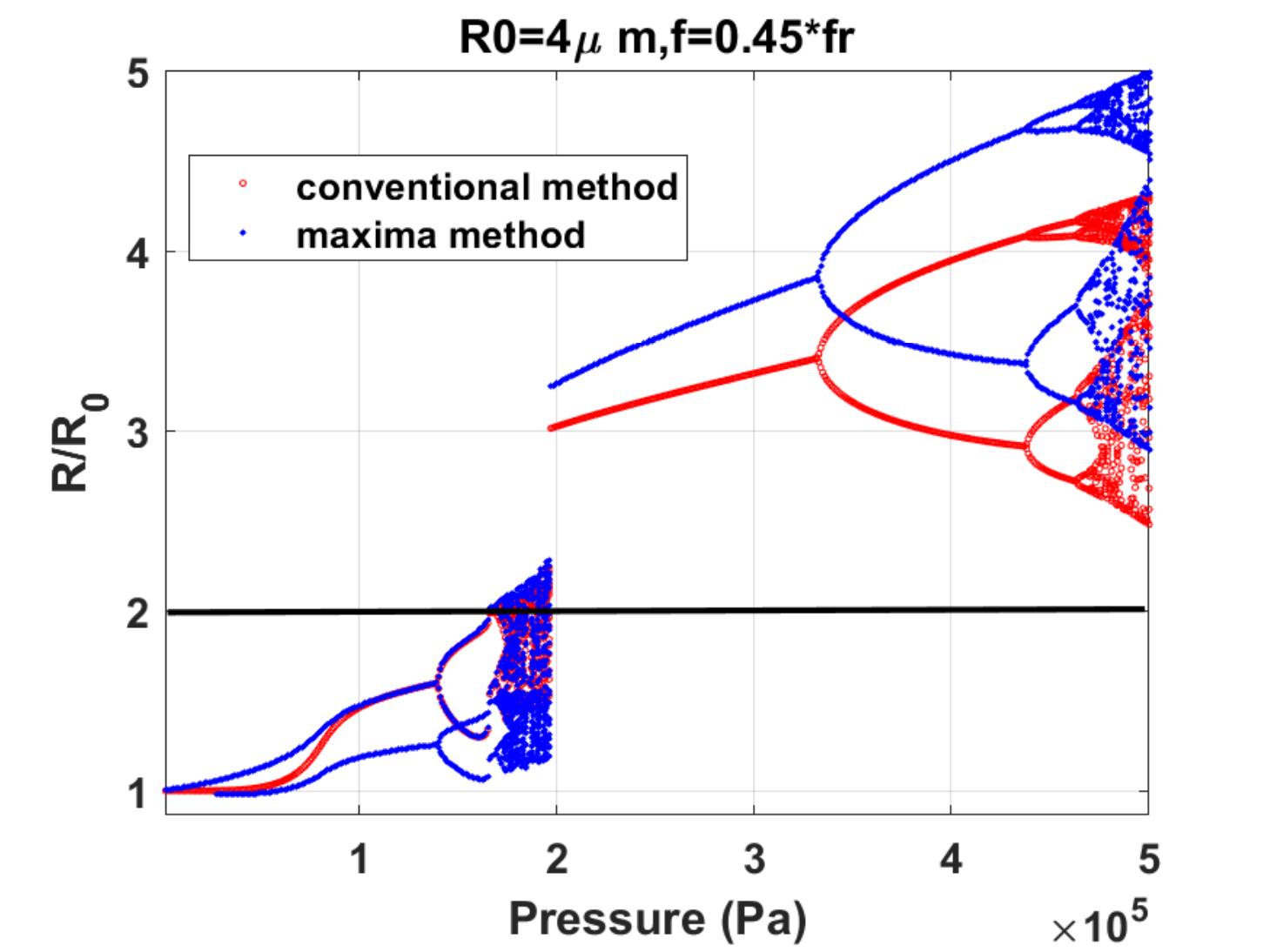}} \scalebox{0.43}{\includegraphics{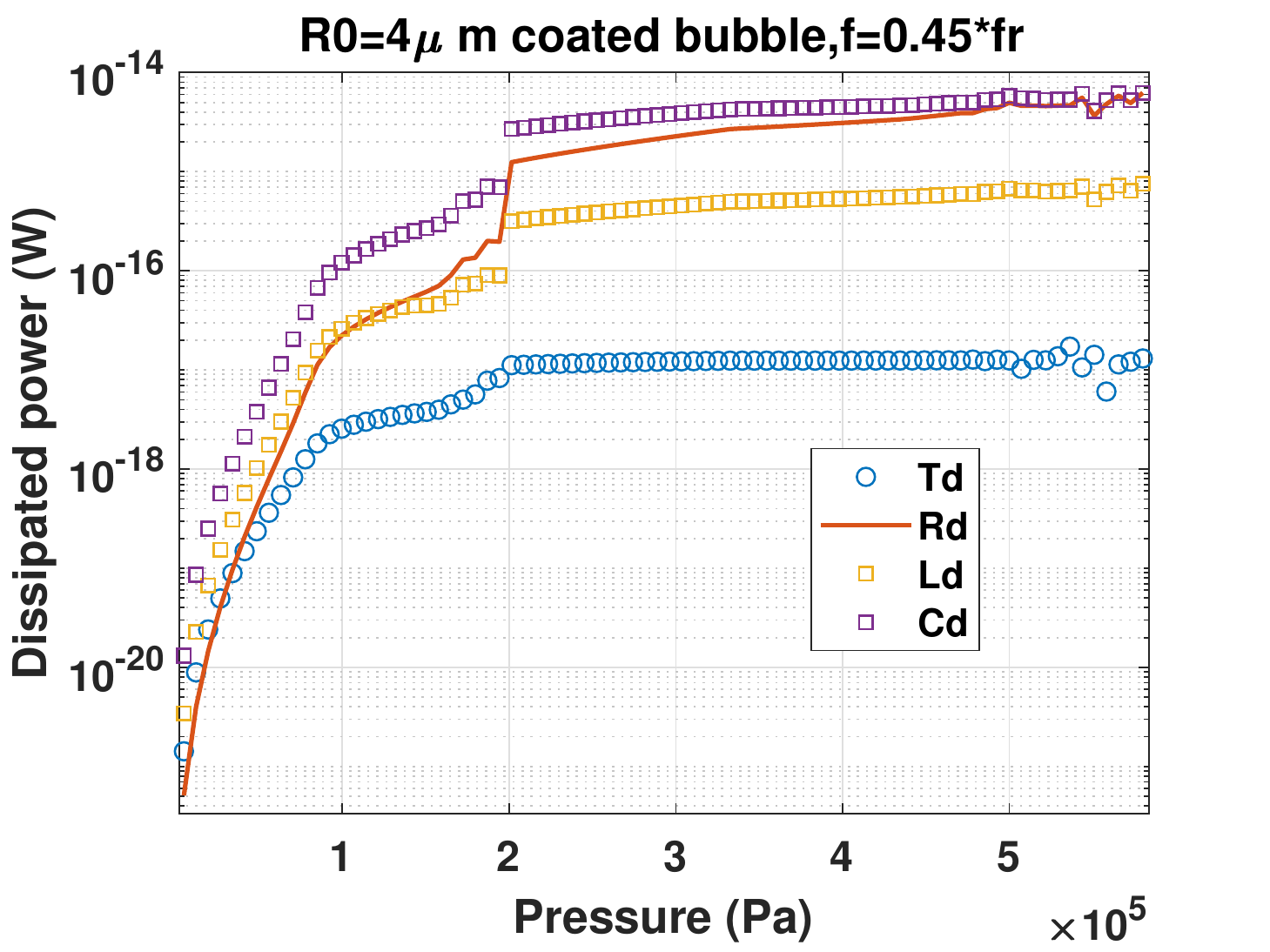}}\\
		\hspace{0.5cm} (c) \hspace{6cm} (d)\\
		\scalebox{0.43}{\includegraphics{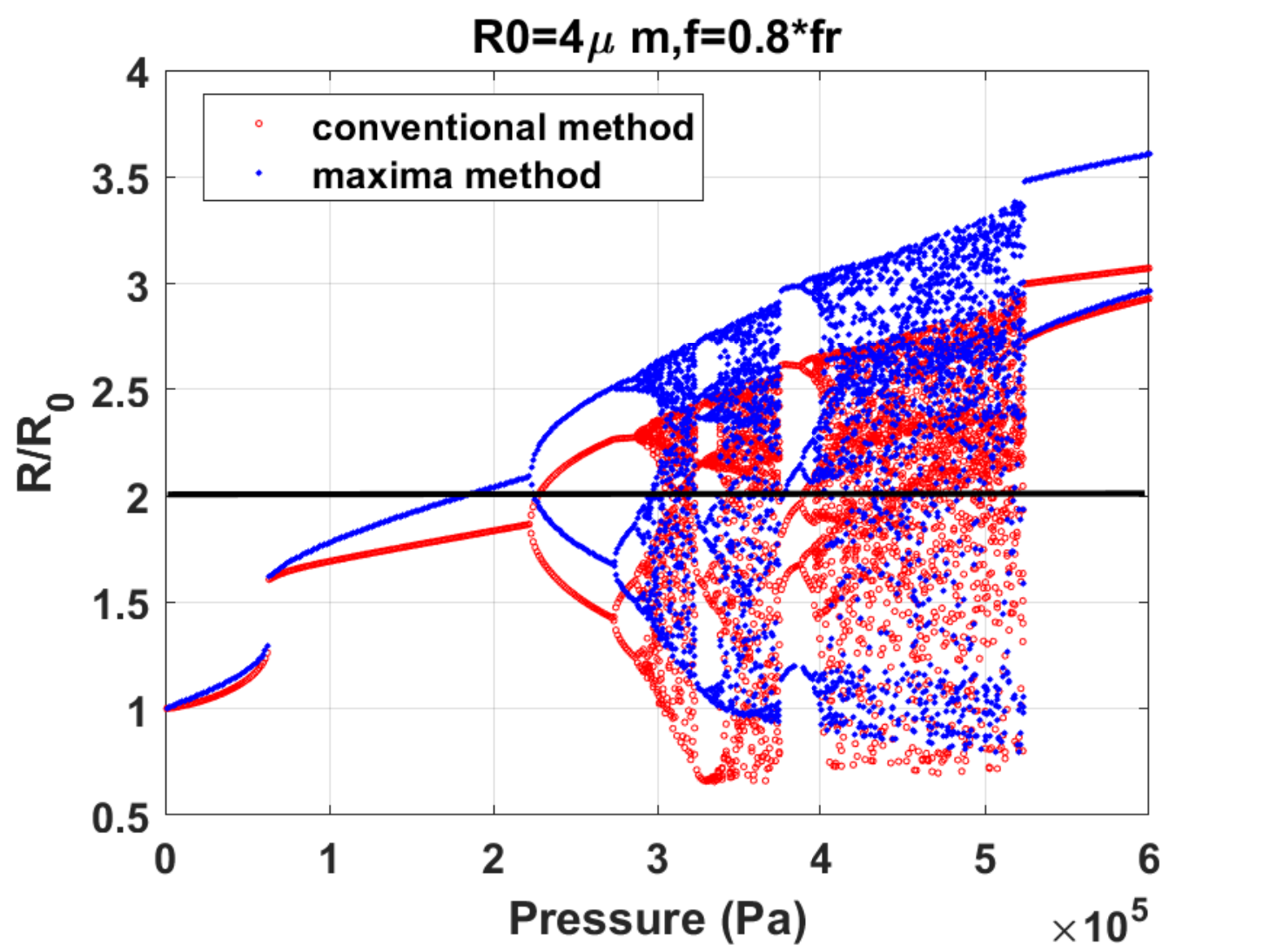}} \scalebox{0.43}{\includegraphics{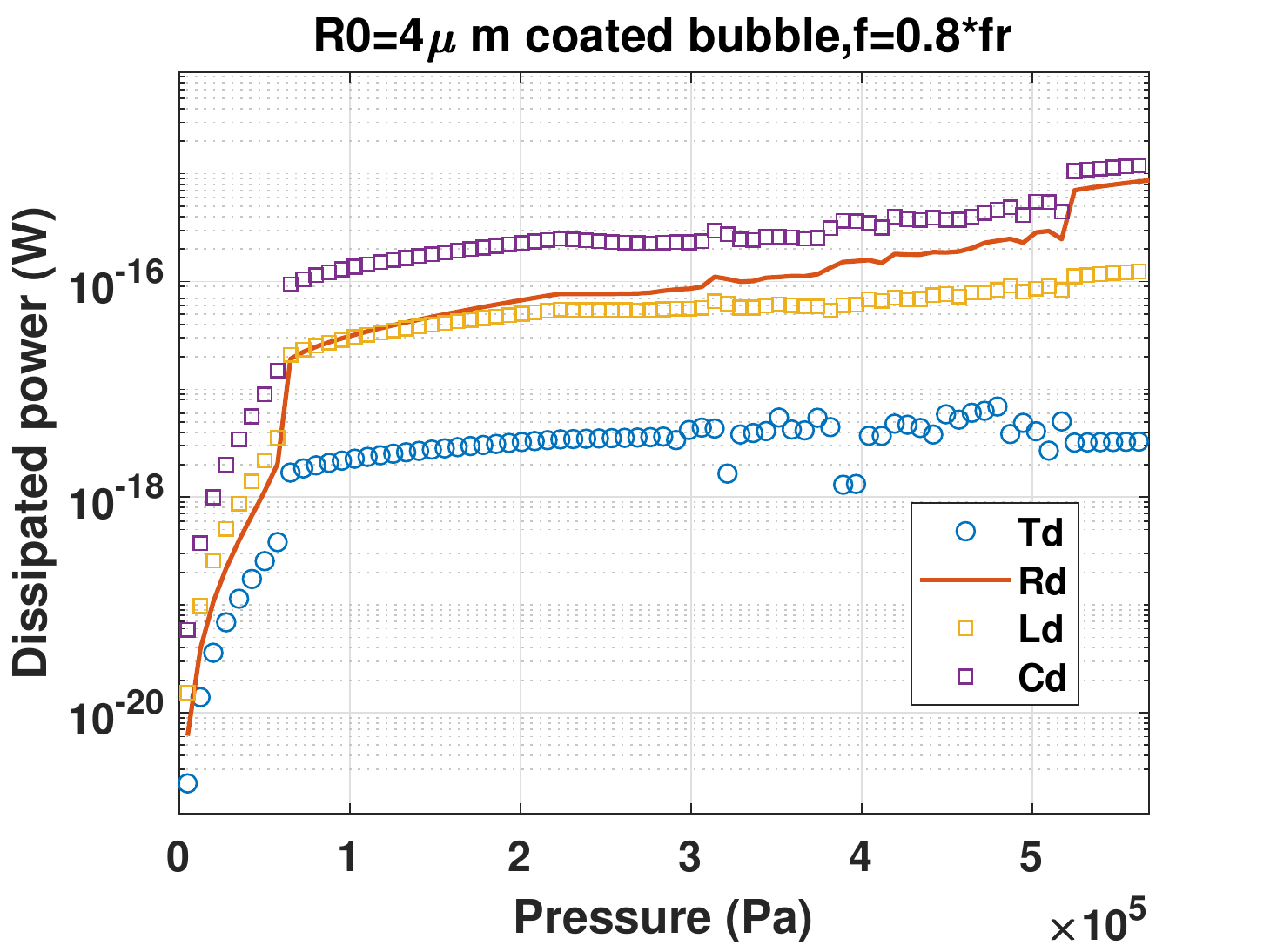}}\\
		\hspace{0.5cm} (e) \hspace{6cm} (f)\\
		\scalebox{0.43}{\includegraphics{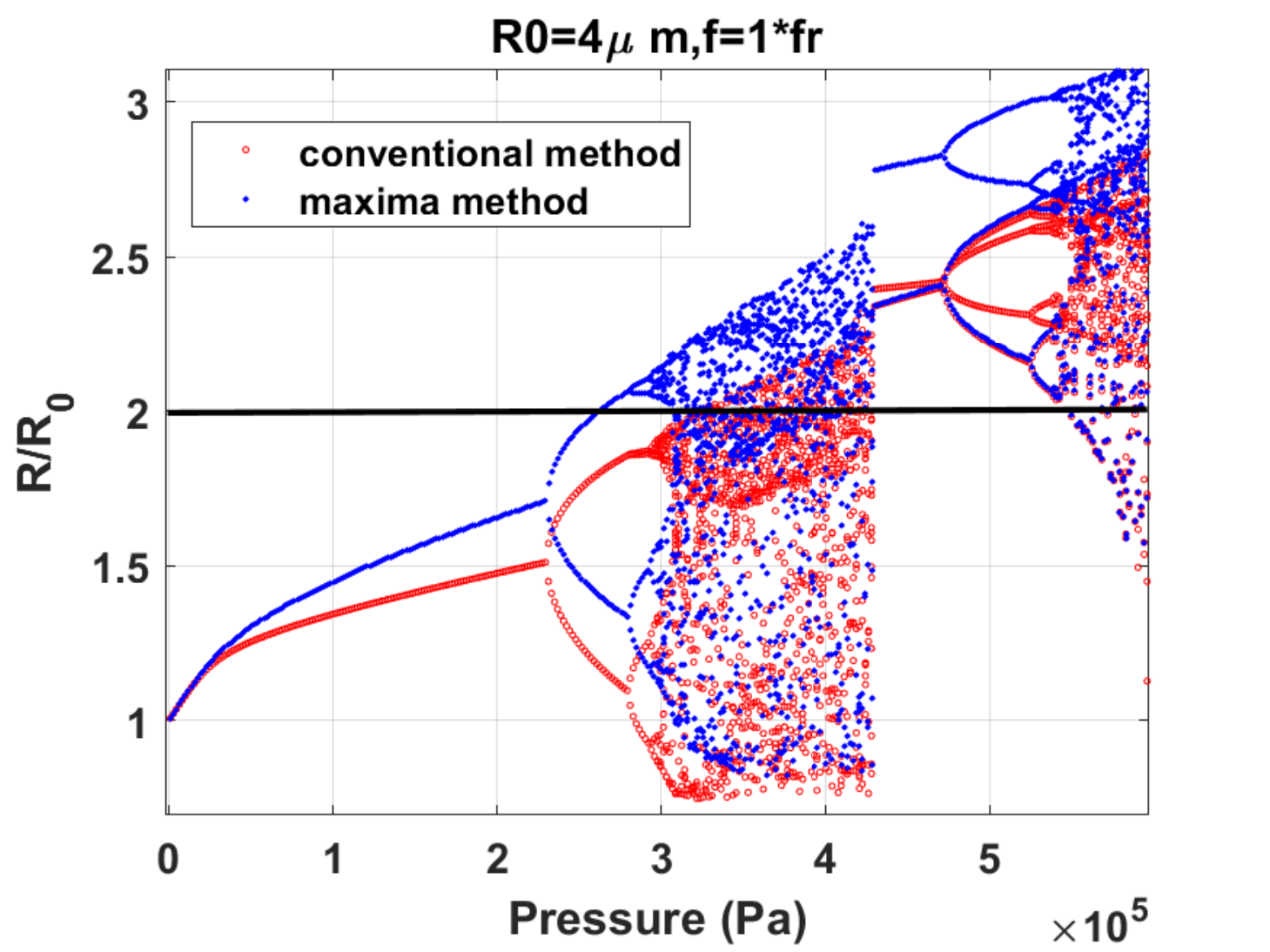}} \scalebox{0.43}{\includegraphics{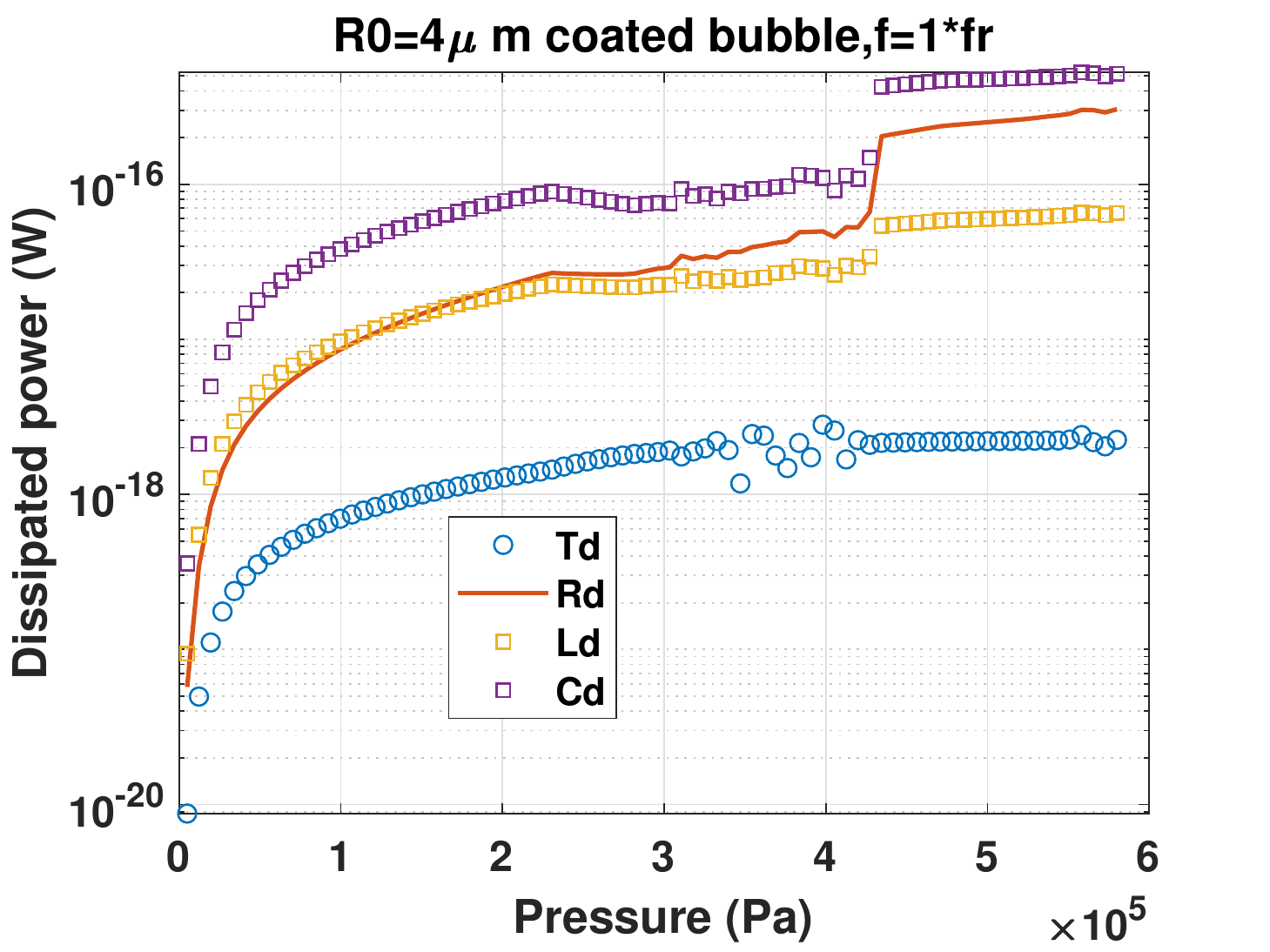}}\\
		\hspace{0.5cm} (g) \hspace{6cm} (h)\\
		\caption{Bifurcation structure (left column) and the dissipated power as a function of pressure (right column) of the oscillations of a coated C3F8 bubble with $R_0=4\mu m$ for $f= 0.25f_r$ (a-b)-$f=0.45f_r$ (c-d)- $f=0.8f_r$ (e-f) \& $f=f_r$ (g-h).}
	\end{center}
\end{figure*}
In this subsection we analyze the dynamics of the coated bubble with $R_0=4 \mu m$ with a C3F8 gas core.
Fig. A3a displays the bifurcation structure of a coated bubble with $R_0= 4 \mu m$ when $f=0.25f_r$. The oscillation amplitude grows with pressure and at $P_a= 80 kPa$, 3 maxima appear in the P1 oscillations (3rd order SuH \cite{50}). The 3rd order SuH undergoes a SN bifurcation to another 3rd order SuH oscillations at $P_a\approxeq 140 kPa$. The signal is still P1 with 3 maxima and at $P_a\approxeq 160 kPa$ Pd takes place and $7/2$ order UH oscillations are generated (P2 oscillations with 6 maxima). A small chaotic window appears before $P_a=200 kPa$ followed by a P1 signal of 2nd order SuH, then $5/2$ UH resonance and later again chaos. At approximately 295 kPa a giant P1 resonance with one maximum emerges out of the chaotic window. The corresponding power dissipation curves are shown in Fig. A3b. Cd is the strongest dissipation mechanism with $Cd>Ld>Td$ for $P_a<80 kPa$. When SuH oscillations occur, Rd becomes equal to Td and keeps growing faster than other dissipation mechanisms until Pd takes place where Rd undergoes the sharpest increase and becomes larger than Ld. Rd keeps growing faster with the SN bifurcation to 2nd order SuH oscillations, Rd undergoes another sharp increase and becomes equal to Cd during chaotic oscillations. Emergence of the giant resonance out of chaos is concomitant with another sharp increase in Rd, making Rd the strongest dissipation mechanism afterwards.\\
Fig. A3c shows the case of sonication with $f=0.45f_r$. As pressure increases, the P1 oscillation amplitude increases and at $\approxeq 25 kPa$ two maxima appear in the P1 oscillations. At approximately 105 kPa; bubble collapses two times in each acoustic cycles with the wall velocity of one of the collapses in phase with the driving acoustic pressure (red and blue line have the same value). Pd takes place at $\approxeq 140 kPa$ ($5/2$ UH oscillations are generated (signal is P2 with 4 maxima)) and $R/R_0$ becomes equal to 2 at the pressure at which chaos is generated at $P_a\approxeq180 kPa$. Slightly below 200 kPa a giant P1 resonance emerges out of the chaotic window which later undergoes successive Pds to chaos at $P_a\approxeq470 kPa$. The corresponding power dissipation curves in Fig. A3d shows that for pressures below the SuH oscillations, $Cd>Ld>Td\approxeq Rd$. After the generation of SuH oscillations Rd supersedes Td and becomes equal to Ld at about 105 kPa (when red curve meets the blue curve in Fig. A3c). Power dissipation curves plateau and when Pd occurs. Rd, Cd, Ld and Td increase. During $5/2$ UH oscillations $Cd>Rd>Ld>Td$. Emergence of giant resonance is concomitant with an increase in all the dissipation mechanisms with Rd exhibiting the sharpest growth. Afterwards Rd grows faster and becomes approximately equal to Cd during chaotic oscillations. Similar to Fig. A3b, Rd and Cd are 3 orders of magnitude larger than Td and about an order of magnitude stronger than Ld.\\
When $f=0.8 f_r$ ($PDf_r$ \cite{32}) a SN bifurcation takes place at $P_a=80 kPa$ and the oscillation amplitude $R/R_0$ exceeds at $P_a \approxeq 190 kPa$ (black line meets the blue curve). At $\approxeq 220 kPa$, Pd takes place and afterwards P2 oscillations undergo successive period doubling to chaos at $\approxeq 280 kPa$. A P2 giant resonance with 2 maxima emerges out of the chaotic window at 520 kPa. The corresponding dissipation curves in Fig. A.3f reveal that when the SN occurs, Cd, Ld, Rd and Td undergo a sharp increase with Td exhibiting the smallest increase. Before the SN, $Cd>Ld>Rd>Td$ and after the SN, $Cd>Ld=Rd>Td$. When $R/R_0$ exceeds 2, Rd becomes stronger than Ld  and power dissipation contribution is in the following order for the rest of the pressure range that studied: $Cd>Rd>Ld>Td$. Emergence of the giant resonance is simultaneous with an increase in Rd, Cd and Ld (with Rd demonstrating the largest increase) and Td decreases.\\
When $f=f_r$ (Fig. A.3g) the oscillation amplitude grows monotonically with pressure. For excitation pressures below 50 kPa, the red curve and blue curves have the same value (wall velocity is in phase with the driving acoustic pressure). The two curves diverge as pressure increases and at 230 kPa Pd takes place. P2 oscillations amplitudes exceed $R/R_0 =2$ at 260 kPa; afterwards successive Pds take place in the bifurcation structure resulting in chaotic oscillations at $\approx 300 kPa$. The chaotic window continues up to 420 kPa; whereby, large amplitude P2 oscillations emerge out of the chaotic window (one of the red solutions is equal to the smallest maxima in blue curve).  Chaos is then generated through successive Pds at 520 kPa. Power dissipation curves in Fig. A.3h show that $Cd>Rd\approxeq Ld>Td$; however, when Pd occurs, Rd slightly exceeds Ld due to the fact that both Cd and Ld undergo a slightly higher decrease compared to Rd. This is possibly due to decrease in the wall velocity concomitant with Pd  when bubble is sonicated with a frequency close to its resonance frequency \cite{32}. Emergence of the giant P2 oscillations is concomitant with a very sharp increase in Rd and Cd, a slight increase in Ld and minimal changes in Td.\\
\begin{figure*}
	\begin{center}
		\scalebox{0.4}{\includegraphics{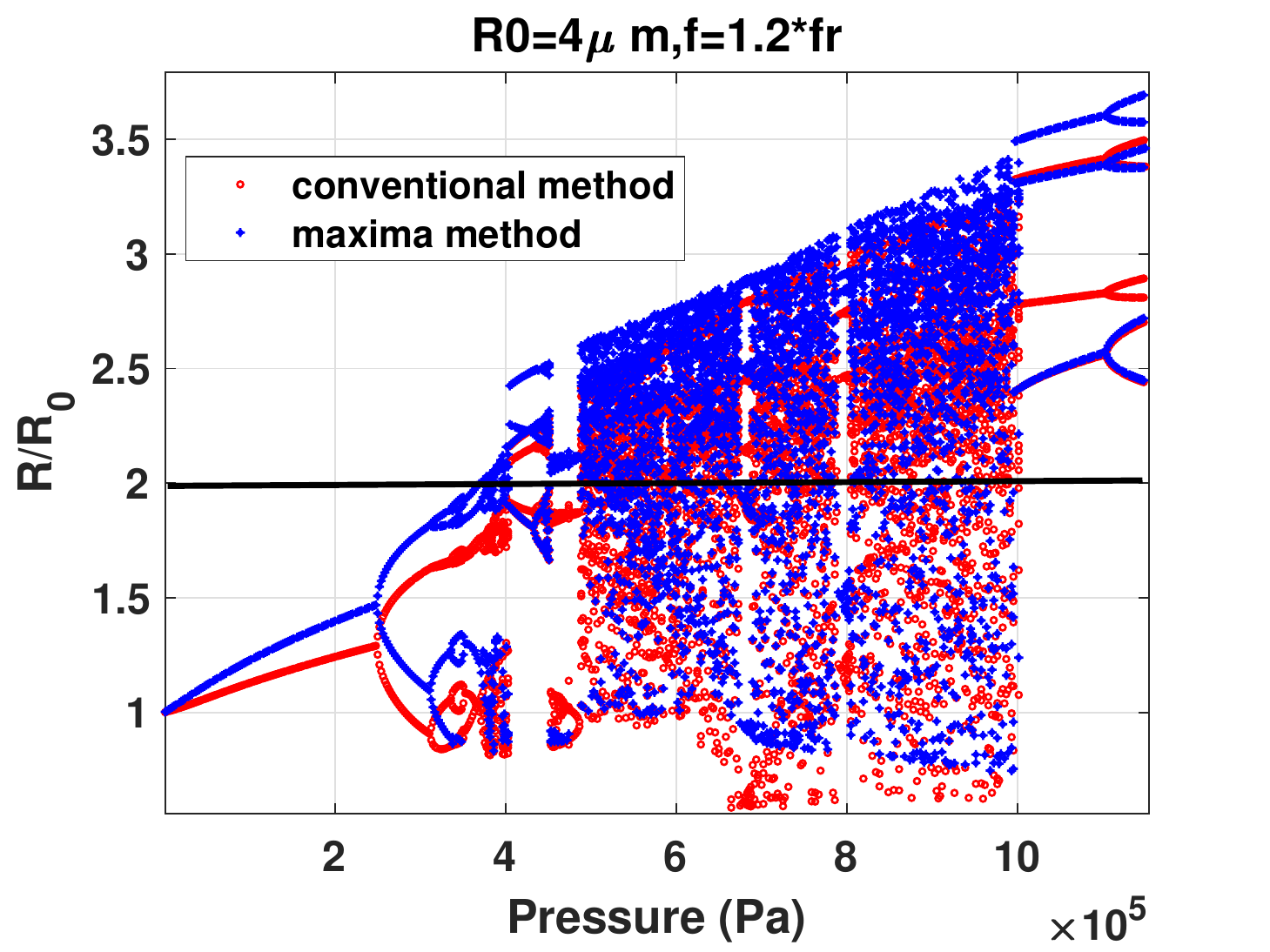}}  \scalebox{0.4}{\includegraphics{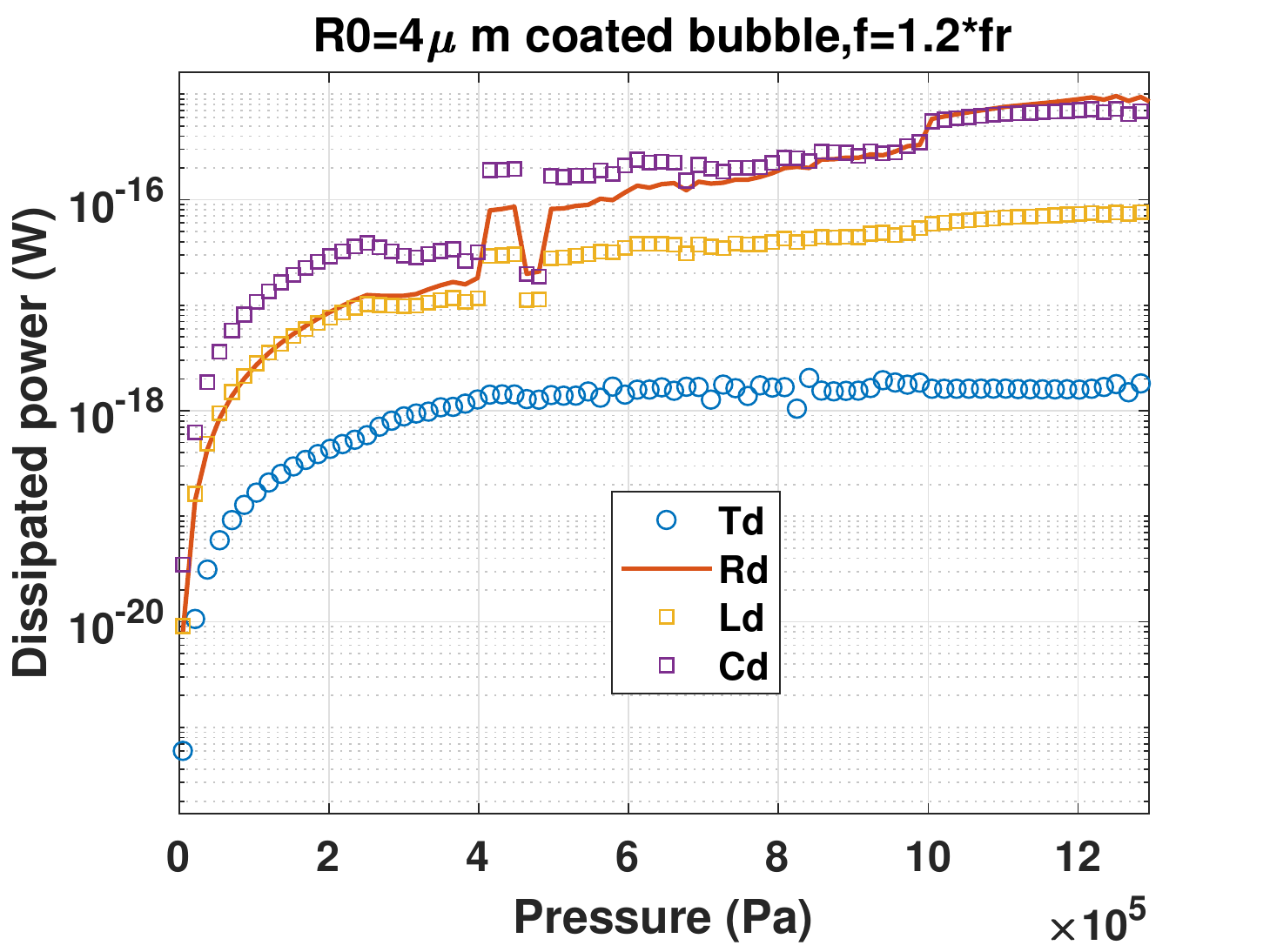}}\\
		\hspace{0.5cm} (a) \hspace{6cm} (b)\\
		\scalebox{0.43}{\includegraphics{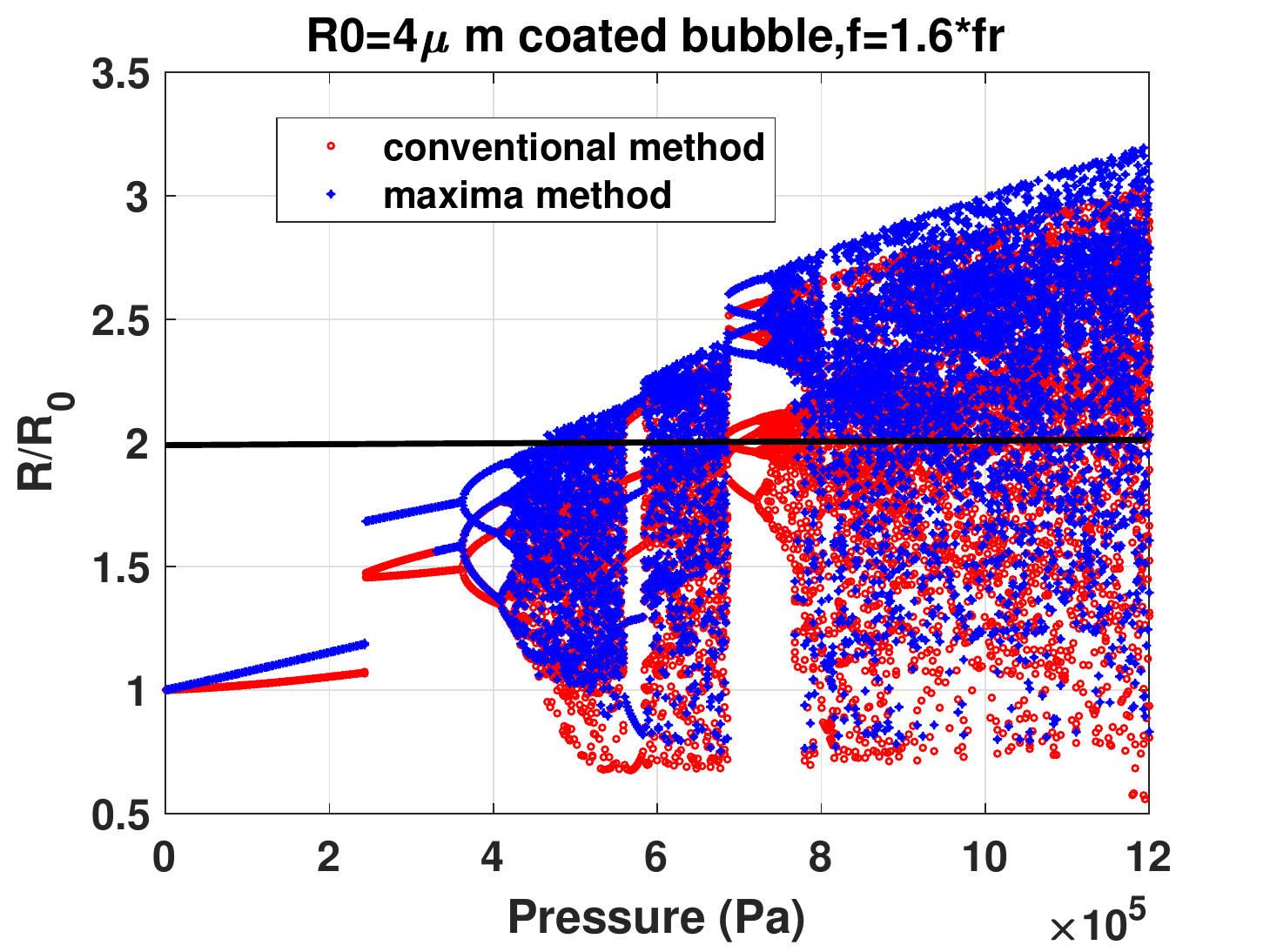}} \scalebox{0.43}{\includegraphics{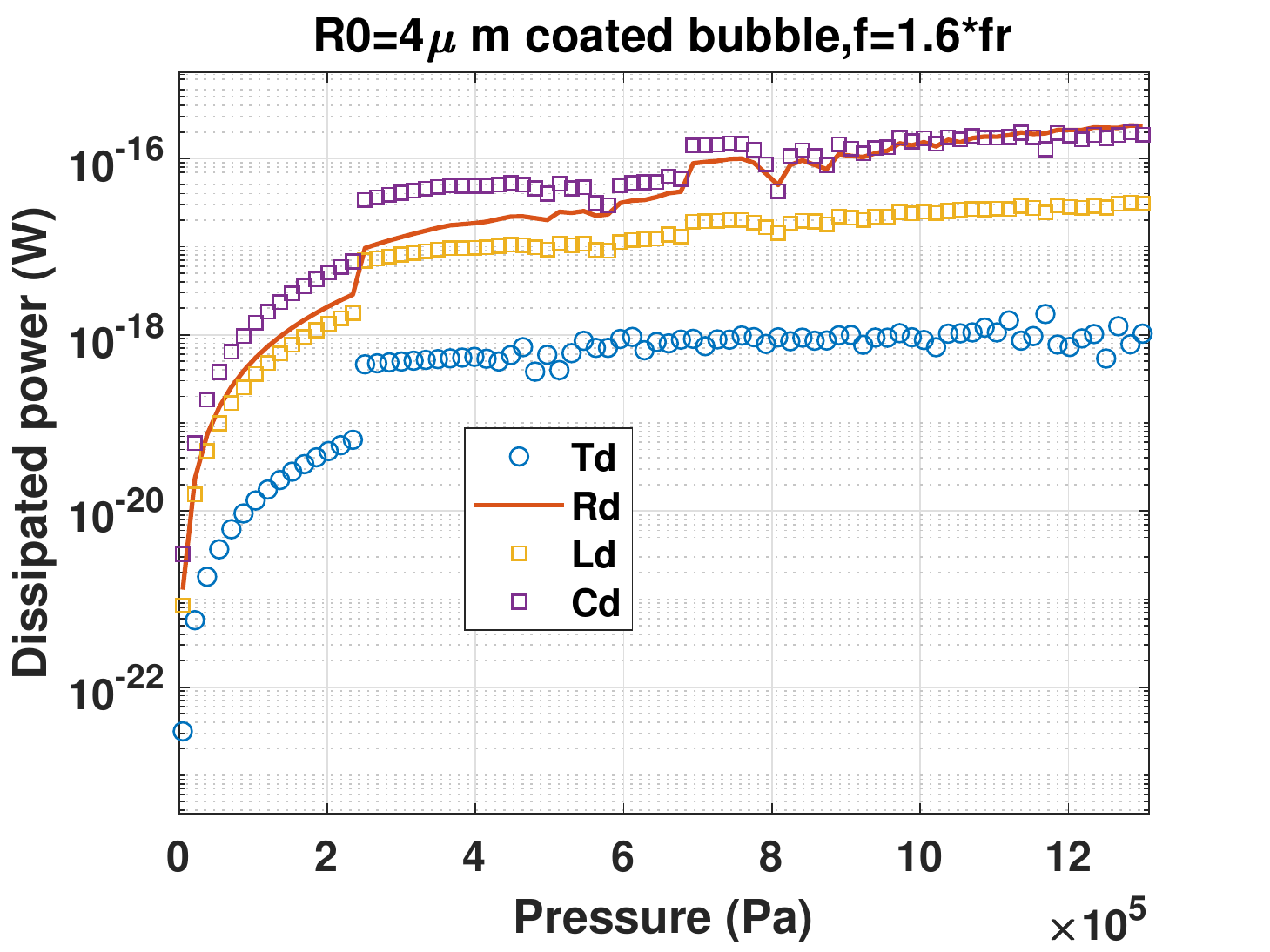}}\\
		\hspace{0.5cm} (c) \hspace{6cm} (d)\\
		\scalebox{0.43}{\includegraphics{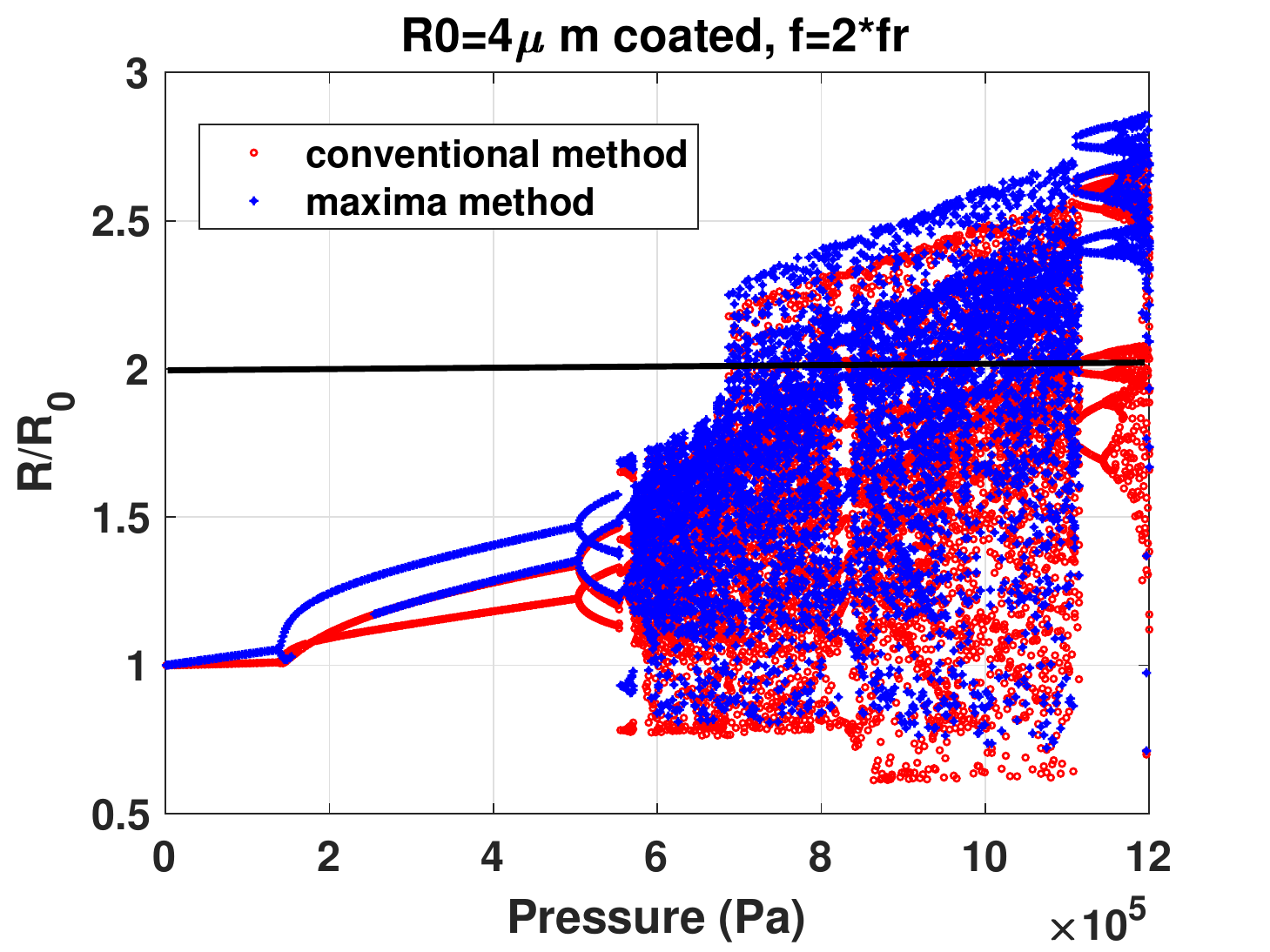}} \scalebox{0.43}{\includegraphics{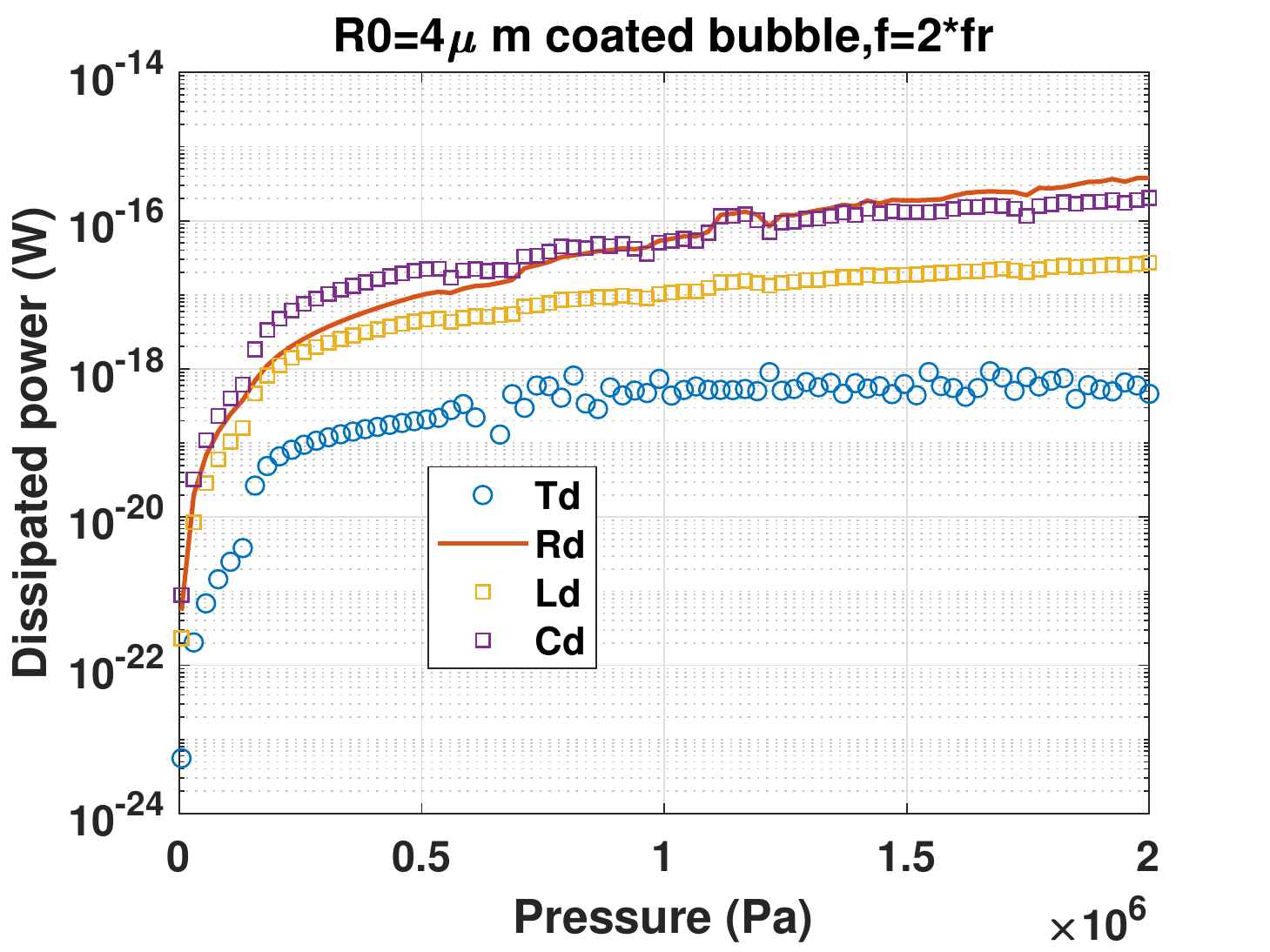}}\\
		\hspace{0.5cm} (e) \hspace{6cm} (f)\\
		\scalebox{0.43}{\includegraphics{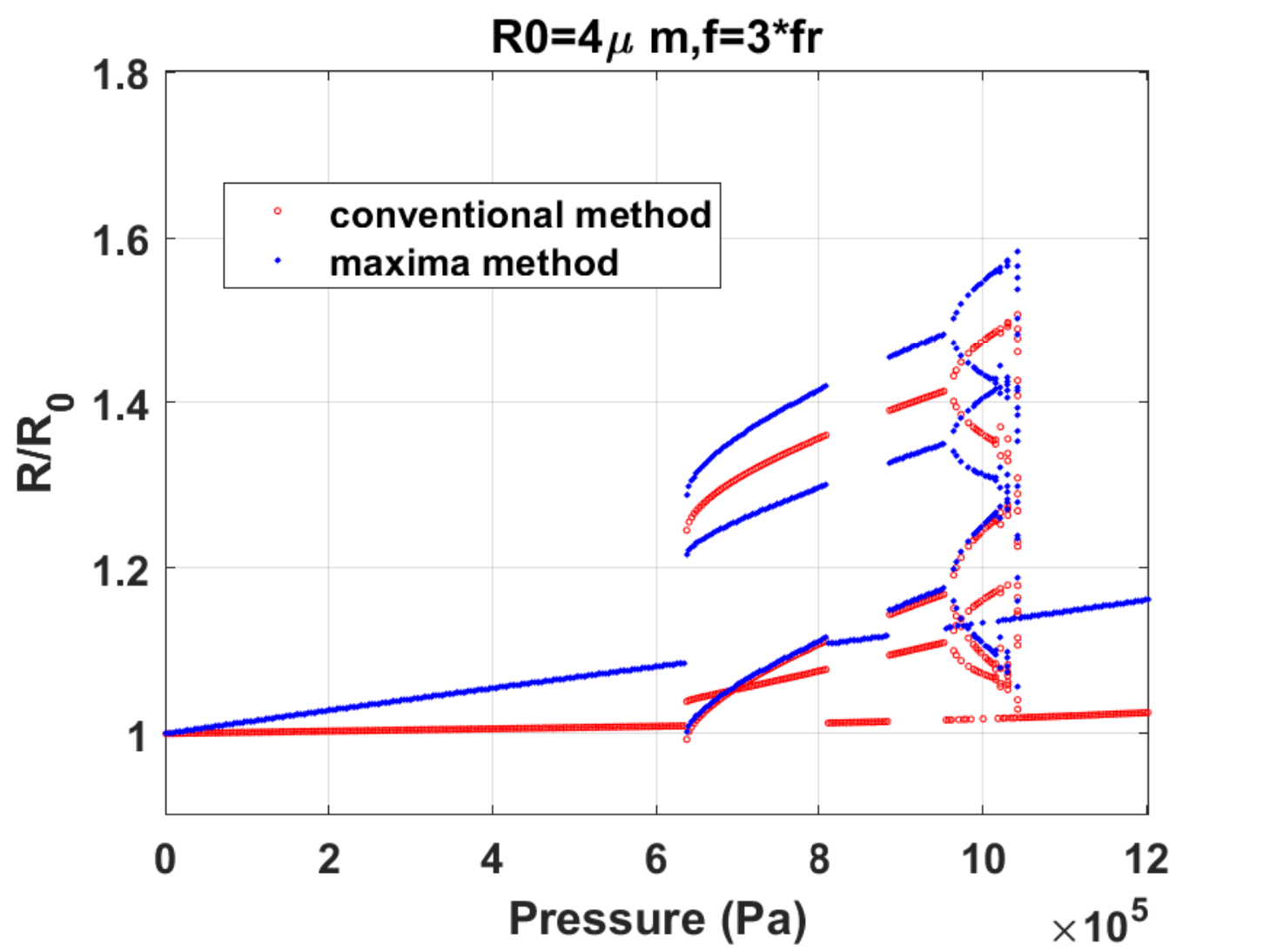}} \scalebox{0.43}{\includegraphics{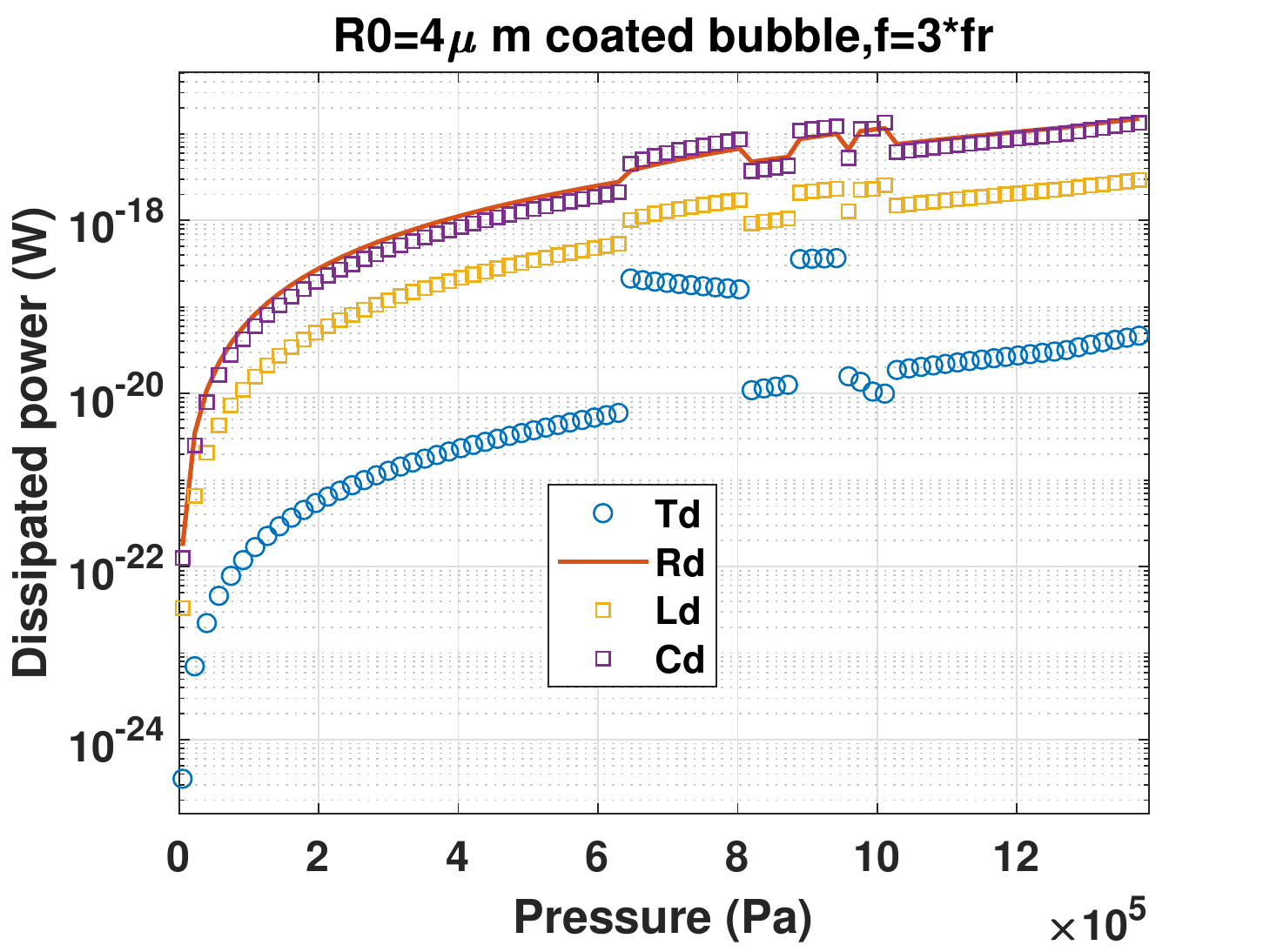}}\\
		\hspace{0.5cm} (g) \hspace{6cm} (h)\\
		\caption{Bifurcation structure (left column) and the dissipated powers (right column) of the oscillations of a coated C3F8 bubble with $R_0=4\mu m$ for $f= 1.2f_r$ (a-b)-$f=1.6f_r$ (c-d)- $f=2f_r$ (e-f) \& $f=3f_r$ (g-h).}
	\end{center}
\end{figure*}
When $f=1.2 f_r$ (Fig. A.4a), the oscillation amplitude grows monotonically with pressure. A Pd takes place at $\approxeq 230 kPa$. Bubbling bifurcation takes place in each of the branches of the P2 regime and a small window of chaos appears followed by a small P3 window which undergoes Pd to P12 after which there is an interesting reverse Pd leading to a sudden onset of chaos. The oscillation amplitude exceeds $R/R_0=2$ at $\approxeq 390 kPa$. The chaotic window extends until $P_a=1 MPa$ where a giant P3 emerges out of the chaotic window which then undergoes Pd to P12 oscillations. The corresponding dissipation curves in Fig. A4b show that $Cd>Ld=Rd>Td$ until Pd after which Rd becomes slightly larger than Ld. Pd results in a notable decrease in Cd and Ld. Td remains two orders of magnitude less than Cd (Fig. A.3b). When the first P3 oscillation occurs, Cd, Rd and Ld undergo a sharp increase with the most notable increase in Rd. Reverse Pd bifurcation results in a decrease in dissipation due to lower oscillation amplitudes. Initiation of chaos leads to a sharp increase in dissipation with $Cd>Rd>Ld>Td$.  Rd grows faster than other mechanisms as pressure increases and becomes equal to Cd at $\approx 820 kPa$. Finally when the P3 giant resonance occurs, Cd, Rd and Ld undergo an increase with Rd experiencing the largest growth. Td undergoes a small decrease during P3 giant resonance oscillations.\\ Fig. A.4c represents the case of sonication with $f=1.6 f_r$ ($PDf_{sh}$ \cite{59}). P1 oscillations grow slowly with pressure and at $\approxeq 230 kPa$ a SN bifurcation from P1 oscillations of lower amplitude to P2 oscillations (with one maximum) of higher amplitude takes place. Second maxima emerges at $\approxeq 320 kPa$; afterwards oscillations undergo Pd at $\approxeq 380 kPa$ which are then followed by successive Pds to chaos at $\approxeq 400 kPa$.  Chaotic window stretches until $\approxeq 700 kPa$ with a small window of P6 oscillations. A P6 oscillation regime with high amplitude emerges out of the chaotic window; later undergoing Pds to P12 and then chaos. The corresponding dissipation curves are shown in Fig. A.4d. For pressures below the SN bifurcation $Cd>Rd>Ld>Td$. Occurrence of SN bifurcation is concomitant with a sharp increase in the dissipated powers. $Cd>Rd>Ld>Td$ until at higher pressures ($>1$ MPa) Rd surpasses Cd.\\ $f=2f_r$ (Fig. A.4e) is the $f_{sh}$ of the bubble \cite{58}. P1 oscillations undergo Pd at $\approxeq 180 kPa$. The P2 oscillations loose one maxima right after the generation of Pd and then evolve in a form of a bow-tie with the second maxima re-emerging with an amplitude equal to the larger branch of the red curve  $\approxeq 280 kPa$. Consistent with previous observations in \cite{58}, sonication with $f=2f_r$ results in the largest pressure range of stable P2 oscillations. Oscillations undergo Pd to P4 oscillations followed by a SN bifurcation to P4 oscillations of higher amplitude at $\approxeq 570 kPa $; before successive Pds to chaos. Amplitude of the chaotic oscillations increases at $\approxeq 710 kPa$ which can possibly lead to bubble destruction as $R/R_0>2$. Chaos continues until 1.1 MPa where a P6 oscillation of large amplitude emerges out of chaos which later undergoes successive Pds to P12 and chaos. The corresponding dissipation curves are depicted in Fig. A.4f. $Cd>Rd>Ld>Td$ with dissipated powers undergoing a fast growth concomitant with Pd. When chaos appears Rd becomes equal to Cd and later at 1.5 MPa onward  $Rd>Cd>Ld>Td$. When $f=2f_r$, Td is in average two orders of magnitude smaller than Cd and Rd.\\ 
The case of sonication with $f=3f_r$ is shown if Fig. A.4g. Oscillations grow slightly with pressure and at $\approxeq 620 kPa$ a SN bifurcation takes place and P1 oscillations turn into P3 oscillations of higher amplitude. P3 then grows in amplitude until it turns to P1 oscillations for a small pressure window and then again re-emerges through a SN bifurcation. P3 oscillations undergo Pds to P6 and then return to P1 oscillations for $P_a\approxeq 1.04 MPa$. Corresponding dissipated power curves are shown in Fig. A.4h. Unlike previous cases here $Rd>Cd>Ld>Td$.  The SN bifurcation results in a large increase in dissipated powers specially in case of Td. This is because during these P3 oscillations the bubble collapses 3 times out of which two are very gentle and thus a large average bubble radius is maintained during oscillations. This increases the surface area for the heat exchange and Td increases. Moreover, bubble collapses strongly only once in every three cycles; thus high velocity and re-radiated pressure are achieved only once in every three acoustic cycles. This is why the average for Ld, Cd and Rd are small. During P3 and P6 oscillations $Cd>Rd>Ld>Td$ and elsewhere $Rd>Cd>Ld>Td$.\\ 
\subsection{Concluding graphs of $|\dot{R(t)}|_{max}$, $|P_{sc}|_{max}$, total dissipated power and STDR} 
\begin{figure*}
	\begin{center}
		\scalebox{0.43}{\includegraphics{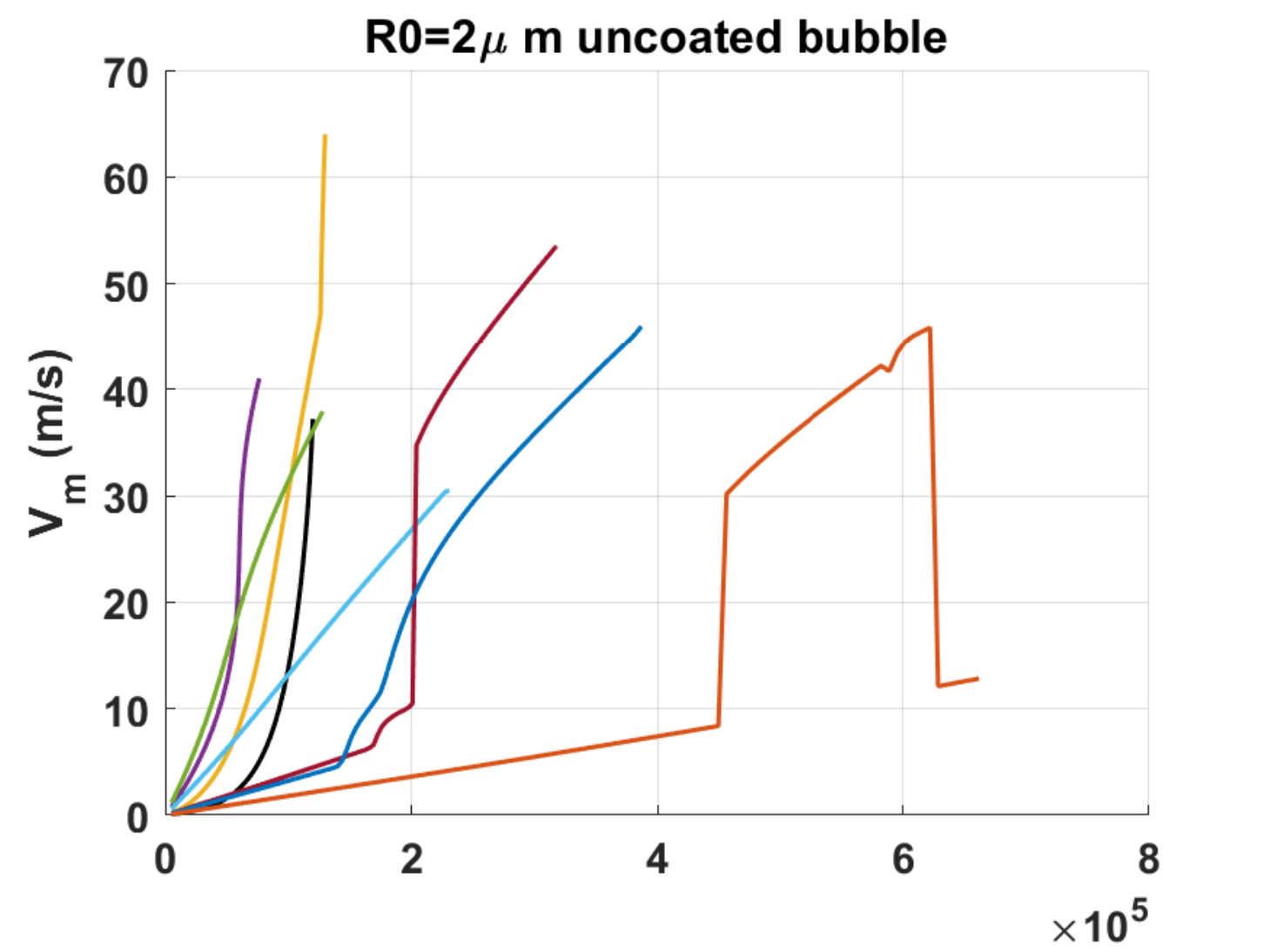}} \scalebox{0.43}{\includegraphics{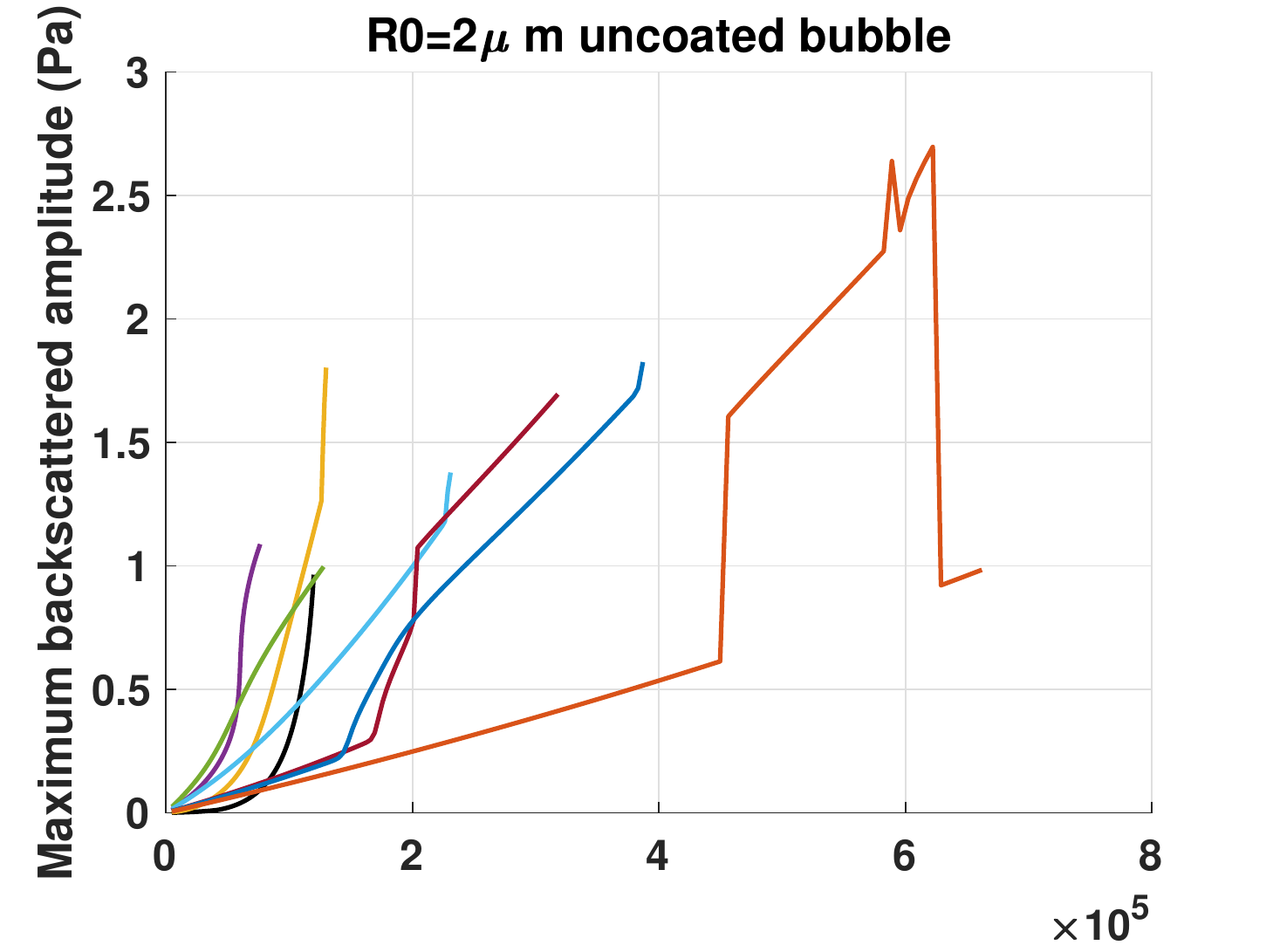}}\\
		\hspace{0.5cm} (a) \hspace{6cm} (b)\\
		\scalebox{0.43}{\includegraphics{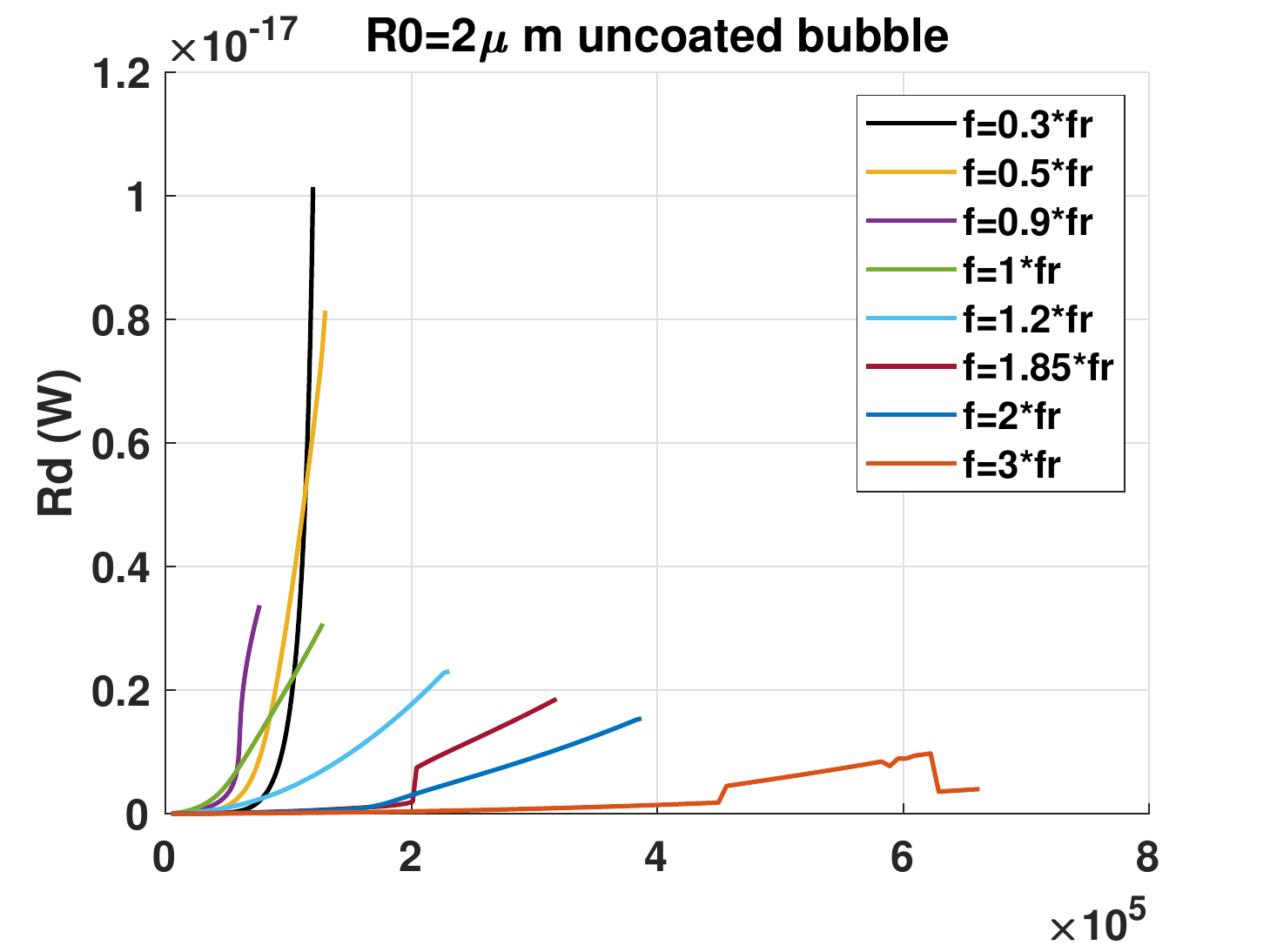}} \scalebox{0.43}{\includegraphics{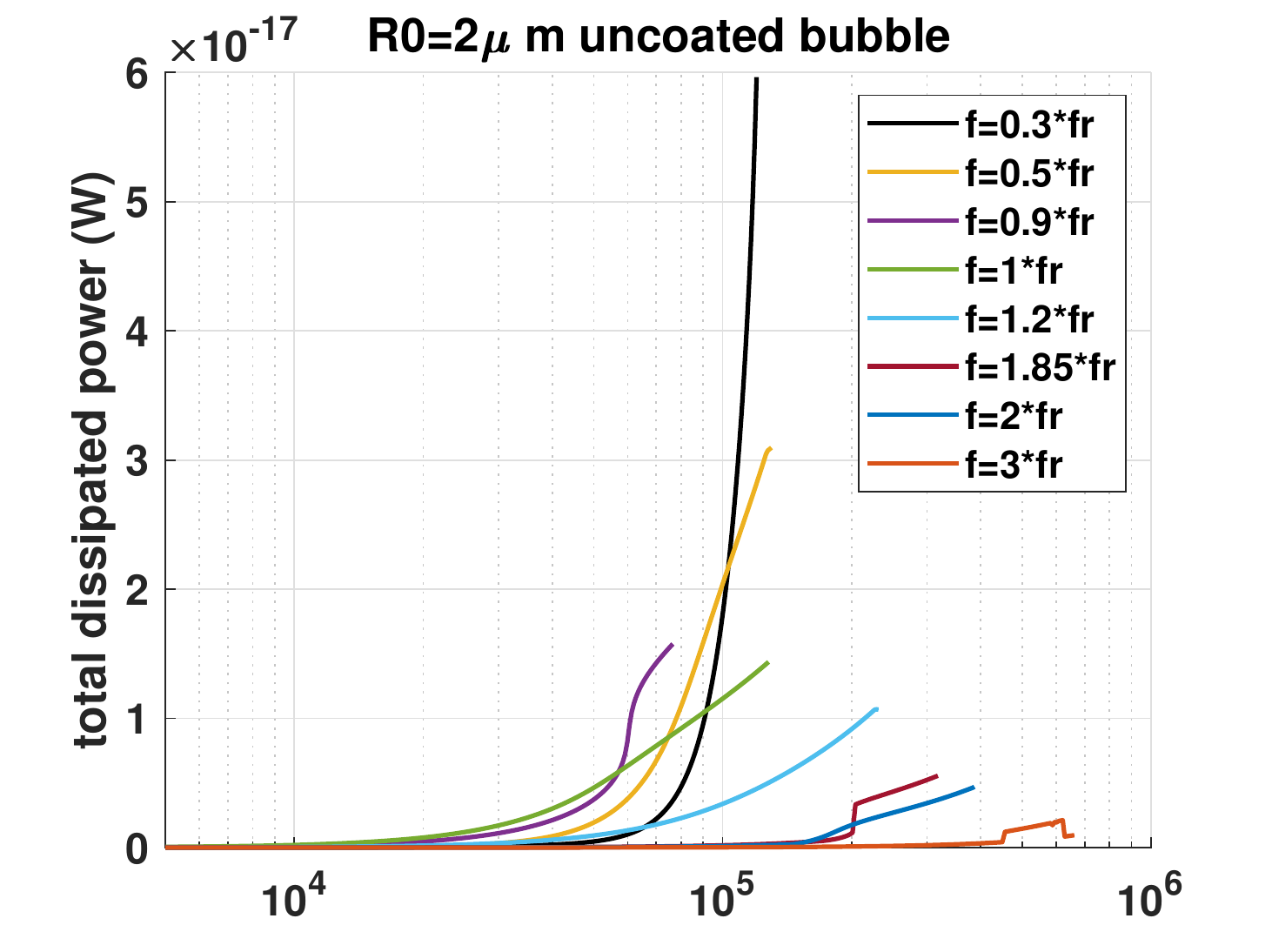}}\\
		\hspace{0.5cm} (c) \hspace{6cm} (d)\\
		\scalebox{0.43}{\includegraphics{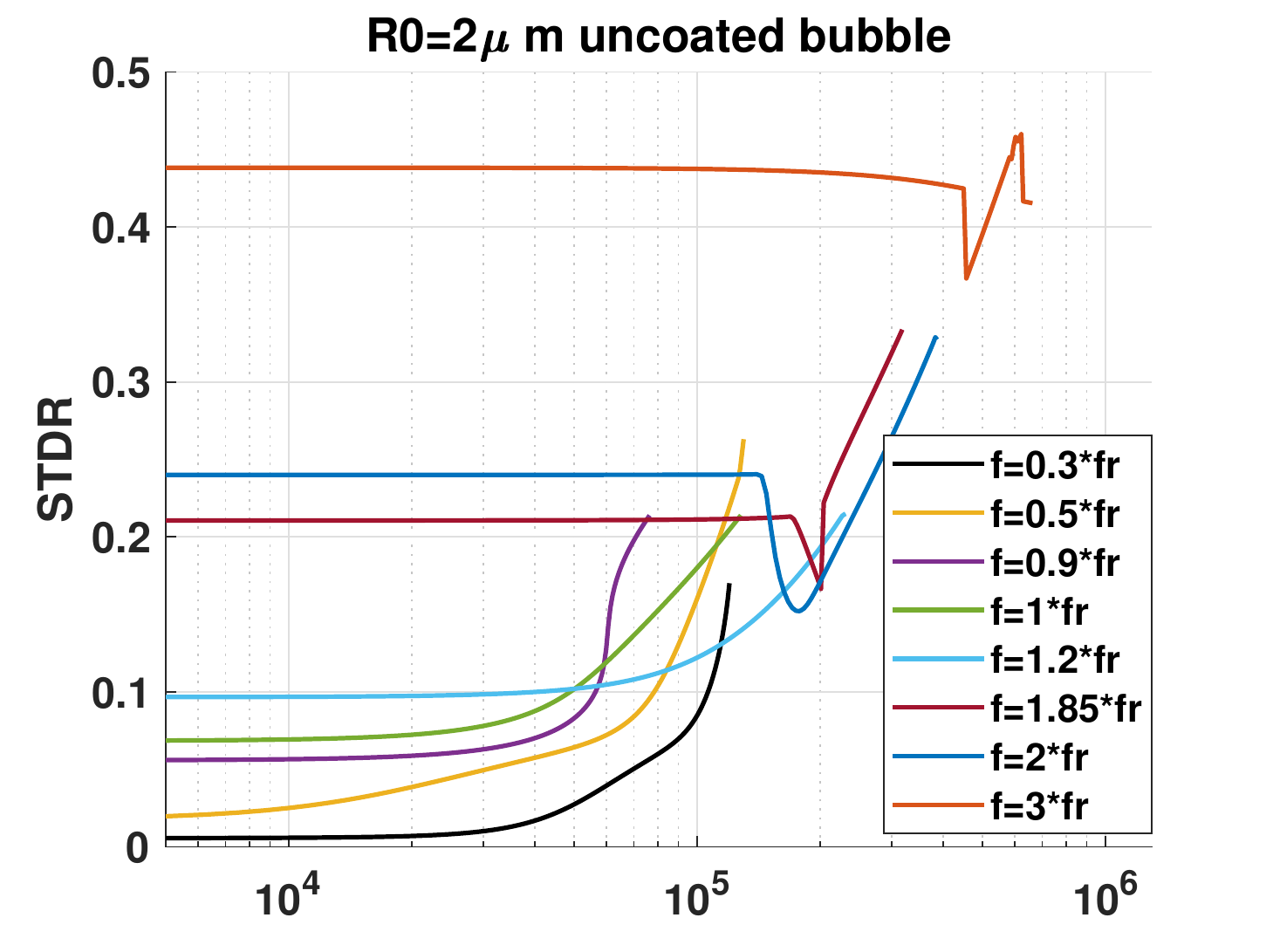}}\\
		\hspace{0.5cm} (e)\\
		\caption{Nondestructive ($R/R_0\leq 2$) values of: a) $|\dot{R(t)}|_{max}$ ($V_m$), b) Maximum backscattered pressure ($|P_{sc}|_{max}$ ($P_m$), c) Rd, d) $W_{total}$ and e) STDR as a function of pressure in the oscillations of an uncoated air bubble with $R_0=2 \mu m$.}
	\end{center}
\end{figure*}
Figure A.5 represents the uncoated air bubble with $R_0=2 \mu m$. The exact same behavior of the case of an uncoated bubble with $R_0=10 \mu m$ (Fig. 5) is observed here. Maximum non-destructive $V_{m}$ and $P_{m}$ occurs for $f=0.5f_r$ (2nd SuH) and $f=3f_r$ respectively. Maximum Rd and $W_{total}$ are achieved when $f=0.25f_r$ (3rd SuH). STDR is higher for higher frequencies with the maximum at $f=3f_r$. Similar to Fig. 5e and Fig. 6e, the onset of nonlinear oscillations results in a decrease and then increase in STDR if the bubble is sonicated above resonance.\\
\begin{figure*}
	\begin{center}
		\scalebox{0.43}{\includegraphics{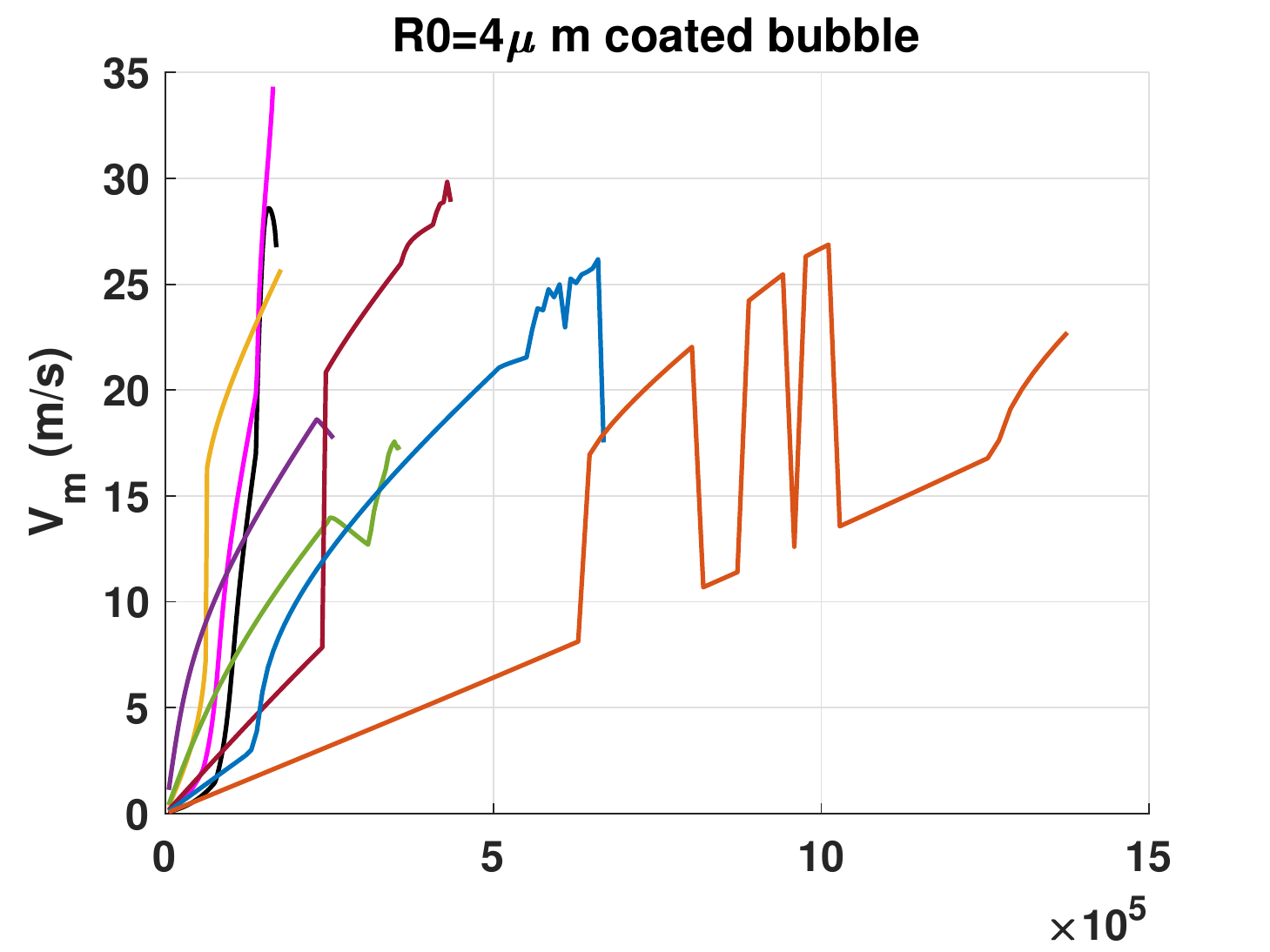}} \scalebox{0.43}{\includegraphics{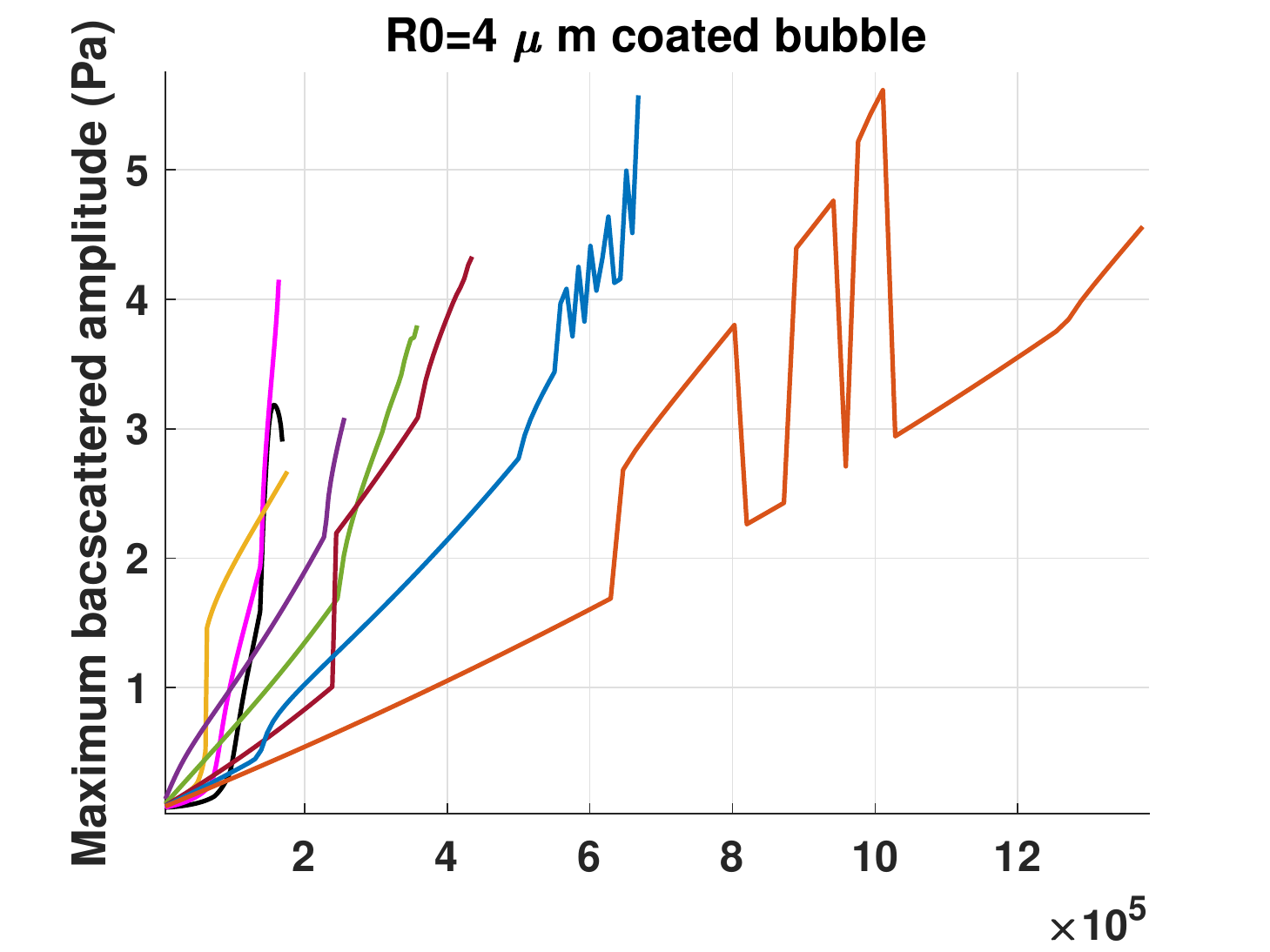}}\\
		\hspace{0.5cm} (a) \hspace{6cm} (b)\\
		\scalebox{0.43}{\includegraphics{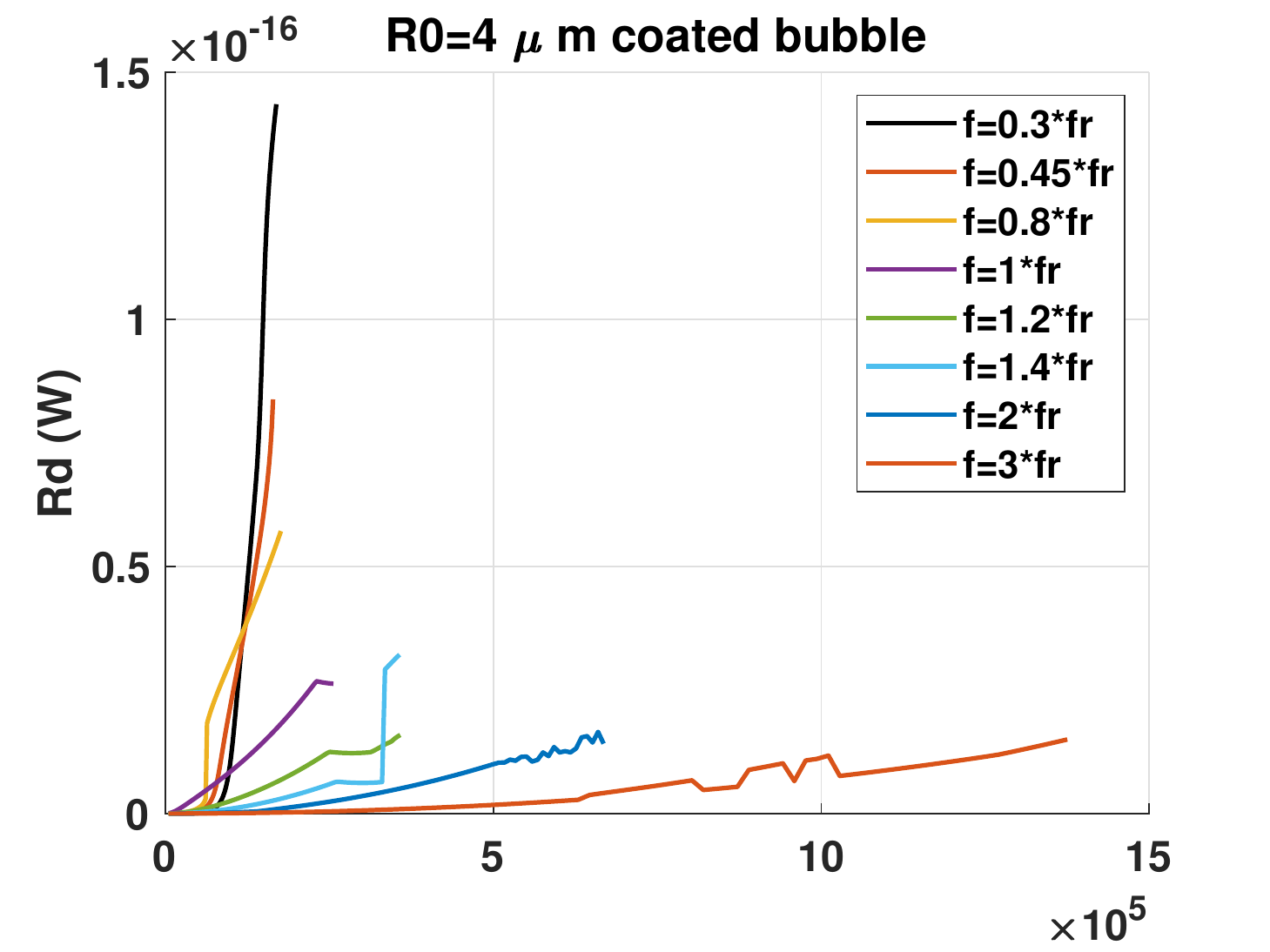}} \scalebox{0.43}{\includegraphics{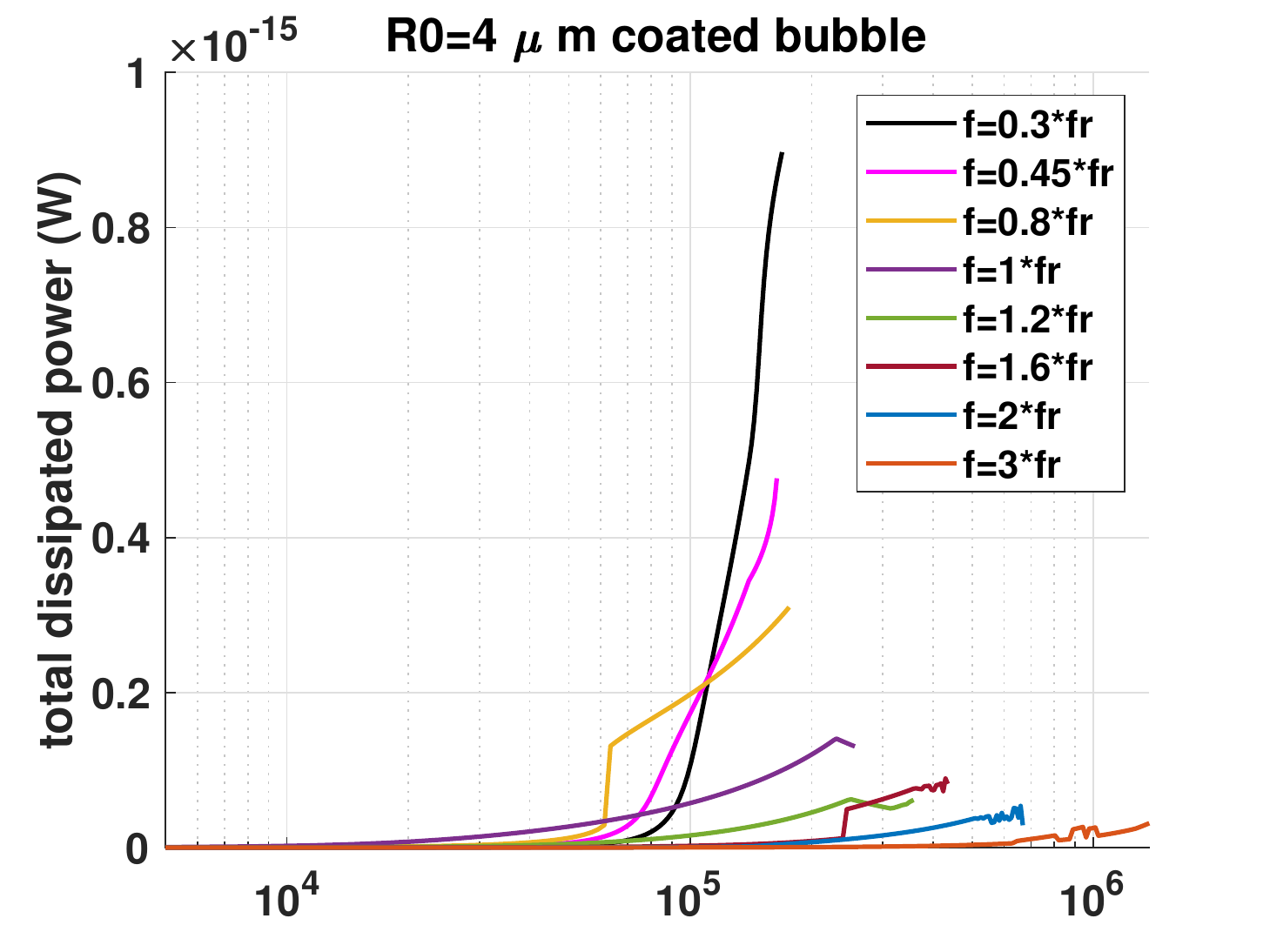}}\\
		\hspace{0.5cm} (c) \hspace{6cm} (d)\\
		\scalebox{0.43}{\includegraphics{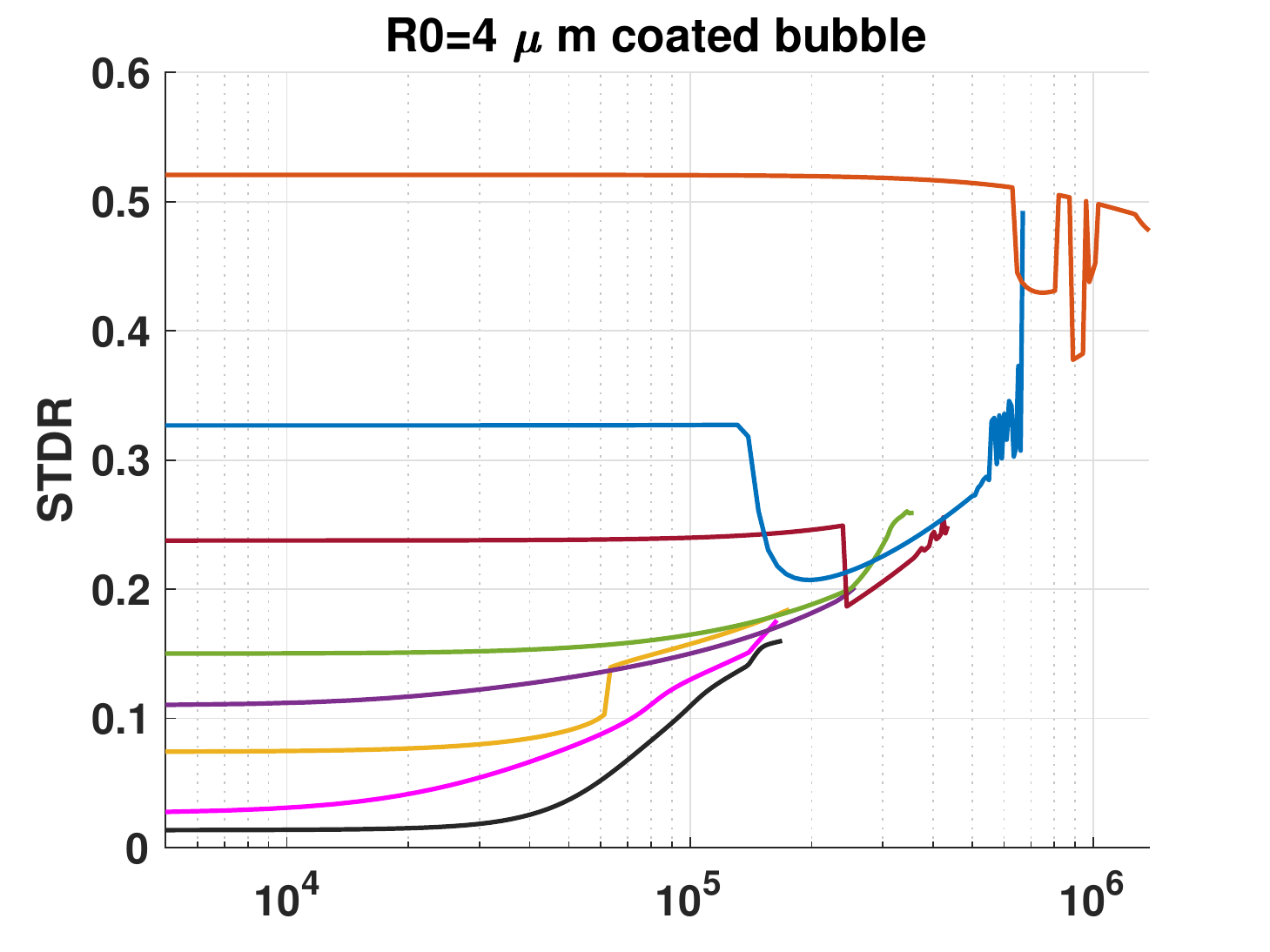}}\\
		\hspace{0.5cm} (e)\\
		\caption{Nondestructive ($R/R_0\leq 2$) values of: a) $|\dot{R(t)}|_{max}$ ($V_m$), b) Maximum backscattered pressure ($|P_{sc}|_{max}$ ($P_m$), c) Rd, d) $W_{total}$ and e) STDR as a function of pressure in the oscillations of a coated C3F8 bubble with $R_0=4 \mu m$.}
	\end{center}
\end{figure*}
The case of the C3F8 coated bubble with $R_0=4 \mu m$ is shown in Fig. A.6. The same conclusions can be drawn as the two previous cases (Figs. 5, 6 and A.5). This indicates a universal behavior of these parameters in the studied cases in this paper.
\end{document}